\documentclass[twocolumn]{aastex631}

\usepackage{amsmath,amstext}
\usepackage[T1]{fontenc}
\usepackage{multirow}
\usepackage[utf8]{inputenc} 
\usepackage{float}
\usepackage{graphicx}
\newcommand{\unit}[1]{\ensuremath{\, \mathrm{#1}}}
\usepackage{tabularx} 

\usepackage{threeparttable}

\shorttitle{TOI-1266 TTVs}
\shortauthors{Greklek-McKeon et al.}

\begin{document}

\title{Tidally Heated Sub-Neptunes, Refined Planetary Compositions, and Confirmation of a Third Planet in the TOI-1266 System}

\correspondingauthor{Michael Greklek-McKeon}
\email{michael@caltech.edu}

\author[0000-0002-0371-1647]{Michael Greklek-McKeon}
\affiliation{Division of Geological and Planetary Sciences, California Institute of Technology, Pasadena, CA 91125, USA}

\author[0000-0003-2527-1475]{Shreyas Vissapragada}
\affiliation{Carnegie Science Observatories, Pasadena, CA 91101, USA}

\author[0000-0002-5375-4725]{Heather A. Knutson}
\affiliation{Division of Geological and Planetary Sciences, California Institute of Technology, Pasadena, CA 91125, USA}

\author[0000-0002-4909-5763]{Akihiko Fukui}
\affiliation{Komaba Institute for Science, The University of Tokyo, Tokyo 153-8902, Japan}
\affiliation{Instituto de Astrof\'{i}sica de Canarias (IAC), 38205 La Laguna, Tenerife, Spain}

\author[0000-0001-9518-9691]{Morgan Saidel}
\affiliation{Division of Geological and Planetary Sciences, California Institute of Technology, Pasadena, CA 91125, USA}

\author[0000-0002-0672-9658]{Jonathan Gomez Barrientos}
\affiliation{Division of Geological and Planetary Sciences, California Institute of Technology, Pasadena, CA 91125, USA}

\author[0000-0002-1422-4430]{W. Garrett Levine}
\affiliation{Department of Astronomy, Yale University, New Haven, CT 06511, USA}

\author[0000-0003-0012-9093]{Aida Behmard}
\affiliation{American Museum of Natural History, 200 Central Park West, New York, NY 10024}

\author[0000-0002-7094-7908]{Konstantin Batygin}
\affiliation{Division of Geological and Planetary Sciences, California Institute of Technology, Pasadena, CA 91125, USA}

\author[0000-0003-1728-8269]{Yayaati Chachan}
\affiliation{Department of Physics, McGill University
3600 Rue University, Montreal, Quebec H3A 2T8, Canada}

\author[0000-0002-1871-6264]{Gautam Vasisht}
\affiliation{Jet Propulsion Laboratory, California Institute of Technology, 4800 Oak Grove Drive, Pasadena, CA 91109, USA}

\author[0000-0003-2215-8485]{Renyu Hu}
\affiliation{Jet Propulsion Laboratory, California Institute of Technology, 4800 Oak Grove Drive, Pasadena, CA 91109, USA}
\affiliation{Division of Geological and Planetary Sciences, California Institute of Technology, Pasadena, CA 91125, USA}

\author[0000-0001-5383-9393]{Ryan Cloutier}
\affiliation{Department of Physics \& Astronomy, McMaster University, 1280 Main St West, Hamilton, ON, L8S 4L8, Canada}

\author[0000-0001-9911-7388]{David Latham}
\affiliation{Center for Astrophysics | Harvard \& Smithsonian, 60 Garden St, Cambridge, MA 02138}

\author[0000-0003-3204-8183]{Mercedes L\'opez-Morales}
\affiliation{Space Telescope Science Institute, 3700 San Martin Drive, Baltimore, MD 21218, USA}

\author[0000-0001-7246-5438]{Andrew Vanderburg}
\affiliation{Department of Physics and Kavli Institute for Astrophysics and Space Research, Massachusetts Institute of Technology, 77 Massachusetts Ave, Cambridge, MA 02139, USA}

\author{Carolyn Heffner}
\affiliation{Division of Physics, Mathematics \& Astronomy, California Institute of Technology, Pasadena, CA 91125, USA}

\author{Paul Nied}
\affiliation{Division of Physics, Mathematics \& Astronomy, California Institute of Technology, Pasadena, CA 91125, USA}

\author{Jennifer Milburn}
\affiliation{Division of Physics, Mathematics \& Astronomy, California Institute of Technology, Pasadena, CA 91125, USA}

\author{Isaac Wilson}
\affiliation{Division of Physics, Mathematics \& Astronomy, California Institute of Technology, Pasadena, CA 91125, USA}

\author{Diana Roderick}
\affiliation{Division of Physics, Mathematics \& Astronomy, California Institute of Technology, Pasadena, CA 91125, USA}

\author{Kathleen Koviak}
\affiliation{Division of Physics, Mathematics \& Astronomy, California Institute of Technology, Pasadena, CA 91125, USA}

\author{Tom Barlow}
\affiliation{Division of Physics, Mathematics \& Astronomy, California Institute of Technology, Pasadena, CA 91125, USA}

\author{John F. Stone}
\affiliation{Division of Physics, Mathematics \& Astronomy, California Institute of Technology, Pasadena, CA 91125, USA}

\author[0000-0003-2102-3159]{Rocio Kiman}
\affiliation{Division of Physics, Mathematics \& Astronomy, California Institute of Technology, Pasadena, CA 91125, USA}

\author[0000-0002-0076-6239]{Judith Korth}
\affiliation{Lund Observatory, Division of Astrophysics, Department of Physics, Lund University, Box 118, 22100 Lund, Sweden}

\author[0000-0002-6424-3410]{Jerome P. de Leon}
\affiliation{Department of Multi-Disciplinary Sciences, Graduate School of Arts and Sciences, The University of Tokyo, 3-8-1 Komaba, Meguro, Tokyo 153-8902, Japan}

\author[0000-0002-9436-2891]{Izuru Fukuda}
\affiliation{Department of Multi-Disciplinary Sciences, Graduate School of Arts and Sciences, The University of Tokyo, 3-8-1 Komaba, Meguro, Tokyo 153-8902, Japan}

\author[0000-0001-8877-0242]{Yuya Hayashi}
\affiliation{Department of Multi-Disciplinary Sciences, Graduate School of Arts and Sciences, The University of Tokyo, 3-8-1 Komaba, Meguro, Tokyo 153-8902, Japan}

\author[0000-0002-5658-5971]{Masahiro Ikoma}
\affiliation{Division of Science, National Astronomical Observatory of Japan, 2-21-1 Osawa, Mitaka, Tokyo 181-8588, Japan}

\author[0000-0002-5978-057X]{Kai Ikuta}
\affiliation{Department of Multi-Disciplinary Sciences, Graduate School of Arts and Sciences, The University of Tokyo, 3-8-1 Komaba, Meguro, Tokyo 153-8902, Japan}

\author[0000-0002-6480-3799]{Keisuke Isogai}
\affiliation{Okayama Observatory, Kyoto University, 3037-5 Honjo, Kamogatacho, Asakuchi, Okayama 719-0232, Japan}
\affiliation{Department of Multi-Disciplinary Sciences, Graduate School of Arts and Sciences, The University of Tokyo, 3-8-1 Komaba, Meguro, Tokyo 153-8902, Japan}

\author[0000-0002-0488-6297]{Yugo Kawai}
\affiliation{Department of Multi-Disciplinary Sciences, Graduate School of Arts and Sciences, The University of Tokyo, 3-8-1 Komaba, Meguro, Tokyo 153-8902, Japan}

\author[0000-0003-1205-5108]{Kiyoe Kawauchi}
\affiliation{Department of Physical Sciences, Ritsumeikan University, Kusatsu, Shiga 525-8577, Japan}

\author[0000-0001-9194-1268]{Nobuhiko Kusakabe}
\affiliation{Astrobiology Center, 2-21-1 Osawa, Mitaka, Tokyo 181-8588, Japan}
\affiliation{National Astronomical Observatory of Japan, 2-21-1 Osawa, Mitaka, Tokyo 181-8588, Japan}

\author[0000-0002-4881-3620]{John H. Livingston}
\affiliation{Astrobiology Center, 2-21-1 Osawa, Mitaka, Tokyo 181-8588, Japan}
\affiliation{National Astronomical Observatory of Japan, 2-21-1 Osawa, Mitaka, Tokyo 181-8588, Japan}
\affiliation{Astronomical Science Program, Graduate University for Advanced Studies, SOKENDAI, 2-21-1, Osawa, Mitaka, Tokyo, 181-8588, Japan}

\author[0000-0003-1368-6593]{Mayuko Mori}
\affiliation{Astrobiology Center, 2-21-1 Osawa, Mitaka, Tokyo 181-8588, Japan}
\affiliation{National Astronomical Observatory of Japan, 2-21-1 Osawa, Mitaka, Tokyo 181-8588, Japan}

\author[0000-0001-8511-2981]{Norio Narita}
\affiliation{Komaba Institute for Science, The University of Tokyo, 3-8-1 Komaba, Meguro, Tokyo 153-8902, Japan}
\affiliation{Astrobiology Center, 2-21-1 Osawa, Mitaka, Tokyo 181-8588, Japan}
\affiliation{Instituto de Astrof\'{i}sica de Canarias (IAC), 38205 La Laguna, Tenerife, Spain}

\author[0000-0002-6510-0681]{Motohide Tamura}
\affiliation{Department of Astronomy, Graduate School of Science, The University of Tokyo, 7-3-1, Hongo, Bunkyo-ku, Tokyo, 113-0033, Japan}
\affiliation{Astrobiology Center, NINS, 2-21-1 Osawa, Mitaka, Tokyo 181-8588, Japan}
\affiliation{National Astronomical Observatory of Japan, NINS, 2-21-1 Osawa, Mitaka, Tokyo 181-8588, Japan}

\author[0000-0002-7522-8195]{Noriharu Watanabe}
\affiliation{Department of Multi-Disciplinary Sciences, Graduate School of Arts and Sciences, The University of Tokyo, 3-8-1 Komaba, Meguro, Tokyo 153-8902, Japan}

\author[0000-0003-0597-7809]{Gareb Fern\'andez-Rodr\'iguez}
\affiliation{Instituto de Astrof\'isica de Canarias (IAC), E-38200 La Laguna, Tenerife, Spain\label{IAC}}
\affiliation{Departamento de Astrof\'isica, Universidad de La Laguna (ULL), E-38206 La Laguna, Tenerife, Spain\label{ull}}

\begin{abstract}

TOI-1266 is a benchmark system of two temperate ($<$ 450 K) sub-Neptune-sized planets orbiting a nearby M dwarf exhibiting a rare inverted architecture with a larger interior planet. In this study, we characterize transit timing variations (TTVs) in the TOI-1266 system using high-precision ground-based follow-up and new TESS data. We confirm the presence of a third exterior non-transiting planet, TOI-1266 d (P = 32.5 d, $M_d$ = 3.68$^{+1.05}_{-1.11} M_{\oplus}$), and combine the TTVs with archival radial velocity (RV) measurements to improve our knowledge of the planetary masses and radii. We find that, consistent with previous studies, TOI-1266 b ($R_b$ = 2.52 $\pm$ 0.08 $R_{\oplus}$, $M_b$ = 4.46 $\pm$ 0.69 $M_{\oplus}$) has a low bulk density requiring the presence of a hydrogen-rich envelope, while TOI-1266 c ($R_c$ = 1.98 $\pm$ 0.10 $R_{\oplus}$, $M_c$ = 3.17 $\pm$ 0.76 $M_{\oplus}$) has a higher bulk density that can be matched by either a hydrogen-rich or water-rich envelope. Our new dynamical model reveals that this system is arranged in a rare configuration with the inner and outer planets located near the 3:1 period ratio with a non-resonant planet in between them. Our dynamical fits indicate that the inner and outer planet have significantly nonzero eccentricities ($e_b + e_d = 0.076^{+0.029}_{-0.019}$), suggesting that TOI-1266 b may have an inflated envelope due to tidal heating. Finally, we explore the corresponding implications for the formation and long-term evolution of the system, which contains two of the most favorable cool ($<$ 500 K) sub-Neptunes for atmospheric characterization with JWST. 

\end{abstract}

\section{Introduction} \label{sec:intro}

Planets transiting nearby M dwarf stars are the best laboratory we have to characterize the properties of sub-Neptune-sized exoplanets. Their relatively large planet-star radius and mass ratios make M dwarf planets amenable to high precision bulk density measurements, and these systems are among the most favorable targets for atmospheric studies. However, the atmospheric compositions of sub-Neptune-sized planets orbiting M dwarfs on close-in orbits may be systematically enriched in water and depleted in hydrogen relative to their counterparts orbiting Sun-like stars.
 
Since the ice lines for M dwarfs are much closer in than for Sun-like stars, close-in sub-Neptune-sized planets orbiting these stars might accrete more water than their counterparts orbiting hotter stars \citep[e.g.,][]{Alibert_bez_2017,ormel_2017,bitsch_2020,kimura_2022}. Even if small planets orbiting M dwarfs initially form with H-rich atmospheres, they may be less likely to retain them even at relatively low equilibrium temperatures \citep[e.g.,][]{Hori_2020}. M dwarfs have much higher flare rates and higher fractional XUV fluxes than Sun-like stars, and this can drive increased atmospheric mass loss \citep{Roettenbacher_2017,Fleming_2020}. 

We can obtain initial constraints on the envelope compositions of sub-Neptune-sized planets by measuring their masses, radii, and corresponding bulk densities. Population-level studies of small planets orbiting M dwarfs \citep[e.g.,][]{Luque_2022} indicate that some planets have high bulk densities consistent with rocky Earth-like compositions, while the densities of others are so low that they can only be matched with a hydrogen-rich envelope, and some planets with intermediate densities are consistent with a wide range of possible envelope compositions \citep{Parc2024}, including water-rich steam atmospheres \citep{Aguichine2021,Luque_2022,Pierrehumbert2023} or more tenuous hydrogen-rich envelopes \citep{rogers_2023}. Some studies have used mass loss models to rule out solar metallicity gas envelopes and provide evidence in favor of water-rich compositions for individual sub-Neptunes \citep[e.g.,][]{Diamond-Lowe2022,Piaulet2023,Castro-Gonzalez2023}. But if  sub-Neptunes have moderately enriched atmospheric metallicities \citep[$\geq50-200\times$ solar;][]{Zhang2022} making atmospheric mass loss much less efficient \citep{Linssen2024}, it is more difficult to rule out the presence of H/He envelopes and break degeneracies in planet composition type. 

We can directly characterize the atmospheric compositions of sub-Neptune-sized planets using transmission spectroscopy, but most warm ($T_{eq}$ $\in$ 500 -- 1000~K) sub-Neptunes observed to date appear to host thick photochemical hazes that obscure the expected molecular absorption features \citep{Wallack2024,Gao2023}. The recent JWST detections of strong atmospheric absorption from TOI-270d and K2-18b \citep{Madhusudhan2023,Benneke2024}, both of which have equilibrium temperatures lower than 400 K, suggests that these obscuring hazes may disappear in cooler atmospheres \cite[e.g.,][]{Brande2024}. Both TOI-270d and K2-18b have hydrogen-rich atmospheres that contain significant quantities of CH$_4$, CO$_2$ and H$_2$O \citep[correspnding to atmospheric metallicities of $\sim100-200\times$ solar;][]{Madhusudhan2023,Wogan2024,Benneke2024}, or the atmosphere of K2-18 b could also be explained with C and N abundances $\sim10\times$ solar and a high H$_2$O/H$_2$ ratio \citep{Yang2024}. Recent JWST observations of the temperate super-Earth LHS 1140 b \citep[$T_{eq}$ = 226 K, $R_p$ = 1.73 $R_{\oplus}$,][]{Cadieux2024} ruled out the presence of an H$_2$-rich envelope, indicating that the planet is likely water rich \citep{Damiano2024}. 

It is unclear whether or not these few planets are representative of the true diversity in the M dwarf sub-Neptune population. Unfortunately, there are currently only a handful of temperate sub-Neptune-sized planets orbiting M dwarfs with well-measured masses and radii that are also favorable targets for atmospheric characterization with JWST. As a result, there are only seven planets with radii between $2-3$ $R_{\oplus}$ and $T_{eq}<500$~K with scheduled JWST observations (TOI-270 d, LP 791-18 c, LTT 3780 c, K2-18 b, TOI-776 c, TOI-1468 c, and TOI-4336 b). The predicted amplitude of the atmospheric absorption signal during transit is inversely proportional to the planet's surface gravity, and well-constrained mass and radius measurements are therefore an essential requirement for JWST observation planning \citep[e.g.,][]{Batalha2017, Batalha2019,DiMaio2023}. For sub-Neptunes, bulk density measurements can also be used in conjunction with transmission spectroscopy to constrain the composition of the planet's envelope \citep[e.g.,][]{Benneke2024,PiauletGhorayeb2024}.

Measuring masses of temperate sub-Neptunes with radial velocities (RVs) can be challenging because the host stars are faint and/or active, and low-mass planets with long orbital periods have correspondingly small RV semi-amplitudes. For planets in dynamically interacting multi-planet systems, the masses can be measured with transit timing variations  \citep[TTVs, e.g.][]{GreklekMcKeon2023}. The size of the TTV signal is maximized in systems where the planets have orbital period ratios close to mean-motion resonance \citep[MMR;][]{Lithwick_2012}, and in systems with larger planet-star mass ratios \citep[e.g., M dwarfs;][]{Agol_2021}. The Transiting Exoplanet Survey Satellite \citep[TESS,][]{Ricker2014} has detected many such systems orbiting M dwarfs and obtained an initial set of transit timing measurements. However, the precision of TTV-based mass estimates depends on the orbital architecture of the system, the sampling of the TTV oscillations, and on our ability to obtain precise transit mid-times. Because of this, planets that are observed at low SNR with TESS require additional high SNR transit observations at later epochs in order to obtain useful mass constraints \citep{ballard_2019}.  

In this study we focus on TOI-1266, a nearby ($d = 36$~pc) M3 star \citep[$T_{*} = 3563 \pm 77$~K,][]{Stefansson_2020} that hosts two sub-Neptune-sized transiting planets with equilibrium temperatures of 415~K and 346~ K, respectively \citep[][hereafter D20 and S20]{Demory_2020,Stefansson_2020}. Systems with multiple planets of different sizes and bulk densities such as TOI-1266 are uniquely advantageous for testing theories of planet formation and evolution \citep[e.g.,][]{Diamond-Lowe2022,Piaulet2023} because the planets formed from the same protoplanetary disk and experienced the same irradiation history scaled to their orbital separations. We can use these multi-planet systems to explore how small differences in initial conditions and irradiation environment can lead to different final planetary composition types. 

The TOI-1266 system has been previously studied in the literature, with two discovery papers characterizing the properties of the inner two transiting planets (D20 and S20), followed by a paper presenting detailed atmospheric modeling for TOI-1266 c \citep{Harman_2022}. This system was also the subject of a separate RV analysis that measured the masses of both transiting planets and reported tentative evidence for a third non-transiting planet candidate \citep[][hereafter C24]{Cloutier_2024}. These three studies find that the inner two TOI-1266 planets ($R_b$ = 2.59 $\pm$ 0.10 $R_{\oplus}$, $R_c$ = 2.04 $\pm$ 0.11 $R_{\oplus}$, $M_b$ = 4.40$^{+0.68}_{-0.70}$ $M_{\oplus}$, $M_c$ = 3.12 $\pm$ 0.76 $M_{\oplus}$, C24) display an ‘inverted’ architecture with a larger interior planet. Most multi-planet systems that span the radius valley have the smaller super-Earth on an interior orbit to the larger sub-Neptune \citep[e.g.,][]{Weiss2018}. The TOI-1266 system therefore runs counter to the canonical narrative that it should be easier for more distant planets to retain larger primordial hydrogen-rich envelopes, even when accounting for the lower mass of TOI-1266 c (C24). Notably, TOI-1266 b has a higher Transmission Spectroscopy Metric \citep[TSM; this indicates the planet's relative favorability for atmospheric characterization with transmission spectroscopy, see][C24, and Table \ref{tab:results}]{Kempton2018} than all currently selected JWST sub-Neptune ($2-3$ $R_\oplus$) targets cooler than 500~K aside from L98-59 d.

The two transiting planets in the TOI-1266 system have period ratios that are 3.5\% away from the 5:3 orbital resonance and 1.3\% away from the 7:4 orbital resonance, making this system a favorable target for TTV mass measurements. \cite{Demory_2020} reported a preliminary detection of a TTV signal from a combination of TESS and ground-based transit observations, suggesting that additional transit timing variations might provide useful constraints on the planet masses. Recent work has revealed that there is a systematic, population-level discrepancy between the densities of sub-Neptune planets whose masses come from TTV measurements versus RV measurements, which is consistent with near-resonant planets being puffier and can be reproduced in the RV sample alone \citep{Leleu2024}. Rare systems such as TOI-1266 which contain both near-resonant and non-resonant planets, and separate RV and TTV mass measurements, enable detailed investigation of this trend.

In this paper, we extend the TTV baseline by 730 days with new space- and ground-based observations, and carry out a joint TTV and RV analysis of TESS photometry including new data from the second extended mission, new ground-based photometric follow-up from the Wide-field InfraRed Camera (WIRC) at Palomar Observatory, the Multicolor Simultaneous Camera for studying Atmospheres of Transiting exoplanets (MuSCAT) instrument series at the National Astronomical Observatory of Japan (NAOJ) and the Las Cumbres Observatory Global Telescope network (LCOGT), a Sinistro imager at LCOGT, and archival RV measurements from the High Accuracy Radial velocity Planet Searcher for the Northern hemisphere (HARPS-N). 

In \S\ref{sec:Observations}, we describe our observations. In \S\ref{sec:Stellar Characterization}, we summarize information on the host star, including new abundance analysis. In \S\ref{sec:transit analysis}, we present our transit analysis. In \S\ref{sec:ttv modeling}, we present our TTV analysis, and in \S\ref{sec:joint modeling} we present our RV and joint RV+TTV analysis. In \S\ref{sec:dynamical analysis} we present our analysis on the dynamics of the system. In \S\ref{sec:planetary compositions} we present results on the planetary compositions. In \S\ref{sec:discussion} we discuss our results. In \S\ref{sec:future obs} we discuss the prospects of future observations before concluding with a summary of our key findings in \S\ref{sec:summary}.

\section{Observations} \label{sec:Observations}

\subsection{TESS}

TOI-1266 has been observed by TESS during the prime mission (PM), extended mission (EM), and second extended mission (SEM). D20 and S20 reported on the discovery of TOI-1266 b and c using PM data from TESS sectors 14, 15, 21, and 22, from UT 2019 July 18 to 2020 March 17. Follow-up work in C24 improved the characterization of the planetary radii with TESS EM data from sectors 41, 48, and 49 between from UT 2021 July 23 to 2022 March 06. In this study, we extend the analysis to SEM data in TESS sectors 75 and 76 between UT 2024 Jan 30 and 2024 March 26.

The TOI-1266 b and c discovery papers (D20 and S20) use the TESS 2-min Presearch Data Conditioning Simple Aperture Photometry (PDCSAP) light curves. C24 performs a more in-depth analysis of the TESS PM and EM data, including PDCSAP light curves and custom light curve extractions from the 2-min cadence TESS Target Pixel Files (TPFs) and the 30- and 10-min cadence Full Frame Images (FFIs). This comprehensive multi-pipeline analysis of the TESS data by C24 was motivated by significant differences in the transit depths of TOI-1266 b and c between the TESS PM and EM data. C24 finds these transit depth discrepancies to be consistent across different photometry sources. We use the 2-min cadence PDCSAP light curves across all TESS sectors. 

\begin{table*}
\caption{Summary of ground-based observations of TOI-1266 b and c.}
\label{tab:Palomar obs}
\centering
\begin{tabular}{ccccccccccc}
\hline
\hline
Instrument & Planet & UT Date & Start & Finish & Transit \% & Baseline \% & z$_{\text{st}}$ &  z$_{\text{min}}$ & z$_{\text{end}}$ & $t_{\text{exp}} (s)$ 
\\ \hline
WIRC & TOI-1266 b & 2022 Feb 28 & 04:37:11 & 11:05:29 & 100\% & 100\% & 1.94 & 1.18 & 1.19 & 15
\\ %\hline
WIRC & TOI-1266 b & 2022 June 6 & 06:04:42 & 10:41:17 & 100\% & 120\% & 1.26 & 1.22 & 2.26 & 12.5 
\\ %\hline
MuSCAT & TOI-1266 b & 2021 April 7 & 10:32:47 & 14:08:56 & 100\% & 80\% & 1.60 & 1.19 & 1.19 & 25, 12, 12
\\
MuSCAT3 & TOI-1266 b & 2021 April 18 & 07:47:49 & 11:12:35 & 100\% & 70\% & 1.54 & 1.41 & 1.46 & 8, 12, 20, 12 
\\
MuSCAT3 & TOI-1266 b & 2022 Jan 26 & 13:35:60 & 15:31:47 & 20\% & 70\% & 1.49 & 1.41 & 1.41 & 8, 12, 20, 12 
\\
MuSCAT3 & TOI-1266 b & 2022 Feb 28 & 08:12:24 & 11:08:16 & 70\% & 80\% & 2.34 & 1.53 & 1.53 & 8, 12, 20, 12 
\\ \hline
MuSCAT3 & TOI-1266 c & 2021 May 13 & 09:32:09 & 12:57:38 & 80\% & 80\% & 1.46 & 1.46 & 2.23 & 8, 12, 20, 12 
\\
MuSCAT & TOI-1266 c & 2022 Jan 31 & 14:07:23 & 20:05:53 & 100\% & $>$200\% & 1.79 & 1.17 & 1.17 & 25, 18, 17 
\\
Sinistro & TOI-1266 c & 2022 March 10 & 06:36:24 & 10:02:06 & 100\% & 90\% & 1.33 & 1.22 & 1.24 & 40\
\\
WIRC & TOI-1266 c & 2022 March 10 & 04:58:26 & 10:51:59 & 100\% & $>$200\% & 1.65 & 1.20 & 1.18 & 12.5 
\\ 
\hline
\end{tabular}
\begin{tablenotes}
      \item[a] Start and Finish columns represent the time of first and last science images in UT time, the transit and baseline fractions are relative to the total transit duration for each planet, z$_{\text{min}}$ is the minimum airmass of the science sequence while z$_{\text{st}}$ and z$_{\text{end}}$ are the starting and ending airmasses, and $t_{\text{exp}}$ is the exposure time used, which is varied on the night of observation based on the sky conditions (Moon fraction and proximity, cloud cover, etc.). The $t_{\text{exp}}$ values for MuSCAT represent the values for the $g$, $r$, and $z_s$ bands, and those for MuSCAT3 represent the values for the $g$, $r$, $i$, and $z_s$ bands, respectively.
\end{tablenotes}
\end{table*}

\subsection{Palomar/WIRC}

We observed three transits each of TOI-1226 b and c in the $J$-band with the Wide-field Infared Camera (WIRC) on the Hale Telescope at Palomar Observatory, California, USA. The Hale Telescope is a 5.08-m telescope equipped with a 2048 x 2048 Rockwell Hawaii-II NIR detector, providing a field of view of 8\farcm7 × 8\farcm7 with a plate scale of 0.''25 per pixel \citep[WIRC,][]{Wilson2003}. Our data were taken with a beam-shaping diffuser that increased our observing efficiency and improved the photometric precision and guiding stability \citep{Stefansson2017,Vissapragada2020}. 

We observed transits of TOI-1266 b on UT February 28, 2022, June 6 2022, and April 7 2023. Transits of TOI-1266 c were observed on UT March 10 2022, June 12 2022, and December 9 2023. The full details of our WIRC observations, including observation dates, exposure times, airmass values, and transit and baseline coverage fractions are included in Table \ref{tab:Palomar obs}. For each night, we obtained calibration images to dark-subtract, flat-field, remove dead and hot pixels, and remove detector structure with a 9-point dithered sky background frame following the methodology of \cite{Vissapragada2020}. We extracted photometry and detrended the light curves with the procedure described in \cite{GreklekMcKeon2023}.

Our diffuser reshapes the stellar PSFs into a top-hat shape with a 3\farcs0 full width at half maximum (FWHM), but for our UT April 7 2023 observations of TOI-1266 b, conditions were cloudy and the seeing was intermittently worse than 3\farcs0, allowing the PSF to spill out beyond the diffuser bounds and introduce a more complicated scintillation noise structure. Since most comparison stars in the field of view on this night also had PSFs interior or exterior to the diffuser bounds at various points, our photometric precision in the final light curve was significantly lower than other nights, and we choose not to include this night of observation in our analysis. We also chose not to include the UT June 12 2022, and December 9 2023 observations in our analysis of TOI-1266 c. Dome closures due to clouds and humidity on these nights meant our coverage was insufficient to provide baseline on both sides of transit, and therefore the precision on our extracted midtimes suffered. 

\subsection{NAOJ 188cm/MuSCAT}

We observed one full transit each of TOI-1266\,b and TOI-1266\,c on UT 2021 April 7 and 2022 January 31, respectively, with the Multiband Simultaneous Camera for studying Atmospheres of Transiting exoplanets \citep[MuSCAT;][]{2015JATIS...1d5001N} mounted on the NAOJ 188\,cm telescope located in Okayama, Japan. MuSCAT has three optical channels for the $g$, $r$, and $z_s$ bands, each having a 1024 $\times$ 1024 pixel CCD camera with a pixel scale of 0\farcs36 and an FoV of 6\farcm1 $\times$ 6\farcm1. Both observations were conducted with the telescope slightly defocused so that the FWHM of the stellar PSF was 8--12 pixels. The exposure times were set at 12--25 s depending on the band and the night.

The raw images were calibrated for dark and flat-field in a standard manner. After that, aperture photometry was performed for the target and several comparison stars to produce relative light curves using a custom pipeline described in \citet{2011PASJ...63..287F}.

\subsection{LCOGT/MuSCAT3}

We observed four (one full and three partial) transits of TOI-1266\,b and two (both partial) transits of TOI-1266\,c with MuSCAT3 \citep{2020SPIE11447E..5KN}, which is mounted on the 2\,m Faulkes Telescope North (FTN) operated by Las Cumbres Observatory Global Telescope \citep[LCOGT,][]{Brown2013}. MuSCAT3 is similar to MuSCAT but has four optical channels for the $g$, $r$, $i$, and $z_s$ bands, each with a 2048 $\times$ 2048 pixel CCD camera with a pixel scale of 0\farcs27 and an FoV of 9\farcm1 $\times$ 9\farcm1. The observations of TOI-1266\,b were conducted on the night of UT 2021 April 18, 2022 January 26, 2022 February 28, and 2022 May 26, while the observations of TOI-1266\,c were conducted on 2021 May 13 and 2022 May 24. During the observations the telescope was defocused, which results in FWHM of stellar PSF of 10--20 pixels depending on the night. The exposure times were set at 8, 12, 20, and 12 s for the $g$, $r$, $i$, and $z_s$ bands, respectively, for all nights.

The raw images were calibrated using the {\tt BANZAI} pipeline \citep{McCully:2018}, and aperture photometry was performed in the same way as for the MuSCAT data. Due to poor photometric quality or insufficient baseline coverage, we decided to omit data from 2022 May 24 and 2022 May 26 from the subsequent analysis.

\subsection{LCOGT/Sinistro}

We observed one full transit of TOI-1266\,c on UT 2022 March 10 with the Sinistro imager on one of the 1\,m telescopes of LCOGT at the McDonald observatory in TX, US. Sinistro is equipped with a 4K $\times$ 4K pixel CCD with a pixel scale of 0\farcs39 and an FoV of 26\farcm5 $\times$ 26\farcm5. The observation was conducted through the $i$ band filter with an exposure time of 40\,s and moderate defocusing. The obtained data were reduced in the same way as the MuSCAT3 data.

\section{Stellar Characterization} \label{sec:Stellar Characterization}

TOI-1266 is a single star located at a distance of 36 pc (C24) that has been characterized extensively with several different spectroscopic instruments and techniques. D20 used multiple methods to independently derive stellar parameters, including the pseudo equivalent-width method described in \cite{Maldonado2015} for TRES spectra, the SpecMatch-Empirical method of \cite{Yee2017} for HIRES spectra, and the broadband spectral energy distribution (SED) together with the Gaia DR2 parallax \citep{Stassun2016,Stassun2017,Stassun2018}. D20 found consistent stellar parameters across all three methods, and conclude based on a HIRES RV measurement and the proper motion and parallax from Gaia DR2 that TOI-1266 has a $\sim$96\% probability of belonging to the galactic thin disc population. D20 also concluded based on TESS prime mission data, both from the PDC pipeline and their own custom reduction from the TESS Target Pixel Files, that TOI-1266 does not exhibit any rotational modulation or flares, consistent with an old, slightly metal-poor early M dwarf with an unspotted photosphere and negligible chromospheric activity. 

\begin{deluxetable*}{cccc}
    \tablecaption{Summary of stellar parameters. \label{tab:stellarparam}}
    \tabletypesize{\scriptsize}
    \tablehead{\colhead{~~~Parameter}                                 &  \colhead{Description}                                            & \colhead{Value}                         & \colhead{Reference}}
    \startdata
    \multicolumn{4}{l}{\hspace{-0.2cm} Main identifiers:}                                                                                                                                                     \\
    TIC                                                               &  -                                                                & 467179528                               & TIC                     \\
    TOI                                                               &  -                                                                & 1266                                    & TIC                     \\
    2MASS                                                             &  -                                                                & J13115955+6550017                       & TIC                     \\
    Gaia DR3                                                          &  -                                                                & 1678074272650459008                     & Gaia                    \\
    \multicolumn{4}{l}{\hspace{-0.2cm} Equatorial Coordinates, Proper Motion and Spectral Type:}           \\
    $\alpha_{\mathrm{J2000}}$                                         &  Right Ascension (RA)                                             & 13:11:59.18                             & Gaia                    \\
    $\delta_{\mathrm{J2000}}$                                         &  Declination (Dec)                                                & +65:50:01.31                            & Gaia                    \\
    $\mu_{\alpha}$                                                    &  Proper motion (RA, \unit{mas\ yr^{-1}})                          & $-150.652 \pm 0.041$                    & Gaia                    \\
    $\mu_{\delta}$                                                    &  Proper motion (Dec, \unit{mas\ yr^{-1}})                         &  $-25.368 \pm 0.039$                    & Gaia                    \\
    Spectral Type                                                     &  -                                                                & M2                                      & S20                     \\
    \multicolumn{4}{l}{\hspace{-0.2cm} Magnitudes:}           \\
    $B$                                                               &  APASS Johnson B mag                                              & $14.578 \pm 0.048$                       & APASS                  \\
    $V$                                                               &  APASS Johnson V mag                                              & $12.941 \pm 0.049$                       & APASS                  \\
    $g^{\prime}$                                                      &  APASS Sloan $g^{\prime}$ mag                                     & $13.811 \pm 0.050$                       & APASS                  \\
    $r^{\prime}$                                                      &  APASS Sloan $r^{\prime}$ mag                                     & $12.297 \pm 0.070$                       & APASS                  \\
    $i^{\prime}$                                                      &  APASS Sloan $i^{\prime}$ mag                                     & $11.246 \pm 0.150$                       & APASS                  \\ 
    $T$                                                 &  TESS magnitude                                          & $11.040 \pm 0.007$                       & TIC                    \\
    $J$                                                               &  2MASS $J$ mag                                                    & $9.706 \pm 0.023$                        & 2MASS                  \\
    $H$                                                               &  2MASS $H$ mag                                                    & $9.065 \pm 0.030$                        & 2MASS                  \\
    $K_s$                                                             &  2MASS $K_s$ mag                                                  & $8.840 \pm 0.020$                        & 2MASS                  \\
    $W1$                                                           &  WISE1 mag                                                        & $8.715 \pm 0.022$                        & WISE                   \\
    $W2$                                                           &  WISE2 mag                                                        & $8.612 \pm 0.019$                        & WISE                   \\
    $W3$                                                           &  WISE3 mag                                                        & $8.504 \pm 0.024$                        & WISE                   \\
    $W4$                                                           &  WISE4 mag                                                        & $8.233 \pm 0.207$                        & WISE                   \\
    \multicolumn{4}{l}{\hspace{-0.2cm} Stellar Parameters$^a$:}           \\
    $T_{\mathrm{eff}}$                                                &  Effective temperature in \unit{K}                                & $3563 \pm 77$                            & S20                    \\
    $\log(g)$                                                         &  Surface gravity in cgs units                                     & $4.826_{-0.021}^{+0.020}$                & S20                    \\
    $M_*$                                                             &  Mass in $M_{\odot}$                                              & $0.437 \pm 0.021$                        & S20                    \\
    $R_*$                                                             &  Radius in $R_{\odot}$                                            & $0.4232_{-0.0079}^{+0.0077}$             & S20                    \\
    $\rho_*$                                                          &  Density in $\unit{g\:cm^{-3}}$                                   & $8.13_{-0.46}^{+0.47}$                   & S20                    \\
    Age                                                               &  Age in Gyrs                                                      & 4.6$^{+1.6}_{-1.2}$                      & This work                    \\
    $L_*$                                                             &  Luminosity in $L_\odot$                                          & $0.02629_{-0.00075}^{+0.00071}$          & S20                    \\
    $A_v$                                                             &  Visual extinction in mag                                         & $0.015_{-0.010}^{+0.011}$                & S20                    \\
    $d$                                                               &  Distance in pc                                                   & $36.011_{-0.030}^{+0.029}$               & Gaia, Bailer-Jones     \\
    $\pi$                                                             &  Parallax in mas                                                  & $27.769_{-0.022}^{+0.023}$               & Gaia                   \\
    \multicolumn{4}{l}{\hspace{-0.2cm} Other Stellar Parameters:}           \\
    $v \sin i_*$                                                      &  Stellar rotational velocity in \unit{km\ s^{-1}}                 & $<1.3$                                   & C24                    \\
    $P_{rot}$                                                         &  Stellar rotational period in \unit{days}                         & $44.6_{-0.8}^{+0.5}$                      & C24                    \\
    $logR'_{HK}$                                                      &  Chromospheric Ca II H and K flux ratio                           & $-5.5_{-0.44}^{+0.35}$                    & C24                    \\
    $RV$                                                              &  Absolute radial velocity in \unit{km\ s^{-1}} ($\gamma$)         & $-41.58 \pm 0.26$                        & C24                    \\
    $U$                                                              &  Galactic $U$ Velocity (km/s)                                     & $-5.8\pm0.2$                             & S20                    \\
    $V$                                                               &  Galactic $V$ Velocity (km/s)                                     & $-40.3\pm0.4$                            & S20                    \\
    $W$                                                               &  Galactic $W$ Velocity (km/s)                                     & $-27.9\pm0.6$                            & S20                    \\
    \multicolumn{4}{l}{\hspace{-0.2cm} Stellar Abundances$^b$:}           \\ 
    $\mathrm{[C/H]}$                                                  &  Carbon metallicity in dex                                        & $-0.132\pm0.130$                         & This work              \\
    $\mathrm{[N/H]}$                                                  &  Nitrogen metallicity in dex                                      & $-0.085\pm0.166$                         & This work              \\
    $\mathrm{[O/H]}$                                                  &  Oxygen metallicity in dex                                        & $-0.114\pm0.102$                         & This work              \\
    $\mathrm{[Mg/H]}$                                                 &  Magnesium metallicity in dex                                     & $-0.052\pm0.105$                         & This work              \\
    $\mathrm{[Al/H]}$                                                 &  Aluminum metallicity in dex                                      & $-0.048\pm0.134$                         & This work              \\
    $\mathrm{[Si/H]}$                                                 &  Silicon metallicity in dex                                       & $-0.104\pm0.101$                         & This work              \\
    $\mathrm{[Ca/H]}$                                                 &  Calcium metallicity in dex                                       & $-0.091\pm0.094$                         & This work              \\
    $\mathrm{[Ti/H]}$                                                 &  Titanium metallicity in dex                                      & $-0.019\pm0.167$                         & This work              \\
    $\mathrm{[Cr/H]}$                                                 &  Chromium metallicity in dex                                       & $-0.077\pm0.132$                         & This work              \\
    $\mathrm{[Fe/H]}$                                                 &  Iron metallicity in dex                                          & $-0.109\pm0.092$                         & This work              \\
    $\mathrm{[Ni/H]}$                                                 &  Nickel metallicity in dex                                        & $-0.105\pm0.096$                         & This work              \\
    \enddata
    \tablenotetext{}{Adapted from \cite{Stefansson_2020}. References are: TIC \citep{Stassun2018,Stassun_2019}, Gaia \citep{gaia2018}, APASS \citep{Henden2015APASS}, 2MASS/WISE \citep{Cutri2014}, Bailer-Jones \citep{Bailer-Jones2018}, S20 \citep{Stefansson_2020}, C24 \citep{Cloutier_2024}.}
    \tablenotetext{a}{Derived using Stellar SED and Isochrone Fits with Gaussian priors on stellar $T_{\mathrm{eff}}$, logg, and [Fe/H] from HPF spectroscopic analysis in \cite{Stefansson_2020}.}
    \tablenotetext{b}{Derived from SDSS-V/APOGEE spectrum using The Cannon (Behmard et al. in prep).}
\end{deluxetable*}

S20 used the empirical spectral matching algorithm described in \cite{Stefansson2020a} on HPF spectra to obtain stellar parameters, along with an independent analysis of the SED.
%they fit the SED with the EXOFASTv2 package (Eastman et al. 2019) using as inputs (a) the available literature photometry, (b) the Gaia distance from Bailer-Jones et al. (2018), and (c) the spectroscopic values discussed above as Gaussian priors. 
Stellar parameters from the two methods of S20 are consistent with D20, with a similar 97.2\% probability that TOI-1266 is a galactic thin disk member. S20 also reported a limit on the projected stellar rotational velocity of $v \sin i_* < 2$~km/s along with a lack of rotational modulation based on Lomb-Scargle periodograms of the TESS PM photometry, ground-based photometry from the All-Sky Automated Survey for SuperNovae \citep[ASAS-SN;][]{Kochanek2017}, and the Zwicky Transient Facility \citep[ZTF;][]{Masci2019}. S20 also do not detect any variability in the Ca II infrared triplet or differential line widths from their HPF spectra, as expected for an inactive star with a moderately long rotation period.

\begin{figure*}
\begin{center}
  \includegraphics[width=18cm]{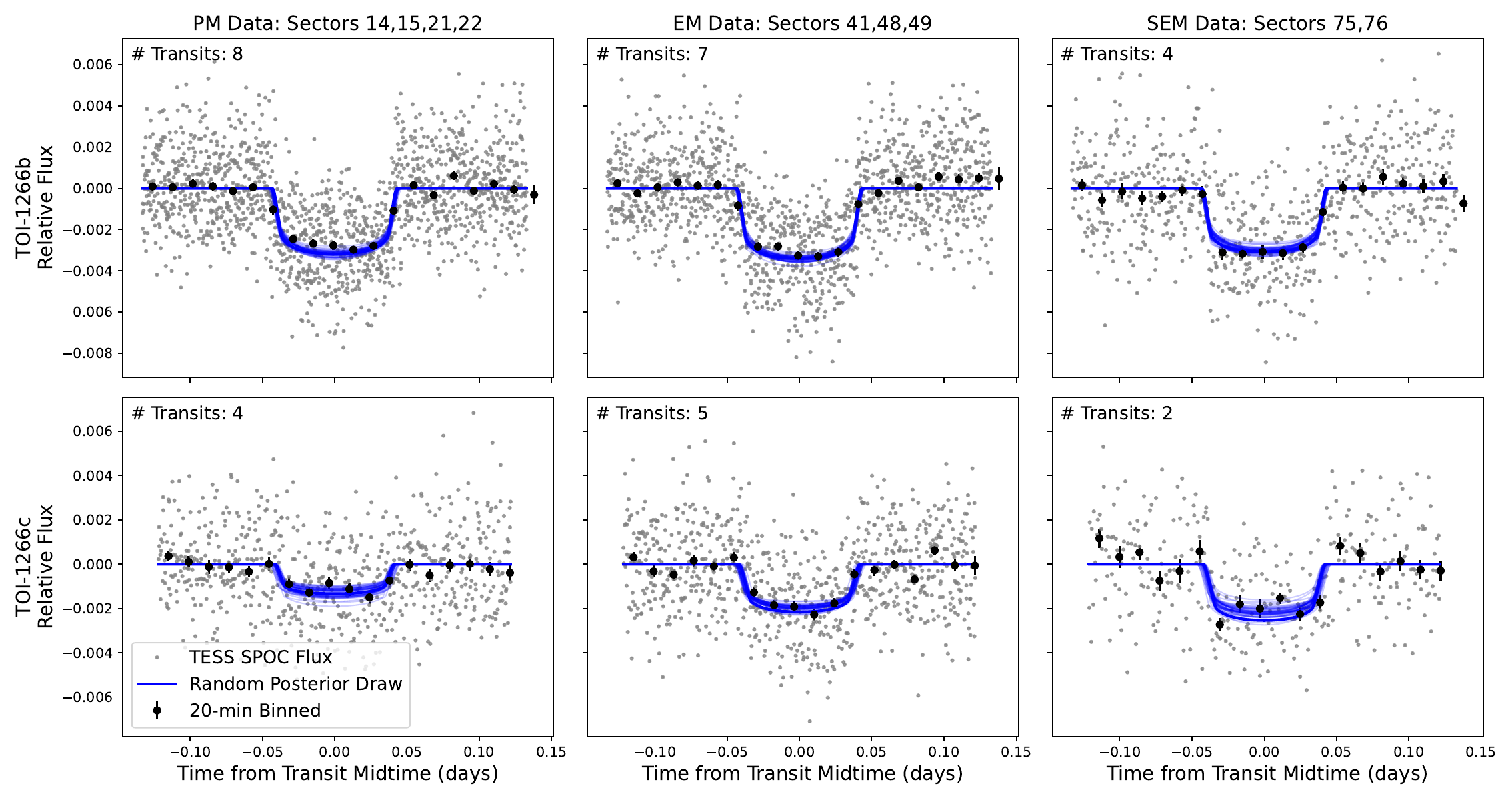}
  \caption{Stacked TESS light curves for TOI-1266 b (top row) and c (bottom row) for the prime mission (left column), extended mission (middle column), and second extended mission (right column). Individual transits are stacked by their best fit midtimes, with the total number of transits per mission noted, and 100 random draws from the transit model posteriors shown in blue.}
  \label{fig:TESS}
\end{center}
\end{figure*}

C24 is likewise unable to recover the stellar rotation period in the TESS PM or EM data. We extend the Lomb-Scargle periodogram search to the TESS PM, EM, and SEM data but find no significant peaks indicative of rotational modulation. However, C24 did find evidence for a stellar rotation period of $\sim$45 days in the spectroscopic HARPS-N time series, and carried out detailed activity modeling to show that this is consistent with bright plages and inconsistent with dark starspots. C24 also reported a log$R'_{HK}$ = -5.5$^{+0.35}_{-0.44}$, in good agreement with D20 and S20's previous conclusion that TOI-1266 is inactive. 

S20 reports the age of TOI-1266 as 7.9$^{+4.2}_{-5.2}$ Gyr based on a fit to the stellar spectral energy distribution. We update this age constraint using the stellar rotation period and log$R'_{HK}$ values from C24 with the M dwarf age-rotation and age-activity relations detailed in \cite{Engle_Guinan_2023} and \cite{Engle2024} assuming an M2 spectral type (S20). We find that the age estimates from the stellar rotation relation and log$R'_{HK}$ relation are consistent with each other within 1$\sigma$, but the rotation relation provides a more precise estimate of 4.6$^{+1.6}_{-1.2}$ Gyr, which we adopt in this work. Future measurement of the stellar X-ray luminosity would provide another independent constraint on the stellar age \citep{Engle2024} and might further improve the measurement (see \S$\ref{sec:future obs}$).

We adopt stellar parameters for our analysis from S20, which are consistent with previously reported values in  D20 and C24, and are based on spectroscopic analysis. D20 also reports stellar parameters from spectroscopic analysis but at lower precision than S20, while C24 adopts stellar parameters from the TESS Input Catalog v8.2 \citep[TIC;][]{Stassun_2019}, which are based on empirical scaling relations from color magnitudes. We comment on the effect of stellar parameter choice on planet properties in \S\ref{sec:planetary compositions}. 

Literature values for the metallicity of TOI-1266 include -0.5$\pm$0.5 dex from D20 and -0.08$^{+0.13}_{-0.10}$ from S20, indicating that TOI-1266 may be moderately metal poor. We report the previously published stellar parameters in Table \ref{tab:stellarparam}, including the parameters we adopt from S20 and newly derived parameters from our work. 

We perform an updated analysis to improve the iron abundance constraint, and measure several other refractory element stellar abundances including [Mg/H] and [Si/H] for the first time. The detailed abundance analysis was carried out with an implementation of \emph{The Cannon} \citep{Behmard2025}, a data-driven method capable of inferring stellar abundances that does not rely on stellar evolution models. This makes \emph{The Cannon} an excellent choice for characterizing M dwarfs, which have notoriously complex spectra due to the presence of molecular features. This \emph{Cannon} implementation was trained on M dwarfs with FGK companions from the Sloan Digital Sky Survey-V/Apache Point Observatory Galactic Evolution Experiment (SDSS-V/APOGEE). SDSS-V/APOGEE is a high resolution ($R$ $\sim$ 22,500), $H$-band (1.51$-$1.7 $\mu$m) spectroscopic survey. Its wavelength coverage is ideal for M dwarfs, which have peak brightness in the near-infrared.

\section{Transit Modeling} \label{sec:transit analysis}

\subsection{TESS} \label{sec:TESS modeling}

We constructed our TESS photometric model using the \texttt{exoplanet} package \citep{exoplanet:joss} for the transit light curve component. Our transit model includes the following free parameters: the stellar radius $R_*$, impact parameter $b$, scaled semimajor axis $a/R_*$, and individual transit mid-center times. We used separate planet-to-star radius ratios $R_p$/$R_*$ for each segment of TESS data (separated by PM, EM, and SEM) to account for possible changes in transit depth (see \S\ref{sec:radius analysis}). In \S\ref{sec:Palomar modeling}, \S\ref{sec:MuSCAT modeling}, and \S\ref{sec:radius analysis}, we refer to b, $a/R_*$, and $R_p$/$R_*$ as ``transit shape parameters''. We allowed the mid-time of each transit to vary individually in our fit using the \texttt{TTVOrbit} module of \texttt{exoplanet}, rather than fitting for a constant orbital period and $t_0$ corresponding to a linear ephemeris. 

We simultaneously modeled systematics in the TESS data from instrumental and stellar variability with a Matern-3/2 kernel GP from \texttt{celerite2} \citep{exoplanet:foremanmackey17,exoplanet:foremanmackey18}. This kernel is commonly utilized for its ability to model quasi-periodic stellar variability signals \citep[e.g.,][]{Demory_2020,Stefansson_2020,Gan2022,Cointepas2024}. For this part of the model, we included a mean offset parameter $\mu$ and GP hyperparameters $\sigma$ and $\rho$ corresponding to the amplitude and timescale of quasi-periodic oscillations, along with an error scaling term added in quadrature to the flux errors.

We fixed the planetary eccentricities to zero in the transit fit, as our TTV modeling indicates that both planets must have orbital eccentricities less than 0.1 (see \S\ref{sec:Joint TTV RV Modeling}) and the corresponding effect on the transit light curve shape is therefore negligible. We also fixed the quadratic limb darkening parameters $u_1$ = 0.3442 and $u_2$ = 0.2006. These are the values predicted by the \texttt{ldtk} \citep{Parviainen_2015} package using the stellar temperature, [Fe/H], and $\log g$ reported in Table \ref{tab:stellarparam}. \texttt{ldtk} uses the library of high-resolution synthetic spectra from \cite{Husser2013}, which are based on the PHOENIX stellar atmosphere code and includes synthetic spectra for stellar temperatures as low as 2300 K. We list the priors for our transit model and GP hyperparameters in Table \ref{tab:results}. Although we used uniform priors for most parameters, we placed a Gaussian prior on $R_*$ and $a/R_*$ based on the stellar mass, radius, and planetary orbital period from S20. We used the \texttt{PyMC3} package \citep{exoplanet:pymc3} to sample the posterior distribution of our model with No U-Turn Sampling (NUTS), with four parallel chains run with 5000 burn-in steps and 3000 posterior sample draws. We confirmed that the chains evolved until the Gelman-Rubin statistic values are < 1.01 for all parameters. The posterior distributions of each parameter are summarized in Table \ref{tab:results}. The detrended transit light curves and their best fit models are shown in Figure \ref{fig:TESS}.

TTV model fits to transit midtimes that have multi-modal or asymmetric posteriors can introduce biases into the retrieved dynamical parameters \citep{Judkovsky2023}. We therefore examined the posterior distributions of the individual TESS midtime parameters to determine whether or not all of the individual transits were detected at high enough significance to yield a unimodal and approximately normally distributed transit midtime constraint. We found that all of the observed TESS transits for both planets satisfied these criteria.

\subsection{WIRC} \label{sec:Palomar modeling}

We fit the WIRC light curves using \texttt{exoplanet} with a combined systematics and transit model. Our systematics model for each night includes a linear combination of comparison star light curve weights, an error inflation term added in quadrature to the flux errors, and a linear slope. We also tested systematics models with linear combinations of weights for the target centroid offset, PSF width, airmass, and local sky background as a function of time. We compared the Bayesian Information Criterion \citep[BIC,][]{Schwarz1978} for all possible combinations of these systematic noise parameters using the same framework as \cite{Jorge2024}. We found that the model that produced the lowest BIC value included weights for the target PSF width and airmass for TOI-1266 b and c on UT 2022-02-28 and UT 2022-03-10, respectively, while our UT 2022-06-06 observation of TOI-1266 c preferred only the PSF width as an additional detrending parameter. The transit midtime posteriors for the UT 2023-04-07 observation of TOI-1266 b and UT 2022-06-12 and UT 2023-12-09 observations of TOI-1266 c were not well constrained by the data due to poor weather and lack of baseline coverage, so we excluded them from our analysis. During our UT 2022-02-28 observation we experienced variable cloud cover and multiple flux drops $>$ 50\% of their peak value, and we excluded images with relative flux decrease $>$ 20\% for this night.

\begin{figure*}
\begin{center}
  \includegraphics[width=18cm]{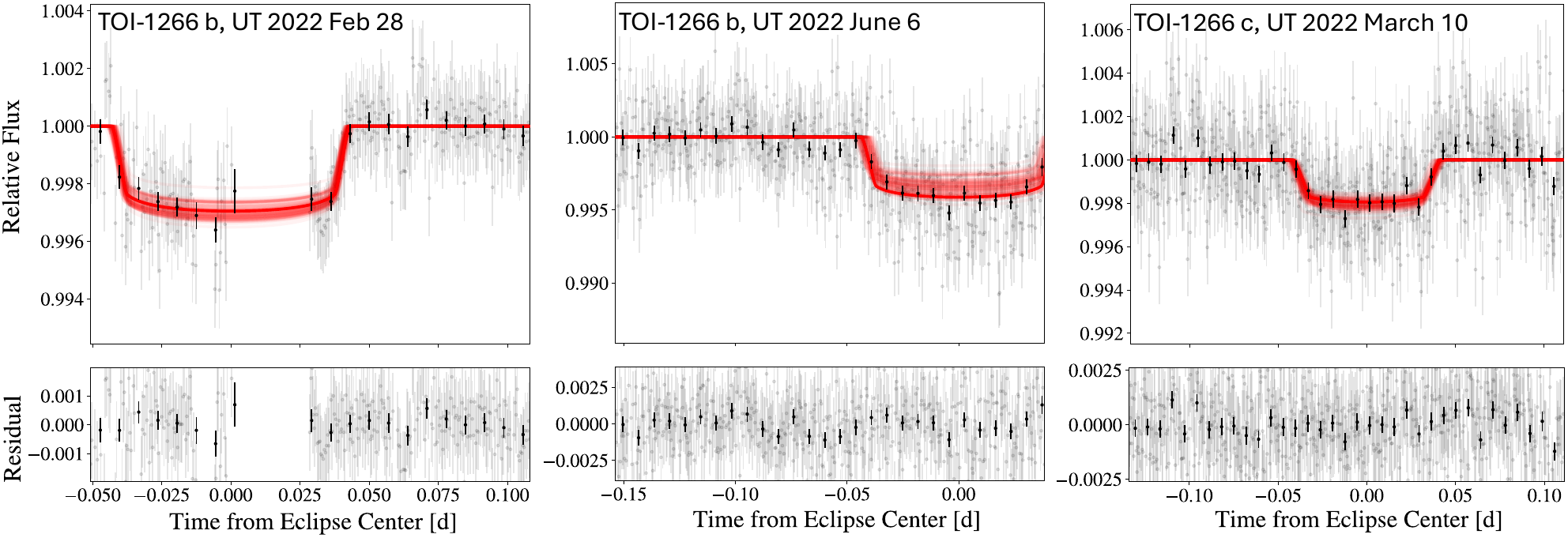}
  \caption{Detrended WIRC light curves for TOI-1266 b night 1 (left) and 2 (middle) and TOI-1266 c (right). The best fit transit models are overplotted as red lines, with red shading to indicate the 1$\sigma$ uncertainties.}
  \label{fig:Palomar}
\end{center}
\end{figure*}

We fit the WIRC transits jointly with the phased TESS transit profile from stacking all transits using their best fit individual midtimes and removing the best fit GP model of out-of-transit variability. We used the same model framework as in \cite{GreklekMcKeon2023}, with a wide uniform prior of $\pm$5 hours on the transit times. We used the same transit shape parameter priors as in \S\ref{sec:TESS modeling}, summarized in Table \ref{tab:results}. As before, we used \texttt{ldtk} to calculate the $J$ band quadratic WIRC limb darkening parameters $u_1=0.167$ and $u_2=0.164$, and held them fixed in our fits. We explored the parameter space with the NUTS sampler in \texttt{PyMC3} for 2000 tune and 2000 draw steps, and confirmed that the chains have evolved until the Gelman-Rubin statistic values are < 1.01 for all parameters. Our measured transit times are listed in Table \ref{tab:Observed transits} in the Appendix, and the final transit light curves are shown in Figure \ref{fig:Palomar}.

\subsection{MuSCAT imagers and Sinistro} \label{sec:MuSCAT modeling}

We fit the MuSCAT series (MuSCAT + MuSCAT3) and Sinistro observations with a similar procedure as described in \S\ref{sec:Palomar modeling}. Since the MuSCATs + Sinistro ground-based observations are not obtained with a beam-shaping diffuser, the noise properties of the light curves are different from the WIRC observations. Consistent with previous MuSCAT and Sinistro transit light curve analysis \citep[eg.][]{Kuzuhara2024,Cointepas2024} and our TESS analysis, we used a Matern-3/2 GP kernel from \texttt{celerite2} to model residual flux variability, with $\sigma$ and $\rho$ parameters corresponding to the amplitude and timescale of variability. We let $\sigma$ vary for each band within each night, while sharing $\rho$ across bands under the assumption that the timescale of the time-correlated noise is common among all bands for a given night. We tested MuSCATs + Sinistro systematics models for each night that also included linear combinations of weights for the target $x$ and $y$ detector positions, total centroid offset, PSF width, and airmass, but found that a pure transit+GP model performed better for all nights.

\begin{figure*}
\begin{center}
  \includegraphics[width=17.5cm]{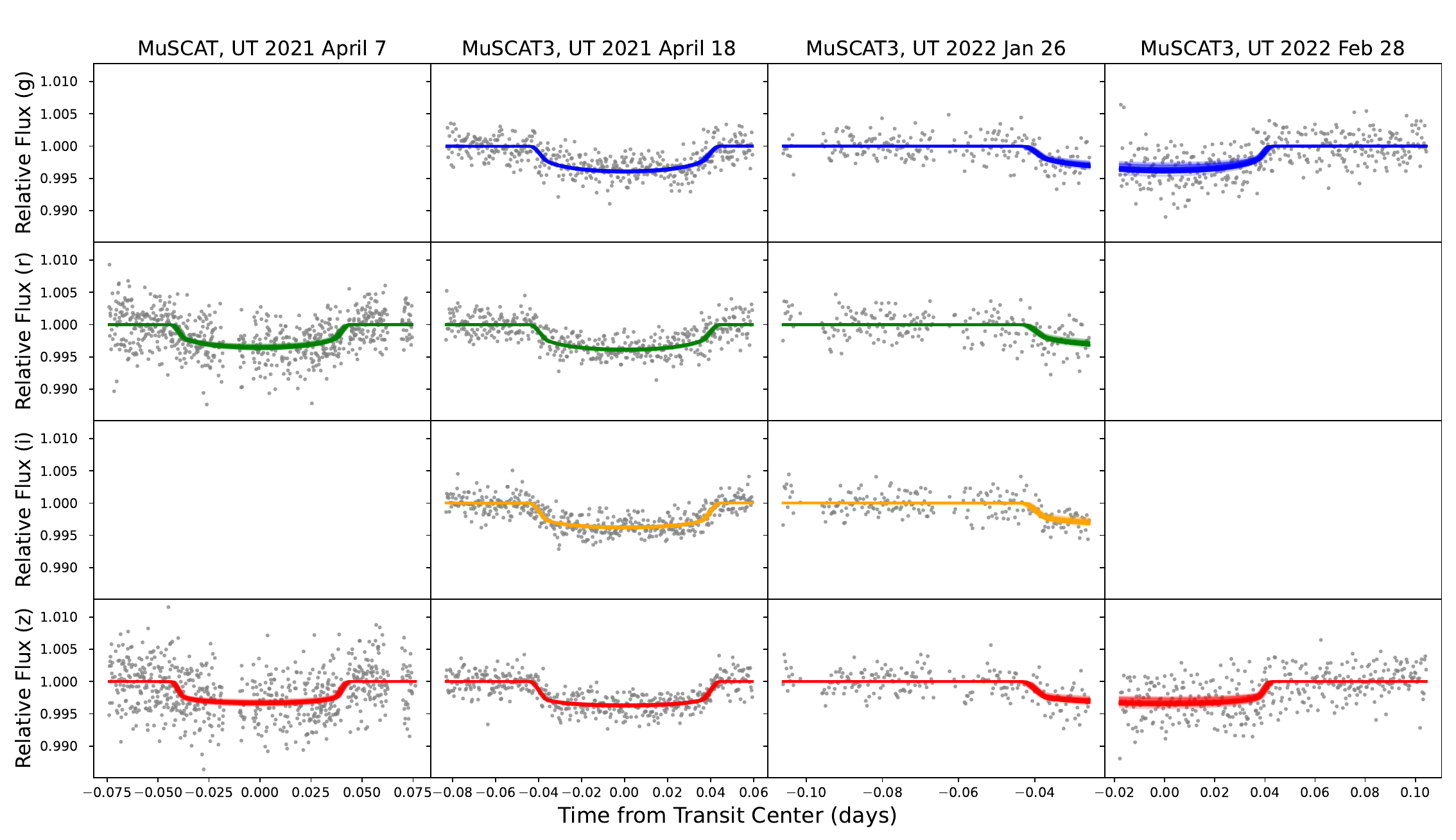}
  \caption{Detrended MuSCAT and MuSCAT3 light curves for TOI-1266 b. Colored lines are transit light curve models generated using 100 random draws from the posterior distribution for each bandpass.}
  \label{fig:MuSCAT b}
\end{center}
\end{figure*}

\begin{figure*}
\begin{center}
  \includegraphics[width=17.5cm]{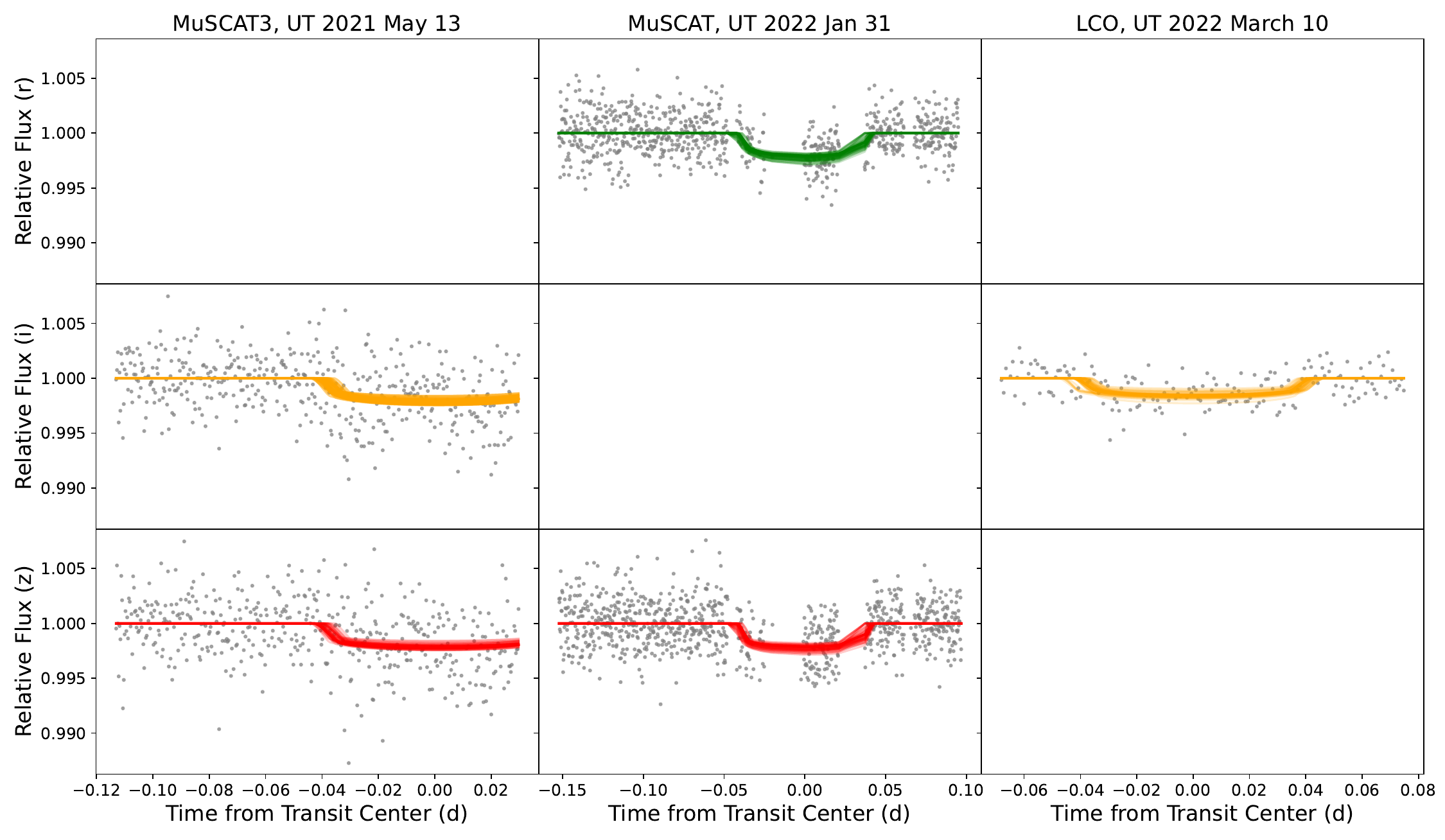}
  \caption{Detrended MuSCAT, MuSCAT3 and Sinistro light curves for TOI-1266 c. Colored lines are transit light curve models generated using 100 random draws from the posterior distribution for each bandpass.}
  \label{fig:MuSCAT c}
\end{center}
\end{figure*}

The MuSCAT and MuSCAT3 observations include simultaneous photometry in three or four bands, respectively, and each night has a different transit coverage fraction. We initially allowed $R_p$/$R_*$ to vary for each band and each night, fitting for individual $t_0$ values shared across all bands for a given night, with all other transit shape parameters shared across bands and nights. We fit the MuSCATs + Sinistro transits jointly with the phased TESS transit profile, used the same priors on the transit parameters as in our WIRC modeling, and fixed the quadratic limb darkening parameters to the predicted values from \texttt{ldtk} for each bandpass. For $g$, we use $u_1$ = 0.5976 and $u_2$ = 0.1739, for $r$ we use $u_1$ = 0.5674 and $u_2$ = 0.1542, for $i$ we use $u_1$ = 0.3900 and $u_2$ = 0.1957, and for $z_s$ we use $u_1$ = 0.3064 and $u_2$ = 0.1924. We explored the parameter space with the NUTS sampler in PyMC3 for 2000 tune and 2000 draw steps and verified that the Gelman-Rubin statistic values are $<1.01$ for all parameters. We then determined the detection significance of the $R_p$/$R_*$ constraint in each band on each night, and excluded any light curves where $R_p$/$R_*$ is consistent with 0 within 3$\sigma$. This ensures that we only include strongly detected transits in our subsequent analysis. This eliminated most partial transit observations and some light curves in $g'$ or $r'$ band with higher noise levels. We then repeated the same fitting procedure with a single $R_p$/$R_*$ parameter shared across bands for a given night in order to account for possible changes in transit depth (see \S\ref{sec:radius analysis}) and to obtain a final set of transit mid-times. As before, we confirmed that all transit mid-times had normally distributed posterior distributions. Our measured transit times are provided in Table \ref{tab:Observed transits} in the Appendix, and the final transit light curves for TOI-1266 b and c are shown in Figures \ref{fig:MuSCAT b} and \ref{fig:MuSCAT c}, respectively. 

\subsection{Planetary Radius Analysis} \label{sec:radius analysis}

C24 reported significant variation (3.5$\sigma$) in the transit depth of TOI-1266 c, and smaller variation in the transit depth of TOI-1266 b (1.9$\sigma$) between the TESS PM and EM observations. These apparent transit depth changes were confirmed by four additional ground-based transits near the TESS PM and EM observations, and were not accompanied by changes in the transit duration, which could have been indicative of orbital variability on short timescales. 

C24 investigated many possible explanations for the transit depth discrepancies, including variable flux dilution, orbital precession, residual artifacts from different TESS extraction methods, changes in stellar activity, and stochastic effects in the TESS data. C24 performed an exhaustive analysis on these possible causes of the transit depth discrepancies, and we refer the reader to \cite{Cloutier_2024} for further details. They concluded that the depth discrepancy is likely caused in part by a small increase in stellar activity and the presence of bright plages from the TESS PM to EM epochs, and by stochastic effects affecting the measurement accuracy of TOI-1266 c's transit depth in the PM data. 

\begin{figure}
%\begin{center}
  \includegraphics[width=8.5cm]{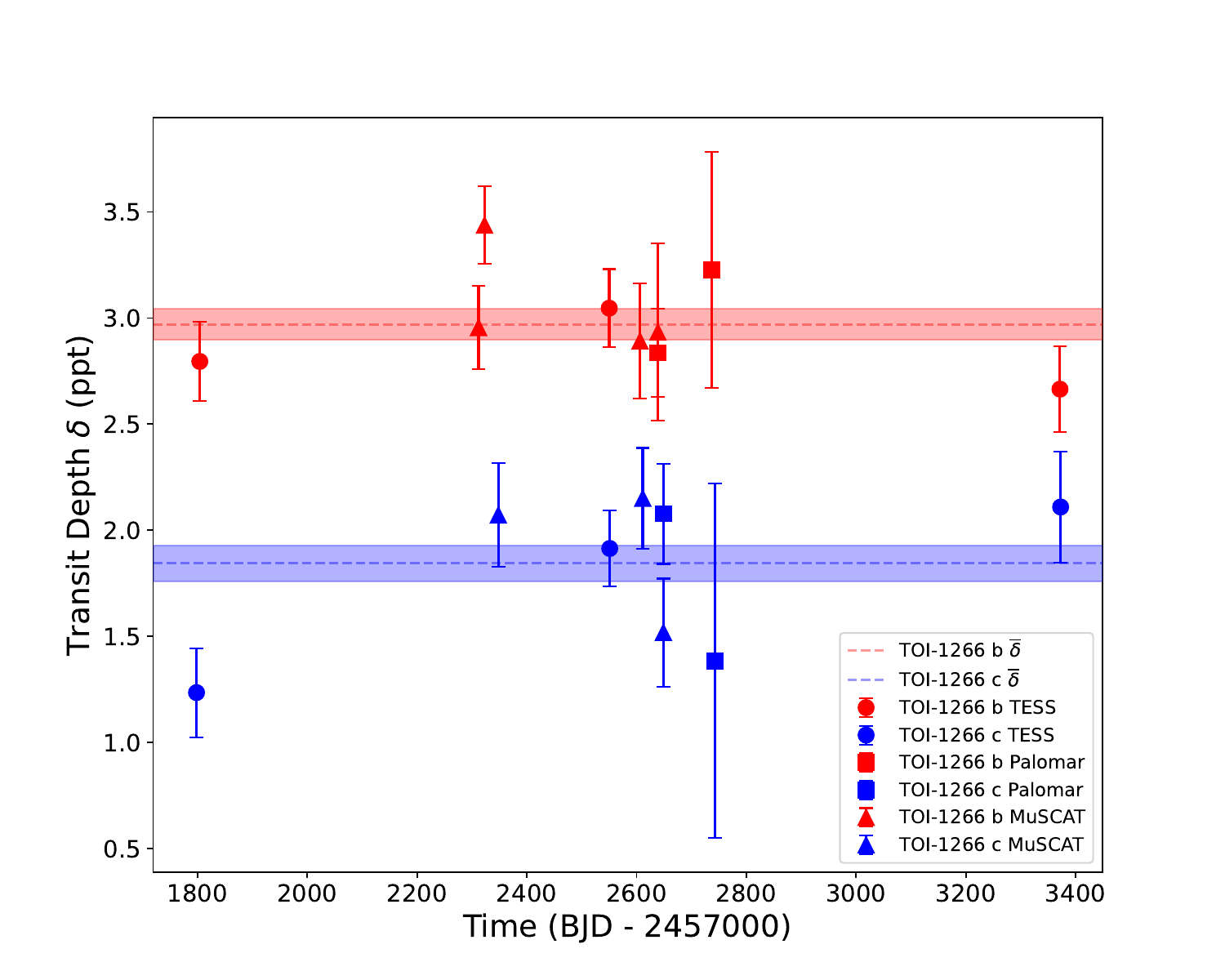}
  \caption{Observed transit depths of TOI-1266 b (blue) and c (red), including TESS PM, EM, and SEM data (circles), WIRC data (squares), and MuSCATs + Sinistro data (triangles). The error-weighted average transit depths are shown with dashed lines along with their associated 1$\sigma$ uncertainties (shaded region).}
  \label{fig:radius analysis plot}
%\end{center}
\end{figure}

Motivated by this unexpected change in transit depth from TESS PM to EM data, we perform modeling of the TESS, WIRC, MuSCAT imagers, and Sinistro photometry assuming different transit depths for each epoch to further investigate this behavior. We model the TESS data with the procedure described in \S\ref{sec:TESS modeling}, including separate transit depths for the PM, EM, and SEM data. We model the WIRC and MuSCATs + Sinistro data with the procedures described in \S\ref{sec:Palomar modeling} and \S\ref{sec:MuSCAT modeling} respectively, and use a separate transit depth parameter for each night of observation per planet. The resulting transit depths from this analysis are shown in Figure \ref{fig:radius analysis plot}. We reproduce the observed discrepancies between TESS PM and EM data and the WIRC observations that originally confirmed the discrepancy as reported in C24. However, we find that the inclusion of additional WIRC transits, the MuSCATs + Sinistro transits, and the new TESS SEM data weaken the statistical significance of the original discrepancy between TESS PM and EM. The depth of TOI-1266 c in the TESS PM data is now only 2.7$\sigma$ from the average measured depth, mostly driven by the addition of two ground-based transits with observed depths close to the PM value. Likewise, the discrepancy for TOI-1266 b from the average observed depth decreases to 0.9$\sigma$ -- mostly driven by the lower observed TESS SEM depth, which is consistent with the PM data. 

We conclude that there is no statistically significant discrepancy in the TESS PM transit depths relative to the remaining TESS and ground-based data, and that the dispersion in measured transit depths is consistent with stochastic variations -- though we cannot rule out the transit depth variability due to evolution of stellar plage coverage described in C24. We therefore assume that the transit depths for each planet are drawn from one underlying distribution, but report planetary radii and uncertainties based on the error-weighted average transit depths. 

\section{TTV Modeling}  \label{sec:ttv modeling}

Three previous studies of the TOI-1266 system (S20, D20, C24) have searched for evidence of TTVs. One of the discovery papers, S20, fit individual transit times from the TESS PM data and two ground-based transits using the \texttt{TTVOrbit} module within \texttt{exoplanet} and found no evidence for statistically significant TTVs. D20 reported marginal evidence for TTVs using the same TESS PM data examined by S20. C24 extended this analysis to the TESS PM and EM data along with several additional ground-based transits and also concluded that there was no compelling evidence for TTVs. C24 then used the lack of TESS TTVs in TOI-1266 b to place upper limits on the combined eccentricities of TOI-1266 b and c \citep[e.g.,][]{Hadden_2019}. 

The addition of the TESS SEM data nearly doubles our observational baseline and allows us to detect TTVs for planet b at high significance. The main oscillation period of this TTV signal is approximately twice the baseline of the data examined in previous studies and still only a fraction of the predicted TTV super-period for TOI-1266 b and d \citep[$P_{TTV} \simeq$ 2000 d, see \S\ref{sec:3-planet TTVs} and Equation 5 of][]{Lithwick_2012}. This explains the previous non-detections: with data covering less than half of the TTV super-period, this signal can be effectively removed by changes to the linear ephemeris of planet b in the TTV retrieval. 

In the following sections we outline our analysis of these TTVs, which exhibit a long-term trend for TOI-1266 b that is driven by near-resonant commensurability with a third external planet, and short-term chopping signatures for both TOI-1266 b and c. The low-significance TTV detection reported in D20 is likely the same second-order short-term chopping signature we detect here. D20 ascribed this signal to near-resonant 2-planet interactions between TOI-1266 b and c, but we now understand it to be part of a more complex 3-planet system in which the inner and outer planets are near the 3:1 mean-motion commensurability.

\subsection{The 2-Planet Case} \label{sec:2-planet TTVs}

We initially attempted to fit the set of transit times summarized in Table \ref{tab:Observed transits} with a 2-planet model, since there are only two transiting planets identified in the system and the RV signal at $\sim$32.3 days reported by C24 is characterized as a tentative detection. We used the \texttt{TTVFast} package to model the observed transit times in Table \ref{tab:Observed transits} in the Appendix. \texttt{TTVFast} \citep{TTVFast} is a computationally efficient $n$-body code that uses a symplectic integrator with a Keplerian interpolator to calculate transit times in multi-planet systems. The modeled transit times are a function of the planetary masses and orbital elements relative to a reference epoch, which we chose to be $T_0 = 1689.0$ (BJD - 2457000) -- shortly before the first transit of TOI-1266 observed by TESS. 

In our TTV modeling, we fixed the planetary orbital inclinations ($i$) to 90$^{\circ}$ 
because our transit fits show a low mutual inclination between planets b and c and are very close to edge-on ($i_b$ = 89.12 $\pm$ 0.08, $i_c$ = 89.17 $\pm$ 0.04). The difference between $M_{p}$ and $M_{p}$sin($i$) is $<$ 0.01\% for both transiting planets. The TTV solution is second-order in mutual inclination, with the strength of the planet-planet gravitational interaction and corresponding TTV amplitudes diminishing rapidly as the planetary mutual inclination increases \citep{nesvorny_2014, hadden_lithwick_2016}. For a purely edge-on orbital inclination the longitude of the ascending node ($\Omega$) becomes undefined, so we arbitrarily set it to 90$^{\circ}$ for both planets. 

Our 2-planet TTV model has ten free parameters in total. These include: the planet-to-star mass ratios, Keplerian orbital periods, mean anomalies reparamaterized with the time of first transit ($t_0$), and the planetary eccentricities and longitudes of periastron.  We reparameterized the latter two quantities as $\sqrt{e}\cos(\omega)$ and $\sqrt{e}\sin(\omega)$ in order to mitigate the degeneracy between e and $\omega$ in our fits while retaining an effective uniform prior on $e$ \citep{exofast}. The orbital periods, mean anomalies, eccentricities, and longitudes of periastron are osculating orbital elements defined at the TTV model start time $T_0$. We fit this model to the data using the affine invariant Markov chain Monte Carlo (MCMC) ensemble sampler \texttt{emcee} \citep{emcee}, and chose wide uniform priors for all parameters: $\textit{U}$(-1, 1) for $\sqrt{e}\cos(\omega)$ and $\sqrt{e}\sin(\omega)$, $\textit{U}$(0, 15$M_{\oplus}$) for the planetary masses sampled in $M_p/M_{*}$ space, $\textit{U}$($P_b$-0.1, $P_b$+0.1) and $\textit{U}$($P_c$-0.2, $P_c$+0.2), and $\textit{U}$($t_{0_b}$-1.0, $t_{0_b}$-1.0) and $\textit{U}$($t_{0_c}$-2.0, $t_{0_c}$-2.0) from the planetary orbital period and $t_0$ values reported in C24.

We initialized the MCMC fit with 2000 walkers (200 per free parameter) randomly distributed across the full prior volume in order to ensure that our MCMC analysis located all of the high likelihood regions of parameter space.  Although nested sampling algorithms utilize a similar approach, they can be very computationally expensive for TTV fits due to the large number of live points required to fully map the large prior volume \citep{dynesty,nestcheck}. Our MCMC walkers achieve a comparable result with fewer function calls.  When proposing new steps for the walkers we randomly selected the DEMove \citep{DEMove} or DESnookerMove\footnote{\href{https://emcee.readthedocs.io/en/stable/user/moves/}{emcee.readthedocs.io}} \citep{SnookerMove} at a rate of 80\% to 20\%, as recommended by \cite{ForemanMackey2019} for potentially multi-modal posterior distributions. This approach is advantageous for TTV fits because TTV model parameters are often degenerate and strongly correlated \citep[e.g., in mass and eccentricity,][]{Lithwick_2012}, their posterior distributions can be non-Gaussian or multi-modal, and the range of parameter combinations that produce a good fit to the observed data are often a very narrow subset of the total prior volume even for well-motivated but non-restrictive priors \citep[e.g.,][]{TOI-1136}. This makes it difficult for conventional MCMC walker evolution methods to identify and map the region of posterior space near the true maximum of the log-likelihood distribution when starting from a static initial guess or the output of a least-squares minimization, as is commonly done in exoplanet modeling. 

\begin{figure*}
\begin{center}
  \includegraphics[width=15.5cm]{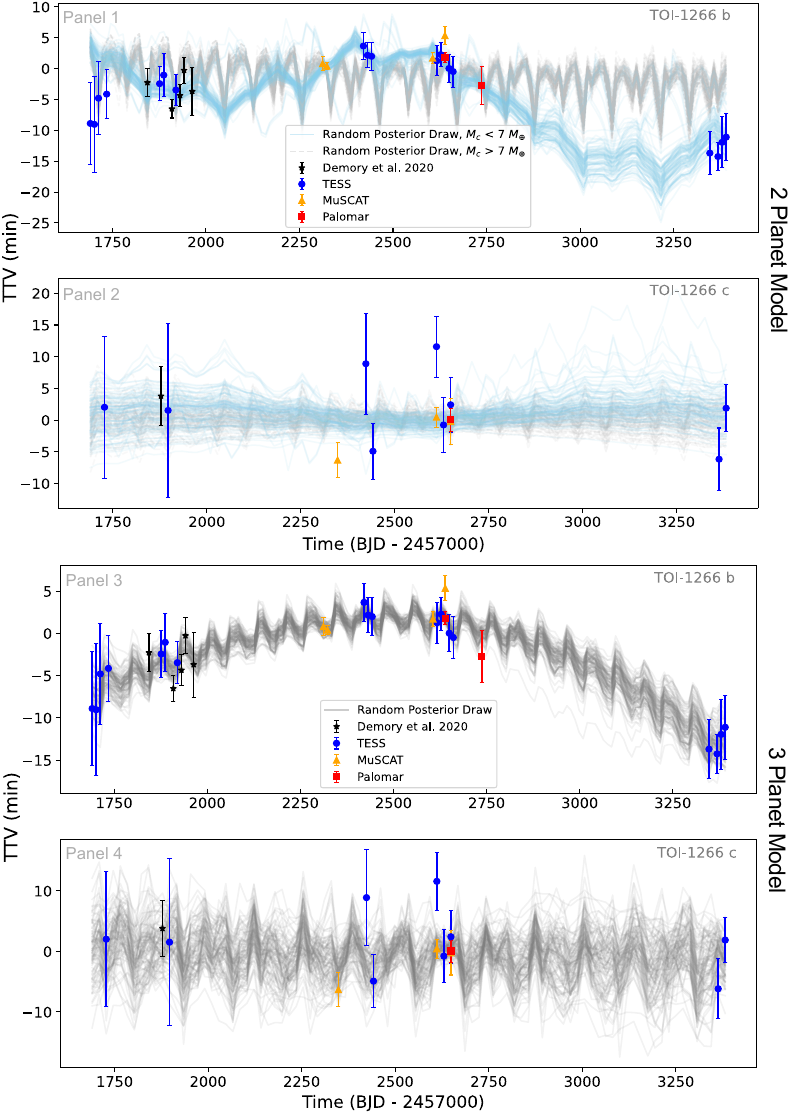}
  \caption{TTVs from TESS (blue circles), WIRC (red squares), and MuSCATs + Sinistro (yellow triangles) for TOI-1266 b (Panels 1 and 2) and TOI-1266 c (Panels 3 and 4) with 100 random posterior draws from our TTV model. Panels 1 and 2 show our 2-planet models with posterior draws from the samples with low $M_c$ and higher eccentricities shown in light blue, while those with higher $M_c$ and lower eccentricities are shown in light gray. Panels 3 and 4 show 100 random posterior draws from our 3-planet TTV model in dark gray. One TESS and one WIRC transit time each for TOI-1266 c are included in TTV fitting but omitted from these plots for clarity as they have timing uncertainties $>$ 15 min.}
  \label{fig:ttvplot_2planet}
\end{center}
\end{figure*}

We initially ran the sampler for 10000 steps in order to allow the 2000 walkers randomly distributed across the prior volume to identify the highest likelihood regions of posterior space. We then sorted the walkers by increasing likelihood and retained the walkers with median likelihoods in the top 20\% from the last 100 steps. Finally, we ran the remaining walkers for $10^6$ steps. We confirmed that the length of the MCMC chain was at least 10 autocorrelation lengths for all parameters. We plot 100 random draws from the resulting TTV posterior distribution and compare them to the observed TTVs in Panels 1 and 2 of Figure \ref{fig:ttvplot_2planet}.

The 2-planet TTV retrieval identifies two distinct families of solutions -- neither of which are a satisfactory fit to the data. One family of solutions prefers very low mass values for both planets, with 95\% upper limits of $M_b < 0.05 M_{\oplus}$ and $M_c < 0.45 M_{\oplus}$, and high eccentricity values, $e_b \simeq 0.15$ and $e_c \simeq 0.4$. This set of models (shown in light blue in Figure \ref{fig:ttvplot_2planet}) is better at reproducing the long-term TTV variability revealed in the TESS SEM data, but does not accurately model the TESS PM data and struggles to reproduce the short-timescale chopping TTV signature. Unlike the dominant long-term TTV trend with a super-period defined by the proximity of the planetary orbital periods to resonance \citep{Lithwick_2012}, the short-timescale chopping signature super-imposed on this trend is defined by the planetary mass and eccentricity vectors \citep{Deck_2015}. The other family of solutions (shown in light gray in Figure  \ref{fig:ttvplot_2planet}) prefers lower planetary eccentricities $e_b \simeq 0.12$ and $e_c \simeq 0.21$ and a very large mass for TOI-1266 c ($M_c = 13.2 \pm 4.1 M_{\oplus}$) that is inconsistent with the RV mass constraint reported in C24, and cannot reproduce the long-term TTV trend revealed by the TESS SEM data. 

In general, any isolated TTV model suffers from inherent degeneracies because it constrains a pair of canonical action/angle coordinates ($e$ and $\varpi$) and not the individual planetary eccentricities. Recent work has shown that orbital solutions with different planetary $\varpi$ values and therefore TTV phases can produce high eccentricity variants of the same TTV signal for a given system \citep{Goldberg2023,Choksi2024}. But in this case the different families of eccentricity solutions do not produce consistent TTV predictions for different combinations of M, e, and $\varpi$. Each of these solutions can reproduce some of the observed TTV data, but neither provide a good fit to the entirety of the observations, and both are in tension with the planetary masses and eccentricities reported in C24 from RV observations.

\begin{figure*}
\begin{center}
  \includegraphics[width=17.5cm]{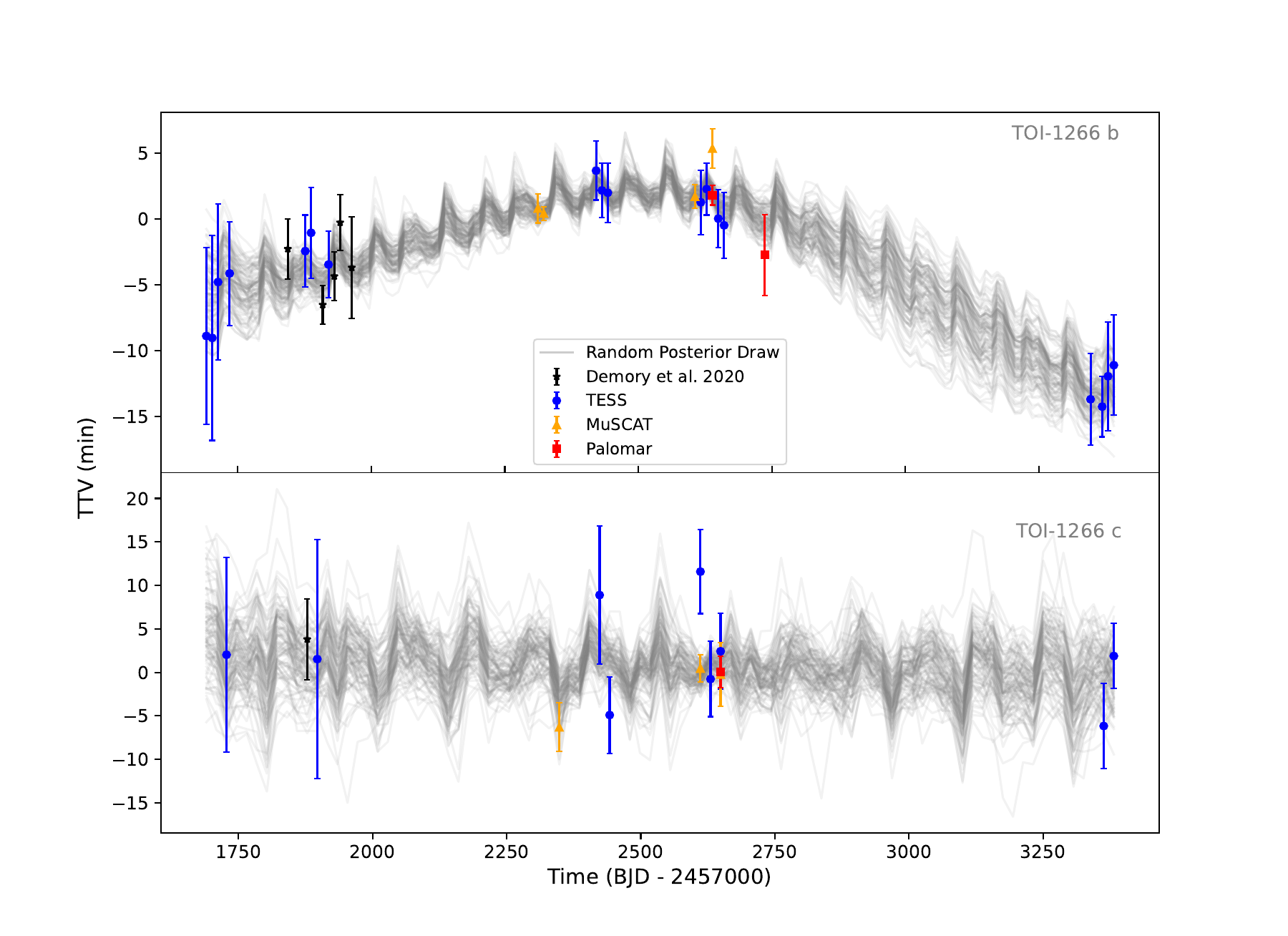}
  \caption{The same as Figure \ref{fig:ttvplot_2planet}, with 100 random posterior draws from our 3-planet joint TTV+RV model (gray curves).}
  \label{fig:ttvplot_joint}
\end{center}
\end{figure*}

\begin{figure*}
\begin{center}
  \includegraphics[width=17.5cm]{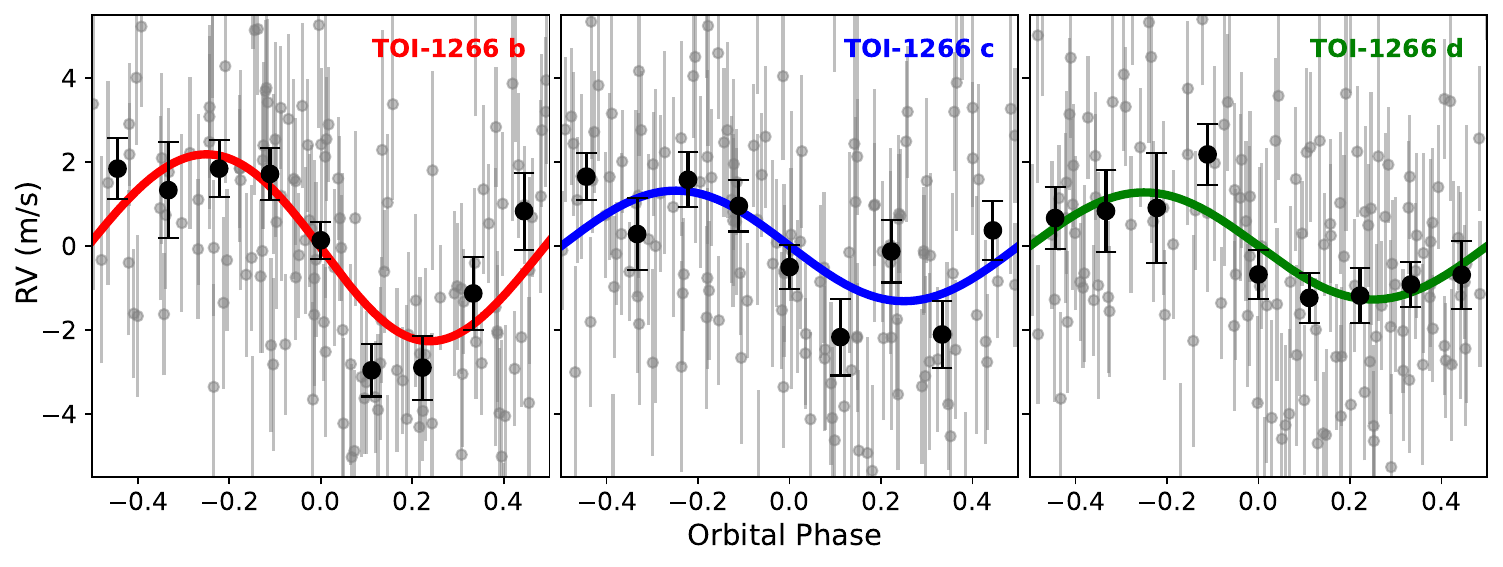}
  \caption{Phase-folded RV data with best fit model results from our joint TTV+RV fit for TOI-1266 b (left), TOI-1266 c (middle), and TOI-1266 d (right).}
  \label{fig:RVplot_joint}
\end{center}
\end{figure*}

\subsection{The 3-Planet Case} \label{sec:3-planet TTVs}

Motivated by the candidate RV planet reported in C24, we used the same framework described in \S\ref{sec:2-planet TTVs} to fit the TTV data with a 3-planet model. This increased the total number of free parameters in the fit from 10 to 15.  Consistent with the 2-planet fits, we fix $i_d$ = $\Omega_d$ = 90 for simplicity. Since the properties of the exterior planet candidate are unconstrained by transit photometry, we performed a brute force search with the same priors as described in \S\ref{sec:2-planet TTVs} for TOI-1266 b and c. We used wider priors for the orbital period of the third planet, with an orbital period range of 19 to 58 days and a corresponding $t_{0_d}$ range from the simulation start time $T_0$ to $T_0$ + 58 days. This allows the third planet's orbit to be anywhere from immediately exterior to TOI-1266 c to just beyond the 3:1 commensurability with TOI-1266 c, including the $\sim$32-day period reported for the tentative planet candidate in C24. 

We ran the MCMC retrieval utilizing the same general approach as in \S\ref{sec:2-planet TTVs}.  For this fit, we found that fewer walkers were required to reach more than 10 autocorrelation lengths for all parameters. We therefore used 600 walkers and ran them for a total of $2\times10^5$ steps.  We then selected the most likely 50\% of walkers and ran them for $10^6$ additional steps. We verified that the number of steps in the resulting chain was at least 10 times the shortest parameter autocorrelation length. The posterior from this blind search strongly favors a planet near the 3:1 commensurability with the inner planet TOI-1266 b, corresponding to an orbital period of 32.509$^{+0.061}_{-0.059}$. The $\Delta$BIC value for the 3-planet TTV model compared to a 2-planet model is $<$-10, indicating very strong preference for the 3-planet case \citep{raftery1995}.

Results from our 3-planet TTV retrieval are shown in Panels 3 and 4 of Figure \ref{fig:ttvplot_2planet}. The new 3-planet model is capable of simultaneously describing both the long-term TTV oscillation that samples part of the super-period of TOI-1266 b, and the short-term chopping signal for planets b and c. The fitted orbital period for TOI-1266 d from our TTV modeling is larger than the reported value from C24 by 3.2$\sigma$ (see \S\ref{sec:Joint TTV RV Modeling} for a discussion on the source of this discrepancy), although our TTV model is consistent with the RV modeling of C24 for the planetary masses and eccentricities. We summarize the prior and posterior distributions for all planets from our 3-planet TTV modeling, the RV modeling described in \S\ref{sec:RV Modeling}, and the TTV+RV joint modeling described in \S\ref{sec:Joint TTV RV Modeling} and in Table \ref{tab:results}.

For completeness, we also consider a scenario in which the third planet has an orbital period intermediate between that of TOI-1266 b and TOI-1266 c.  We fit this model to the data using the same brute-force TTV retrieval as above, but find that there are no combinations of parameters that can adequately describe the complete set of observations -- similar to the results of the 2-planet model fit in \S\ref{sec:2-planet TTVs}. Lastly, we consider a case in which the third planet is located interior to TOI-1266 b instead of exterior to TOI-1266 c. We find that the posterior distribution for this fit also prefers a third planet near the 3:1 commensurability with TOI-1266 b, corresponding to an orbital period near 3.6 days. The overall quality of this fit is similar to that of the external 3:1 scenario shown in Panels 3 and 4 of Figure \ref{fig:ttvplot_2planet}. 

We used several factors to rule out the interior 3:1 scenario as the true solution. First, we phased the TESS photometry up using the predicted individual transit times from the TTV model in order to search for transits from planet d, but we did not find any transit signal.  The short orbital period of planet d in this model means that if it is not transiting, it must have a mutual inclination of $> 2.6^{\circ}$ relative to TOI-1266 b -- much higher than the mutual inclination between TOI-1266 b and c ($0.04^{\circ}$). Population-level studies of mutual inclinations in other compact multi-planet systems \citep[e.g.,][]{mutual_incs} suggest that this degree of misalignment is unlikely. Additionally, C24 used RV observations to rule out the presence of any planets $\gtrsim 1 M_{\oplus}$ with orbital periods at 3.6 days. The preferred TTV mass for a planet at this location is $2.3 \pm 1.3 M_{\oplus}$, which would have a $\sim$84\% recovery rate in the RV data. This leaves the exterior 3:1 solution as the more plausible explanation for the observed TTV signal. This conclusion is additionally supported by C24's RV analysis, which independently reported the detection of a candidate planet with an orbital period very close to 32 days, though slightly below our TTV-based constraints ($P_{d_{RV}}$ = 32.340 $\pm$ 0.099 d, $P_{d_{TTV}}$ = 32.706$^{+0.057}_{-0.049}$ d).  
 
In the section below, we jointly model the TTV and RV data in order to obtain improved constraints on the orbital parameters of all three planets and to reconcile the slight tension between orbital periods for the planet candidate in TTV versus RV data. 

\section{TTV and RV Joint Modeling} \label{sec:joint modeling}

\subsection{RV-Only Modeling} \label{sec:RV Modeling}

Before performing a joint fit on the available TTV and RV data, we first sought to independently reproduce the RV results reported in C24 using our \texttt{TTVFast} retrieval framework, which utilizes an $n$-body integrator rather than Keplerian orbital models to predict the stellar RV at a specified time. We constructed our RV-only \texttt{TTVFast} model using the same 15 free parameters described in \S\ref{sec:3-planet TTVs} to model the planet signals, along with an additional seven free parameters to describe non-planetary structure in the RV data. These include a scalar jitter term $\sigma_{RV}$ added in quadrature to the RV observation uncertainties, the systematic RV $\gamma_{RV}$, and the same Gaussian Process (GP) regression model used in C24 to fit the stellar activity signal. The power spectral density of this GP kernel is the superposition of two damped simple harmonic oscillators, and it is therefore well-suited to modeling stellar activity signals. The GP model consists of five free parameters: $\Sigma$, $P_{rot}$ $Q_0$, $dQ$, and $f$, corresponding to the standard deviation of the process, the primary variability period, the quality factor of the secondary oscillation, the difference in quality factor for the two oscillation modes, and the fractional amplitudes of primary and secondary modes, respectively. C24 trained the GP on the H$\alpha$ time series of TOI-1266, which exhibits a periodic signal at $\sim$44.6 d attributed to stellar rotation. We adopted Gaussian priors for the four GP hyperparameters that were trained on the H$\alpha$ timeseries ($P_{rot}$, $Q_0$, $dQ$, $f$) from Table 4 of C24, with the same wide uniform priors on the other RV model parameters as in C24.

We use \texttt{emcee} to fit our \texttt{TTVFast} model to the RV data from Table 2 of C24, with the same fitting procedure, number of walkers, and step numbers as in \S\ref{sec:3-planet TTVs}, though we impose Gaussian priors on the orbital period and $t_0$ values from our TTV analysis because these are poorly constrained by the RV data and typically assigned Gaussian priors from transit fits where available. We ensure that the total number of steps is more than 10 times the autocorrelation length of all parameters. We find that our TTVFast model agrees with the planetary masses, orbital parameters, and RV noise model parameters reported in C24 at better than $1\sigma$ for all parameters and has similar uncertainties ($<$ 10\% difference). 

\subsection{Joint TTV + RV Modeling}   \label{sec:Joint TTV RV Modeling}

Next, we performed a joint fit of the TTV observations from Table \ref{tab:Observed transits} and the RV observations from C24 using \texttt{TTVFast} and \texttt{emcee}, with the same 22 free parameters as described in \S\ref{sec:RV Modeling}. We initialized the 15 planetary parameters close to the location of the best fit solution from the 3-planet TTV fit, and the 7 RV parameters close to the best fit solution from the 3-planet RV fit. We sampled the posterior distribution with DEMCMC for $8\times10^5$ steps, which corresponded to $>$ 10 autocorrelation lengths for all parameters after discarding the first $10^5$ steps as burn-in. We summarize the prior and posterior distributions for all parameters in Table \ref{tab:results}. 

This new joint fit resolves the previous tension between the orbital periods for the third planet candidate from the separate TTV and RV fits, with a best fit solution that is in good agreement with both the TTV and RV data. This confirms the existence of a third planet in the TOI-1266 system, and from here on we refer to the third non-transiting planet as TOI-1266 d. The joint fit model results for the TTV data are shown in Figure \ref{fig:ttvplot_joint}, while the RV model results are shown in Figure \ref{fig:RVplot_joint}. We include corner plots for the final planetary mass and eccentricity distributions along with orbital period and $t_0$ distributions in Appendix \ref{appendix}.

The final planetary mass constraints from the joint fit are consistent with those reported in D20, S20, and C24. Our new joint fit decreases the  fractional mass uncertainties for TOI-1266 b and c by 1\% and 5\%, respectively, relative to the RV-only constraints from C24. The joint model also prefers a slightly larger mass for TOI-1266 c and a slightly smaller mass for TOI-1266 d than reported in C24, making all three planets more uniform in mass. This intra-system mass uniformity is consistent with previously reported results for Kepler multi-planet systems \citep[``peas-in-a-pod''; e.g.,][]{Millholland_2017,Weiss2018,Millholland2022,Otegi2022,Goyal2023,Goyal_Wang2024,Rice2024}. The dispersion in planet masses is also consistent with the level of intra-system mass uniformity exhibited by the subset of Kepler multi-planet systems that are in resonance, which typically display even greater mass uniformity \citep{Goldberg_2022}. 

The joint TTV-RV model fit also results in large improvements in the planetary eccentricity constraints relative to the RV-only fit. Notably, the joint fit indicates that TOI-1266 b and d must have statistically significant nonzero eccentricities. This is due to the proximity of TOI-1266 b and d to the 3:1 resonance and the detection of a short-term chopping signal superimposed on the long-term TTV trend of TOI-1266 b. This chopping signal is strongly dependent on the masses and eccentricities of the planets driving the main TTV super-period (b and d). Unfortunately, we have no TTV observations for TOI-1266 d and the RV observations alone cannot constrain the planetary eccentricities to better than 0.1. 

\begin{figure}
%\begin{center}
  \includegraphics[width=8.5cm]{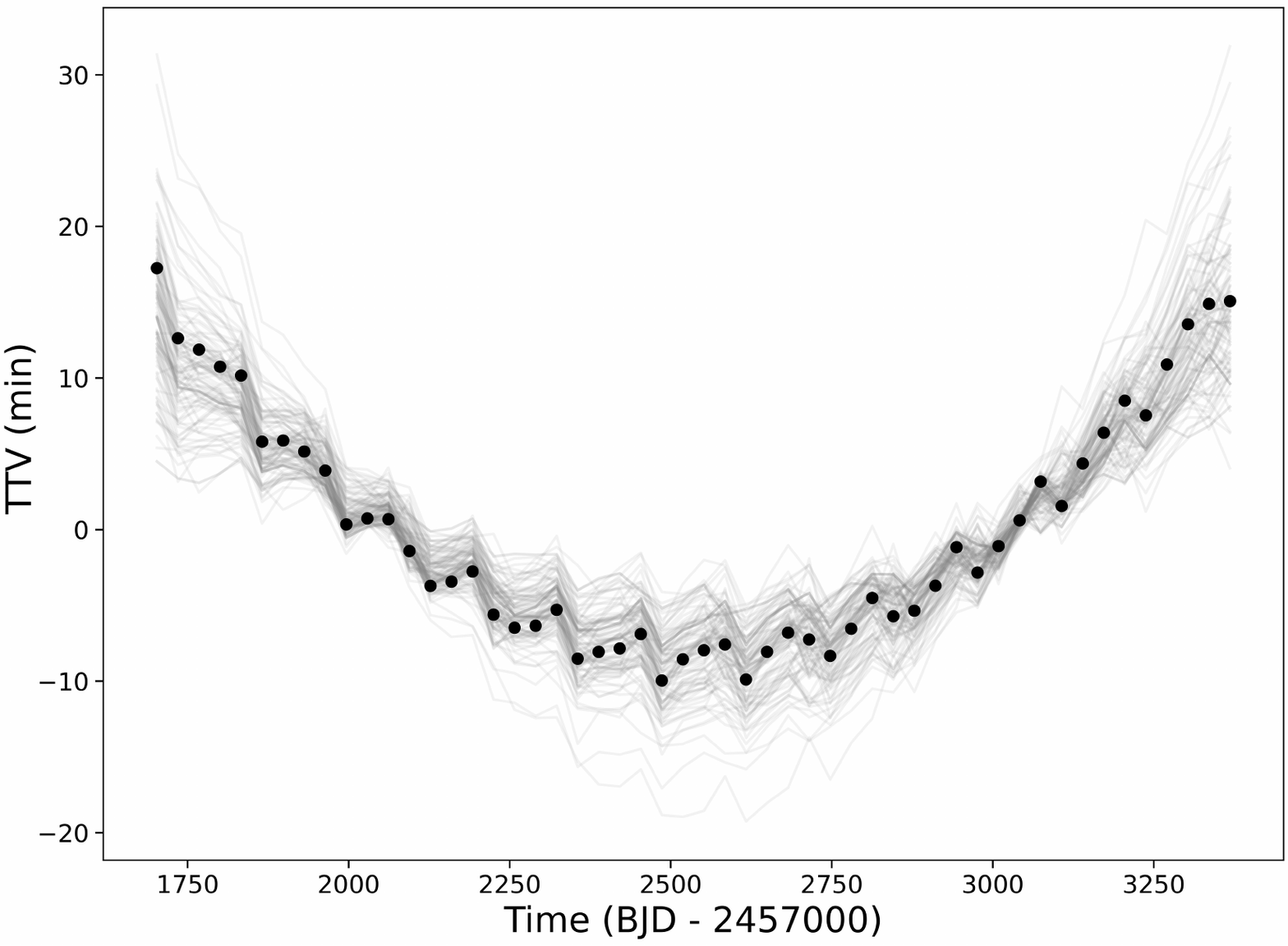}
  \caption{Predicted TTVs for TOI-1266 d from our best fit TTV+RV joint model (black points) spanning the same time range as in the previous TTV plots, with 100 random draws from the posterior distribution (gray). The TTV spread of $\sim$ 30 minutes could prohibit the detection of shallow transits in a grazing configuration.}
  \label{fig:d TTVs}
%\end{center}
\end{figure}

\begin{figure}
%\begin{center}
  \includegraphics[width=9.4cm]{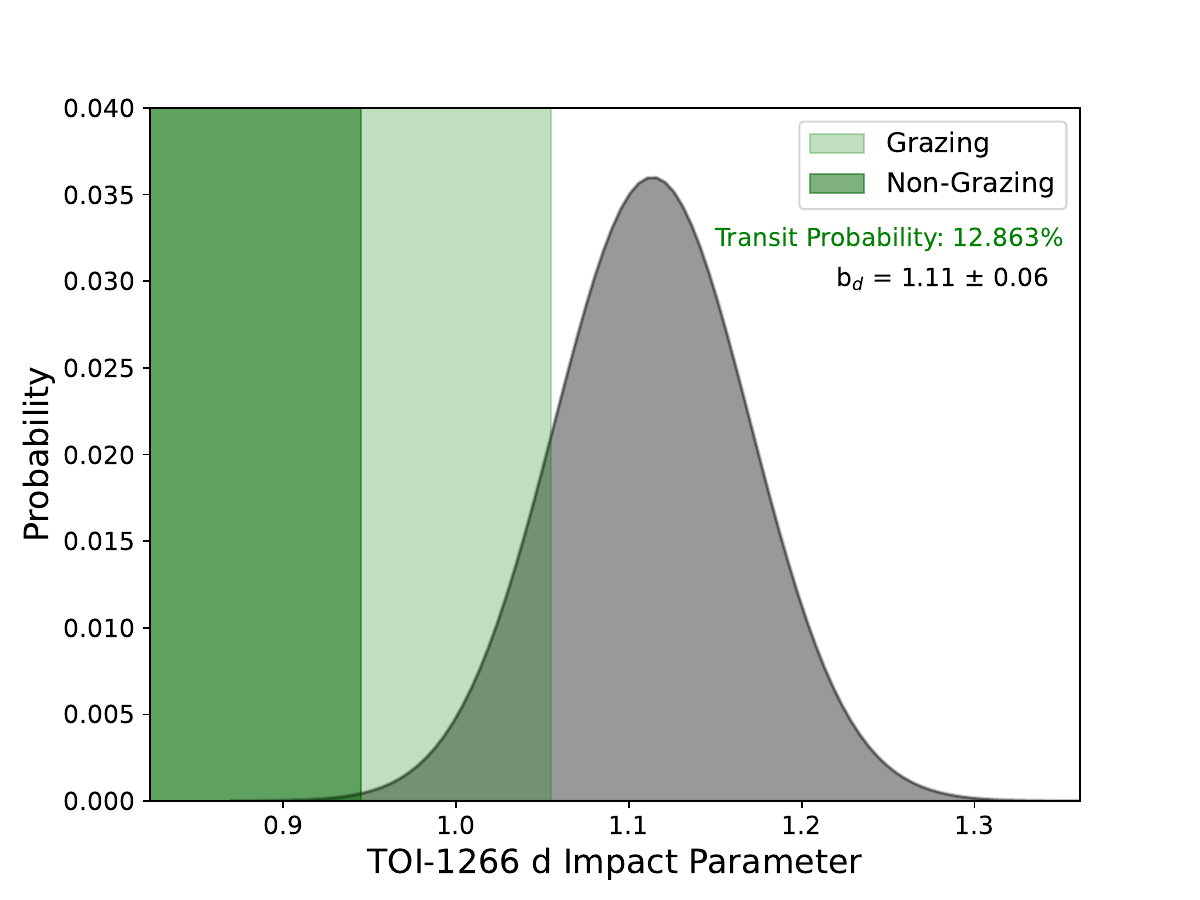}
  \caption{Geometric transit probability for TOI-1266 d when projecting out the orbital inclination distribution of TOI-1266 c, for both grazing configurations where $b < 1 + R_p/R_*$ and non-grazing configurations where $b < 1 - R_p/R_*$.}
  \label{fig:d transit prob}
%\end{center}
\end{figure}

\begin{figure}
\begin{center}
  \includegraphics[width=8.5cm]{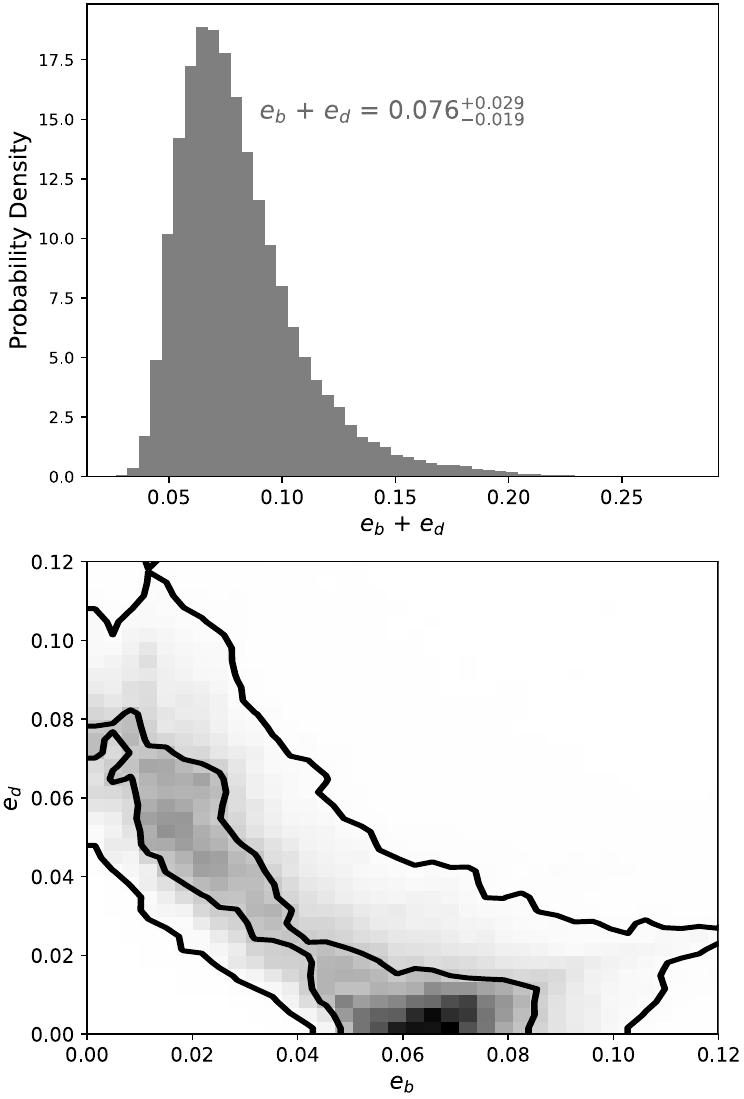}
  \caption{Posterior distribution for the eccentricities of TOI-1266 b and d with 1$\sigma$ and 2$\sigma$ contour lines (bottom) from TTV+RV joint fit, and the posterior probability distribution for the combined eccentricities of TOI-1266 b and d (top). TOI-1266 b prefers a moderate eccentricity if TOI-1266 d is not eccentric, or a small but nonzero eccentricity if TOI-1266 d is moderately eccentric.}
  \label{fig:bd_ecc_plot}
\end{center}
\end{figure}

This means that the TTV+RV solution is ambiguous as to the true eccentricity of TOI-1266 d, and there are two families of solutions for TOI-1266 b that depend on the eccentricity of d. If $e_d$ is small ($<$ 0.02), $e_b$ must be larger ($e_b$ = ${0.061}^{+0.021}_{-0.017}$), while if $e_d$ is larger ($>$ 0.02), $e_b$ is smaller ($e_b$ = ${0.025}^{+0.025}_{-0.015}$). The distribution of $e_b$ therefore has two peaks. The bulk of the posterior volume prefers larger $e_b$ values while $e_d$ is consistent with 0, but there is a second peak where $e_d$ is large and $e_b$ is smaller but still prefers a nonzero eccentricity. Another possible but less likely scenario is where both planets simultaneously have moderate eccentricities $\simeq$ 0.03. The joint fit retrieval rules out solutions in which $e_d$ + $e_b$ $<$ 0.03, and constrains $e_b$ + $e_d$ = ${0.076}^{+0.029}_{-0.019}$. These results are visualized in Figure \ref{fig:bd_ecc_plot}. The fact that the inner planet in this compact multi-planet system is eccentric has significant implications for the dynamical history and long-term evolution of this system, which we discuss further in \S\ref{sec:dynamical analysis}.

\begin{table*}
  \centering
  \caption{Priors and posteriors for TOI-1266 model parameters.\label{tab:results}}
  \begin{tabular}{lcccc}
    \hline
    \hline
    Parameter & Prior & \multicolumn{3}{c}{Posterior} \\
    \hline
    \multicolumn{5}{c}{\emph{TESS systematics parameters}} \\
    \hline
    Mean flux, $f_{0}$ (ppt) & $\mathcal{U}(-\infty,\infty)$ & \multicolumn{3}{c}{$0.000035\pm 0.000026$} \\
    Log jitter, $\ln{s^2}$ (ppt$^2$) & $\mathcal{U}(-18,2)$ & \multicolumn{3}{c}{$-14.54\pm 2.04$} \\
    $\sigma$ (ppt) & $\mathcal{U}(10^{-7},10^{-2})$ & \multicolumn{3}{c}{$0.000298^{+0.000019}_{-0.000017}$} \\
    $\rho$ (days) & $\mathcal{U}(10^{-3},10^{2})$ & \multicolumn{3}{c}{$0.644^{+0.078}_{-0.071}$} \\
    \multicolumn{5}{c}{\emph{RV systematics parameters}} \\
    \hline
    $\ln{\Sigma}$ (m/s) & $\mathcal{U}(-3,3)$ & \multicolumn{3}{c}{$-0.36^{+0.58}_{-1.61}$} \\
    $P_{\text{rot}}$ (days) & $p(P_{\text{rot}}\vert \mathrm{H}\alpha)^\textsc{1}$ & \multicolumn{3}{c}{$44.33^{+0.97}_{-1.01}$} \\
    $\ln{Q_0}$ & $p(\ln{Q_0}\vert \mathrm{H}\alpha)^\textsc{1}$ & \multicolumn{3}{c}{$0.50^{+0.54}_{-0.51}$} \\
    $\ln{dQ}$ & $p(\ln{dQ}\vert \mathrm{H}\alpha)^\textsc{1}$ & \multicolumn{3}{c}{$-0.65^{+1.68}_{-1.80}$} \\
    $\ln{f}$ & $p(\ln{f}\vert \mathrm{H}\alpha)^\textsc{1}$ & \multicolumn{3}{c}{$-0.18^{+0.25}_{-0.24}$} \\
    Log jitter, $\ln{s}_{\rm RV}$ (m/s) & $\mathcal{U}(-3,3)$ & \multicolumn{3}{c}{$0.57^{+0.14}_{-0.17}$} \\
    Mean velocity, $\gamma_{\rm RV}$ (m/s) & $\mathcal{U}(-41650,-41630)$ & \multicolumn{3}{c}{$-41639.96016^{+0.28}_{-0.25}$} \\    
    \multicolumn{5}{c}{\emph{Measured planetary parameters, transit fit}} \\
    \hline
    && \emph{TOI-1266 b} & \emph{TOI-1266 c} & \emph{TOI-1266 d} \\
    $a/R_\star$ & $p(a/R_\star\vert P,M_{\star}, R_{\star})$ & $35.97^{+1.21}_{-1.20}$ & $51.63^{+1.72}_{-1.71}$ & - \\
    $R_p/R_\star$ & $\mathcal{U}(0.0,0.2)$ & $0.0545^{+0.0007}_{-0.0007}$ & $0.0429^{+0.0010}_{-0.0010}$ & - \\
    Impact parameter, $b$ & $\mathcal{U}(0,1+R_p/R_\star)$ & $0.549^{+0.043}_{-0.047}$ & $0.747^{+0.027}_{-0.032}$ & - \\
    \multicolumn{5}{c}{\emph{Measured planetary parameters, TTV + RV joint fit (adopted)}} \\
    \hline
    && \emph{TOI-1266 b} & \emph{TOI-1266 c} & \emph{TOI-1266 d} \\
    $P$ (days) & $\mathcal{U}(P-0.2,P+0.2)^\textsc{1}$ & $10.89450^{+0.00023}_{-0.00026}$ & $18.80270^{+0.00126}_{-0.00116}$ & $32.509^{+0.062}_{-0.049}$ \\
    $t_0$ (BJD - 2,457,000) & $\mathcal{U}(t_{0}-2.0,t_{0}+2.0)^\textsc{1}$ & $1691.0068^{+0.0017}_{-0.0017}$ & $1689.9588^{+0.0028}_{-0.0026}$ & $1729.03^{+1.97}_{-2.25}$ \\
    $\sqrt{e}\cos{\omega}$ & $\mathcal{U}(-1,1)$ & $-0.116^{+0.123}_{-0.092}$ & $0.116^{+0.087}_{-0.122}$ & $0.032^{+0.108}_{-0.134}$ \\
    $\sqrt{e}\sin{\omega}$ & $\mathcal{U}(-1,1)$ & $0.038^{+0.123}_{-0.135}$ & $0.020^{+0.151}_{-0.150}$ & $-0.032^{+0.135}_{-0.143}$ \\
    $M_p/M_\star$ x $10^{-5}$ & $\mathcal{U}(0,10)$ & $3.02^{+0.47}_{-0.48}$ & $2.15^{+0.52}_{-0.52}$ & $2.50^{+0.71}_{-0.75}$ \\
    \multicolumn{5}{c}{\emph{Measured planetary parameters, TTV only fit (3-planet case)}} \\
    \hline
    && \emph{TOI-1266 b} & \emph{TOI-1266 c} & \emph{TOI-1266 d} \\
    $P$ (days) & $\mathcal{U}(P-0.2,P+0.2)^\textsc{1}$ & $10.89442^{+0.00021}_{-0.00019}$ & $18.80543^{+0.00254}_{-0.00195}$ & $32.706^{+0.057}_{-0.049}$ \\
    $t_0$ (BJD - 2,457,000) & $\mathcal{U}(t_{0}-2.0,t_{0}+2.0)^\textsc{1}$ & $1691.0048^{+0.0012}_{-0.0009}$ & $1689.9568^{+0.0020}_{-0.0021}$ & $1723.09^{+1.71}_{-2.35}$ \\
    $\sqrt{e}\cos{\omega}$ & $\mathcal{U}(-1,1)$ & $-0.080^{+0.131}_{-0.098}$ & $0.073^{+0.085}_{-0.112}$ & $0.114^{+0.111}_{-0.179}$ \\
    $\sqrt{e}\sin{\omega}$ & $\mathcal{U}(-1,1)$ & $0.126^{+0.073}_{-0.084}$ & $0.015^{+0.106}_{-0.110}$ & $-0.111^{+0.132}_{-0.107}$ \\
    $M_p/M_\star$ x $10^{-5}$ & $\mathcal{U}(0,10)$ & $4.03^{+2.67}_{-2.34}$ & $2.28^{+0.90}_{-0.81}$ & $1.13^{+0.93}_{-0.65}$ \\    
    \multicolumn{5}{c}{\emph{Measured planetary parameters, RV only fit}} \\
    \hline
    && \emph{TOI-1266 b} & \emph{TOI-1266 c} & \emph{TOI-1266 d} \\
    $P$ (days) & $p(P\vert TTV)$ & $10.89428^{+0.00032}_{-0.00032}$ & $18.80469^{+0.00292}_{-0.00293}$ & $32.686^{+0.048}_{-0.046}$ \\
    $t_0$ (BJD - 2,457,000) & $p(t_0\vert TTV)$ & $1691.0053^{+0.0017}_{-0.0017}$ & $1689.9564^{+0.0033}_{-0.0033}$ & $1723.41^{+1.16}_{-1.14}$ \\
    $\sqrt{e}\cos{\omega}$ & $\mathcal{U}(-1,1)$ & $-0.271^{+0.192}_{-0.126}$ & $-0.277^{+0.290}_{-0.212}$ & $0.110^{+0.293}_{-0.335}$ \\
    $\sqrt{e}\sin{\omega}$ & $\mathcal{U}(-1,1)$ & $0.220^{+0.230}_{-0.306}$ & $0.226^{+0.293}_{-0.427}$ & $0.169^{+0.284}_{-0.336}$ \\
    $M_p/M_\star$ x $10^{-5}$ & $\mathcal{U}(0,10)$ & $2.95^{+0.48}_{-0.48}$ & $2.02^{+0.63}_{-0.64}$ & $3.24^{+0.76}_{-0.76}$ \\
    \multicolumn{5}{c}{\emph{Derived planetary parameters}} \\
    \hline
    Inclination, $i$ (deg) &-& $89.13^{+0.08}_{-0.08}$ & $89.17^{+0.04}_{-0.04}$ & - \\
    Eccentricity, $e$ &-& $0.039^{+0.035}_{-0.025}$ & $<0.071^\textsc{2}$ & $<0.087^\textsc{2}$ \\
    Planet radius, $R_{p}$ (R$_{\oplus}$) &-& $2.52\pm 0.08$ & $1.98\pm 0.10$ & - \\
    Planet mass, $M_{p}$ (M$_{\oplus}$) & - & $4.46\pm 0.69$ & $3.17\pm 0.76$ & $3.68^{+1.05}_{-1.11}$ \\ 
    Bulk density, $\rho$ (g/cm$^3$) &-& $1.54\pm 0.28$ & $2.25\pm 0.64$ & - \\
    Semimajor axis, $a$ (au) &-& $0.0730^{+0.0011}_{-0.0011}$ & $0.1050^{+0.0017}_{-0.0017}$ & $0.1513^{+0.0024}_{-0.0024}$ \\
    Equilibrium temperature, T$_{\text{eq}}$$^\textsc{3}$ (K) &-& $415^{+7}_{-7}$ & $346^{+6}_{-6}$ & $288^{+5}_{-5}$ \\
    TSM$^\textsc{4}$, & - & $120^{+26}_{-20}$ & $69^{+25}_{-16}$ & - \\
    \hline
    \multicolumn{5}{l}{\footnotesize{$^\textsc{1}$ Marginalized posterior distribution from \cite{Cloutier_2024}}} \\
    \multicolumn{5}{l}{\footnotesize{$^\textsc{2}$ 95\% upper limit}} \\
    \multicolumn{5}{l}{\footnotesize{$^\textsc{3}$ Equilibrium temperature assuming zero albedo and perfect heat redistribution}} \\
    \multicolumn{5}{l}{\footnotesize{$^\textsc{4}$ Transmission Spectroscopy Metric \citep{Kempton2018}.}} \\
\end{tabular}
\end{table*}

\subsection{Is TOI-1266 d Transiting?} \label{sec:non-transiting search}

In \S\ref{sec:3-planet TTVs} and \S\ref{sec:Joint TTV RV Modeling} we confirm the existence of a third non-transiting planet, TOI-1266 d. This planet was originally identified as a candidate signal based on RV observations reported in C24. C24 visually checked for transits of TOI-1266 d using the assumed linear ephemeris from the RV data and also carried out a broader search extending to nearby orbital periods using the Transit Least Squares algorithm \citep[TLS,][]{tls}, but did not find any evidence for transits.

Standard transit search algorithms such as Box Least Squares \citep[BLS,][]{bls} and TLS assume a constant orbital period and can therefore fail to detect transits for planets exhibiting TTVs \citep{ttv_detection_bias}, especially when the signal-to-noise of individual transit events is low \citep[e.g.,][]{Leleu_2021}. Our joint fit to the TTV data of TOI-1266 b and c described in \S\ref{sec:Joint TTV RV Modeling} suggests that TOI-1266 d should indeed exhibit significant TTVs, as shown in Figure \ref{fig:d TTVs}. If TOI-1266 d has a transit depth comparable to or smaller than that of TOI-1266 c, it would be difficult to identify with standard transit search methods.  

We search for the transit of TOI-1266 d by phase-folding all available TESS data by individual transit midtimes from our joint fit model prediction, ensuring that we account for the effect of TTVs in the phased light curve. We do not find any transit signature, which is not unexpected. Our fits indicate that planets b and c have an extremely low mutual inclination ($\Delta i < 0.1^\circ$).  If we assume that planet d is also coplanar with TOI-1266 c, we find that TOI-1266 d has a relatively low geometric transit probability. We plot the geometric transit probability of TOI-1266 d under the coplanar assumption -- including grazing transit configurations -- in Figure \ref{fig:d transit prob}.

\section{Dynamical Analysis}  \label{sec:dynamical analysis}

\subsection{System Stability} \label{sec:rebound sims}

We investigated the long-term dynamical stability of the TOI-1266 system using the WHFast integrator in the \texttt{rebound} $n$-body code \citep{rebound}. We initialized $n$-body integrations with planetary masses, semimajor axes, eccentricities, mean anomalies, and longitudes of periastron drawn randomly from the chain of TTV+RV joint fit posterior 
samples described in \S\ref{sec:Joint TTV RV Modeling}, and converted from the joint fit sampling parameter basis to 
\texttt{rebound} parameterizations and units where appropriate. The orbital inclinations were initialized from a random normal distribution 
based on the transit fit results described in Table \ref{tab:results}, while the longitudes of the ascending node were randomly drawn from 0 to 2$\pi$. We integrated with a time step of 0.5 days, corresponding to less than 5\% of the orbital period of the inner planet, and evolved the system for a total of 92 kyr, or $\sim 10^6$ orbits of the outermost planet, repeating this process for 20 independent simulations. 

\begin{figure}
%\begin{center}
  \includegraphics[width=8.5cm]{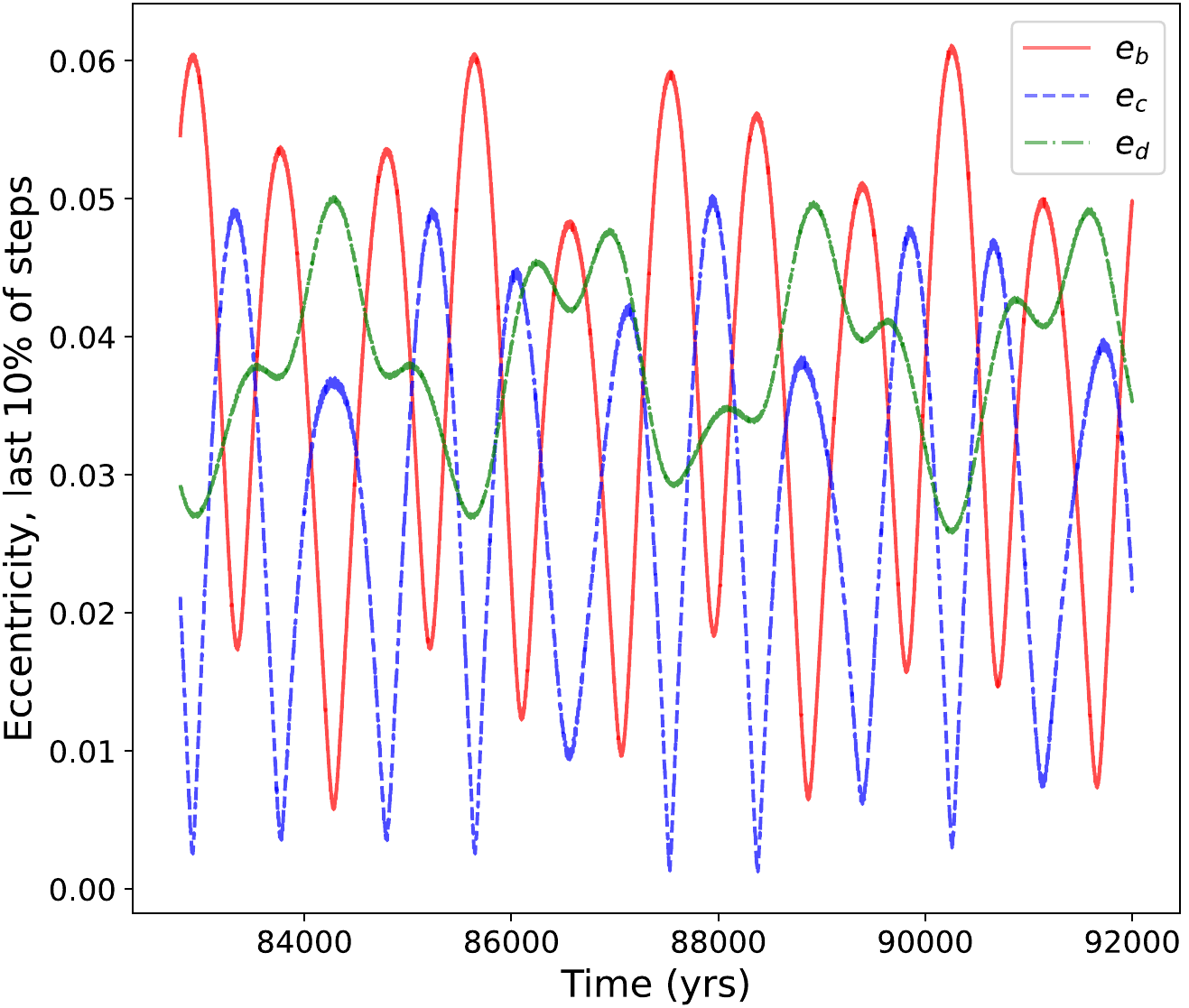}
  \caption{Evolution of planetary eccentricities during the last 9.2 kyr (final 10\%) of a \texttt{rebound} simulation with orbital parameters initialized from the TTV+RV joint fit posterior. All simulations displayed similar behavior: eccentricity oscillation around stable equilibria for the full simulation duration, with oscillation amplitudes ranging from $\sim$ 0.01 to 0.1.}
  \label{fig:ecc_oscillation}
%\end{center}
\end{figure}

We found that in all of our simulations, the planetary semimajor axes, eccentricities, and inclinations oscillate around stable equilibria for the duration of the model. The semimajor axes oscillate with an amplitude of $\sim0.01\%$ of their initial values over the course of the simulation due to planet-planet interactions, but the equilibria do not appear to vary. We conclude that the system is likely stable over the full range of orbital parameters identified by the TTV+RV joint model. The eccentricities in these simulations also oscillate around stable equilibria near their initial values with amplitudes ranging from 0.01 to 0.1. A representative figure from one simulation illustrating the eccentricity evolution is shown in Figure \ref{fig:ecc_oscillation}, and this behavior was common across all simulations. 

This confirms that, despite the compact nature of this system, we cannot use stability constraints to obtain tighter constraints on the orbital eccentricities of the three planets. 
We investigate the more realistic scenario of orbital evolution under the effect of tides in \S\ref{sec:eccs and tides}.

\subsection{Will TOI-1266 d Become a Transiting Planet?} \label{sec:future d transits?}

TOI-1266 b and c are coplanar ($\Delta_i$ = 0.04$^{\circ}$), and TOI-1266 d is unlikely to be transiting if it also orbits in this plane (\S\ref{sec:non-transiting search}). If we assume TOI-1266 d has a similar radius to TOI-1266 c, then an inclination increase of 0.11$^{\circ}$ would shift the orbit into a fully transiting, non-grazing configuration. In the $n$-body simulations described in \S\ref{sec:rebound sims}, we observe the orbital inclination of the outer planet librating around stable equilibria with libration amplitudes that vary between $\sim$0.02$^{\circ}$ and $\sim$0.2$^{\circ}$ depending on the initial orbital parameters drawn from the TTV+RV joint fit posterior distribution. In some simulations, the inclination variability timescale is as short as tens of years, while in most simulations it is hundreds of years. 
 
These simulations indicate that TOI-1266d may become observable in transit on decadal timescales. Any future search for transits of planet d will need to account for TTVs (see Fig. \ref{fig:d transit prob}), so we include in Table \ref{tab:Predicted transits} in the Appendix the predicted transit times for all planets from our best fit TTV solution extending 4 years from the present, which is when the uncertainty in our predicted transit times becomes larger than the transit duration. Continued follow-up of TTVs to refine these timing predictions may enable the transit detection of TOI-1266 d, though if  TOI-1266 d is already misaligned $>$ 0.2$^{\circ}$ relative to the orbital plane of TOI-1266 b and c then this scenario is unlikely.

\subsection{A Search for Orbital Resonances}

C24 analyzed the orbital period ratios of the TOI-1266 planets and determined that the b/c and c/d planet pairs do not lie close to any low-order two body MMRs, and that the three planets do not lie close to any three-body resonances -- concluding that the planets are likely not in a resonance chain. D20 analyzed TTVs from the TESS PM data and several ground-based transit observations and concluded that the b/c pair was weakly influenced by the stronger first-order 2:1 resonance, although it is nearly 14\% away from that resonance with a resonance proximity parameter \citep{Lithwick_2012} of $\Delta_{bc_{2:1}} = 0.137$. This conclusion may have been based on the tentative detection of the chopping TTV in TOI-1266 b, which D20 misattributed at the time to dynamical interactions with planet c. Based on our updated orbital parameter constraints from TTV fitting and detailed dynamical n-body modeling, we reassess whether any resonances exist in the TOI-1266 system. 

We first examine possible resonant states for the two adjacent planet pairs. Both TOI-1266 b and c and TOI-1266 c and d are somewhat close to the second-order 5:3 MMR. Our updated mean orbital period constraints place the period ratios at 1.729 for the d/c pair and 1.726 for the c/b pair, respectively.  We calculate $\Delta_{bc_{5:3}} = 0.037$ and $\Delta_{cd_{5:3}} = 0.036$, both within 4\% of the 5:3 resonance. For a planet pair to be in resonance, there must be a critical resonant angle that librates around a fixed point, rather than circulating from 0 to 2$\pi$. We tested resonant angles of the form $\theta = p\lambda - q\lambda' + \varpi$, where $\lambda$ and $\lambda$' are the mean longitudes of the inner and outer planets, $\varpi$ is the longitude of periastron for either planet, and $p$ and $q$ are the integers describing the MMR with $p - q = 1$ for a first-order resonance, $p - q = 2$ for second-order, etc. Given the proximity of the orbital period ratios for the adjacent planet pairs to various MMR commensurabilities, we perform \texttt{rebound} simulations to track the evolution of all possible two-body critical angles of the 2:1, 3:2, 5:3, 7:4, 8:5, and 9:5 first through fourth-order resonances for each adjacent planet pair. We draw initial planetary parameters from the TTV+RV joint fit as described in \S\ref{sec:rebound sims}, and evolve the system for 100 yr in 50 independent random simulations. We identify circulation in all resonant angles, confirming that neither adjacent planet pair are in a resonance.

Next, we investigate whether the system might be in a three-body resonance by combining multiple two-body resonant angles to determine if there are any three-body critical resonant angles that librate \citep[e.g. Kepler-223,][]{Mills16}. These three-body resonances take the form $\phi = p\lambda_{b} - (p+q)\lambda_{c} + q\lambda_{d}$ or $\phi = p\lambda_b + (p+q)\lambda_c - q\lambda_d$. In some systems, TTVs have revealed that \textit{only} these three-body angles librate, and the system is in a three-body resonance without any planet pairs in a two-body resonance \citep{Godziewski15,Mills16,MacDonald16}. These three-body resonances can be active even when the system is very far from any two-body resonances \citep[e.g., Kepler-221,][]{Goldberg_2021}.

We check for libration of all three-body resonant angles with $|p+q| < 15$, integrating the system in \texttt{rebound} for 100 years with initial parameters randomly drawn from the joint fit posterior and find circulation in $\phi$ for all independent simulations. We therefore conclude that the system is not in a three-body resonance.

The only possible resonance configuration left to consider is for the nonadjacent b/d planet pair. Our updated constraint on the period of TOI-1266 d moves it within 1\% of the 3:1 resonance with planet b, with $\Delta_{bd_{3:1}} = 0.0053$. We again take 100 random draws from the joint fit posterior distribution and integrate them for an initial 100 years. For a small number of the total draws, we observe a librating $\phi$ for the 3:1 MMR between planets b and d. This initially suggested that resonant interactions between TOI-1266 b and d might dynamically excite the planetary eccentricities to a level consistent with the TTV observations. To investigate whether the resonant configurations we identify are stable long-term and can maintain the observed nonzero eccentricities against the forces of tidal damping, we evolve these parameter combinations with a librating $\phi_{3:1_{bd}}$ for an additional 100 Myr. 

We find that the system is eventually knocked out of resonance and then becomes unstable for all initially resonant simulations. For simulations including tidal effects consistent with the modeling procedure described in \S\ref{sec:eccs and tides}, the inner planet's orbital period shrinks slightly due to tidal circularization. This causes the resonant angle to begin circulating and destabilizes the system, typically scattering the other two planets within 100 kyr. Resonant simulations not including tidal effects also went unstable, but on longer timescales ($1 - 10$ Myr), with the middle planet typically undergoing initial rapid orbital variations due to perturbations from the inner or outer planet which then destabilizes the entire system.

We conclude that although there is a small region of our posterior probability space where planets b and d are located in the 3:1 resonance, stability constraints indicate that this configuration is unstable on timescales much shorter than the age of the system ($\leq$ 100 Myr). The TOI-1266 system is therefore in a unique dynamical configuration, with two planets near a second-order near-resonant commensurability and a third planet located in-between them. There are no other known compact multi-planet systems in this kind of configuration, and it is an open question as to how such a system could have originated.

\begin{figure}
%\begin{center}
  \includegraphics[width=8.5cm]{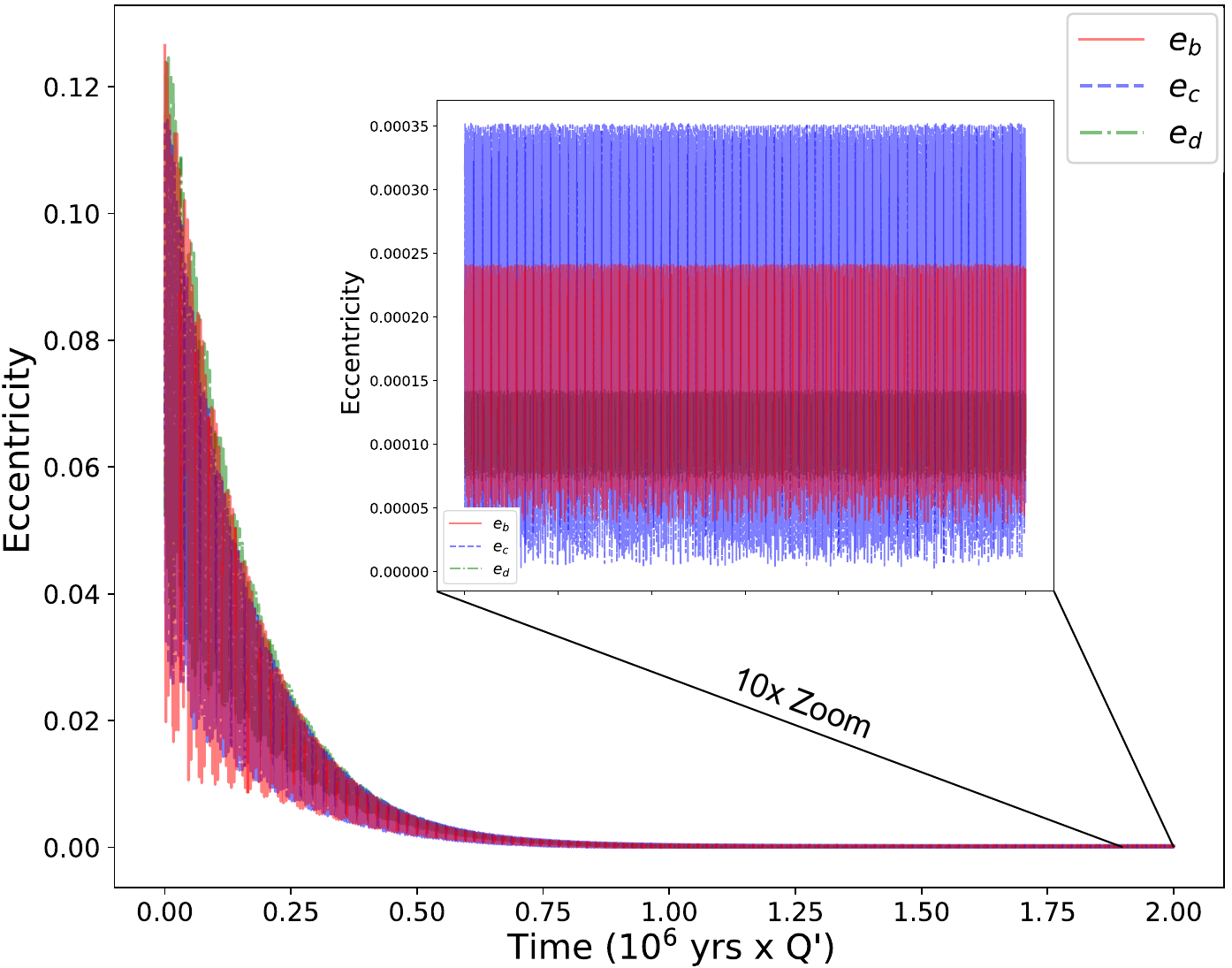}
  \caption{Orbital eccentricities of TOI-1266 b (red), c (blue) and d (green) under the effects of tidal forces in a \texttt{REBOUNDx} model with initial planet parameters drawn from the TTV+RV joint posterior distribution. Time units are normalized by the uncertain Q/$k_2$. The inset plot shows the last 10\% of simulation steps after the eccentricities have damped to their stable long-term equilibria. For Q/$k_2$ $>$ 1200, the damping timescale is longer than 2 Gyr, consistent with the stellar age.}
  \label{fig:ecc damping plot}
%\end{center}
\end{figure}

\subsection{Planetary Eccentricities and Tides}  \label{sec:eccs and tides}

TOI-1266 is a compact multi-planet system with close-in planetary orbits ($<$ 0.15 AU), indicating that tidal forces may play an important role in the dynamical evolution of the system. To investigate the potential for tidal damping of the planetary spins and eccentricities, we perform long-term dynamical simulations of the system incorporating a constant time-lag tidal model \citep{Eggleton1998} using the \texttt{tides-spin} implementation of \cite{Lu_2023} in \texttt{REBOUNDx 1.6.1} \citep{Tamayo2020}, which includes the effects of tides raised on the star and the orbiting planets and consistently tracks both spin and orbital evolution. In this model framework, the planets and star have physical structure and distortion parameterized by their radii, moment of inertia, potential Love number $k_2$, spin angular rotation frequency (denoted as $S$ to avoid confusion with the previously described longitude of ascending node, $\Omega$), and the tidal quality factor $Q$. 

Both the planetary $k_2$ and $Q$ values are highly uncertain parameters, and are therefore sometimes combined as $Q$/$k_2$ and referred to as the reduced tidal quality factor $Q'$. The $k_2$ value depends sensitively on the internal composition of the planet, rheology, presence of ice/silicate/metal layers, their extent and melting, and other parameters -- all of which are unconstrained for the TOI-1266 planets. \cite{Tobie2019} predicts the $k_2$ values of rocky and ice-rich planets from multi-layer interior structure modeling, and finds values that range from $0.05 - 0.5$ that primarily depend on the relative size of the iron core for rocky planets and the presence of ice-rich outer layers for water-rich planets. \cite{Nettelmann2011} model the interior of the sub-Neptune planet GJ 1214 b, which is similar in radius and mass to TOI-1266 b -- assuming a rocky core with a H/He-rich envelope, a pure water envelope, or a mixture of H/He and water envelopes. This study finds $k_2$ values ranging from $\sim0.02 - 0.7$, depending on the bulk composition.  \cite{Kellermann2018} arrives at similar conclusions from structure modeling of $\sim$2 $R_{\oplus}$ planets with masses between $1-8$ $M_{\oplus}$ approximated with an iron core, silicate mantle, and solar metallicity envelope, or with significant bulk water content. This study finds $k_2$ values from $0.01 - 0.8$. We adopt the latter range -- which is the most generous of the three -- for our simulations.

\begin{figure}
%\begin{center}
  \includegraphics[width=8.5cm]{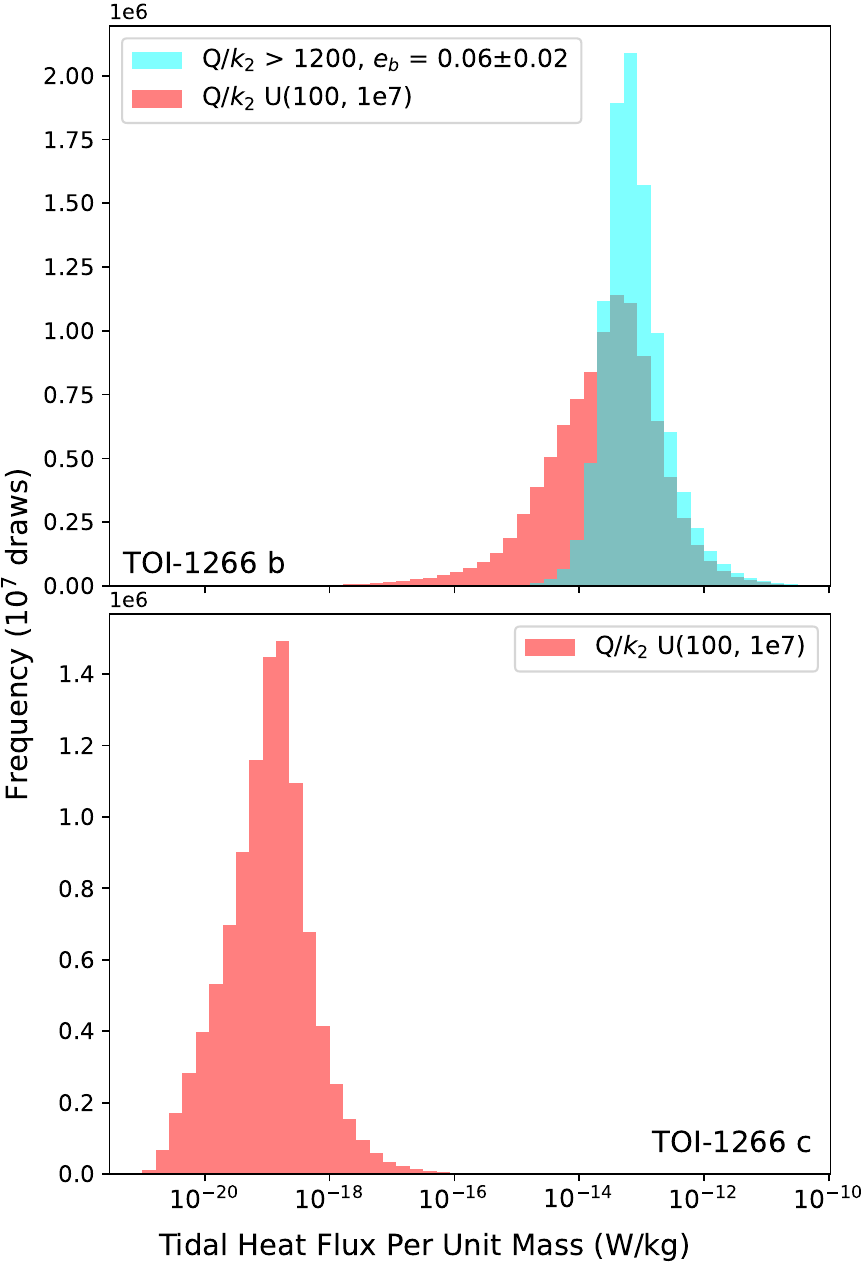}
  \caption{The tidal heat flux per unit mass of TOI-1266 b (top) and c (bottom) calculated with a wide range of possible Q (100 - 10000) and $k_2$ (0.01 - 0.8) values, with $e$ values drawn from the TTV+RV joint posterior distribution (red). Tidal heat fluxes with $e_b$ drawn from the higher eccentricity peak where $e_d$ is small and where Q/$k_2$ is $>$ 1200 such that the eccentricity damping timescale is $>$ 2 Gyr for TOI-1266 b are shown in blue. For both planets, the lower bound on heat flux is set by the post-damping eccentricity equilibrium e $\simeq$ 0.0001.}
  \label{fig:tidal heat plot}
%\end{center}
\end{figure}

The dissipation rate $Q$ is sensitive to the forcing frequency and internal viscosity, which depends on the thermal evolution of the planet, and is unknown. For Earth-like rocky planets, $Q$ could be as low as 10 \citep{Clausen2015,Goldreich1996}, while a population-level analysis of compact multi-planet super-Earth and sub-Neptune systems suggests that Neptune-like $Q$ values as large as 10$^5$ may be common \citep{Louden2023}. We adopt a range of $100 - 10000$ for $Q$ in our simulations, since our measured bulk densities for TOI-1266 b and c indicate that they are less dense than Earth but their gas envelopes likely comprise a relatively small fraction of their total masses.

Next, we confirm that TOI-1266 b and c are expected to be tidally locked by calculating the tidal spin locking timescale \citep{Gladman1996}. For all plausible $Q$ and $k_2$ values, the spin locking timescales for both planets are $<$ 400 Myr -- much less than the estimated age of the system. However, in a recent population-level analysis of radius inflation for compact multi-planet sub-Neptune systems, \cite{Millholland2024} suggested that it may be common for planets that migrate inward to be locked into higher obliquity spin-orbit resonance Cassini states. If TOI-1266 b or c are in a nonsynchronous spin-orbit resonance, the tidal dissipation could be orders of magnitude larger. The analysis we present here therefore represents a lower limit on the tidal heat flux for each planet.

We ran the \texttt{REBOUNDx} simulations for $2\times10^6\times(Q/k_2)$ years, so that the damping timescale can easily be mapped onto various possible values of $Q$ and $k_2$. We initialized the planet orbital parameters and eccentricities by sampling from the joint TTV+RV fit posterior, and randomly selected $Q$ and $k_2$ values from uniform distributions with ranges of $100 - 10000$ and $0.01 - 0.8$, respectively. The planetary spins are initialized from a synchronous state. We performed 100 random parameter draws and evolved the system for $2\times10^6\times(Q/k_2)$ years. 

We found that all three planetary eccentricities initially oscillated around a constant value, consistent with the results of \S\ref{sec:rebound sims}, but were then quickly damped until they reached stable equilibria. The changes in planetary semimajor axis in all of the simulations were $< 1\%$, and all of the systems remained stable for the simulation duration. Even after the damping was complete, we found that the planets still maintained slightly nonzero eccentricities at late times, which oscillated around stable values. We show an example of this in Figure \ref{fig:ecc damping plot}. The final ratios of planetary eccentricities in our simulations depends on the initial configuration, with final $e_b$ equilibrium values typically ranging from $0.0001 - 0.001$. This remains true across all random orbital parameter draws and spans the full range of $Q$ and $k_2$ values tested.

We can leverage the fact that our TTV+RV joint fit prefers a moderately nonzero eccentricity for TOI-1266 b and the smaller eccentricity values from our tidal damping simulations to constrain the value of $Q$/$k_2$. Our dynamical simulations indicate that all of the eccentricity values allowed by our TTV+RV posterior probability distribution are stable long-term, and that there are no orbital resonances that can pump up the planetary eccentricities. However, our long-term dynamical simulations incorporating tidal damping indicate that any primordial eccentricities should be damped to values of e $\simeq$ 0.0001 in less than $2\times10^6\times(Q/k_2)$, while TOI-1266 b has a retrieved orbital eccentricity of $\sim0.06$ in a majority of our TTV+RV posterior. Since this system is likely older than $\sim$ 2 Gyr, this implies that $Q$/$k_2$ for TOI-1266 b must be greater than 1200, which is the value that sets $\tau_{circ} >$ 2 Gyr. An alternate possibility is that TOI-1266 b had its eccentricity excited by a recent dynamical disturbance, for example by a widely separated outer planet that is undetected in the RV observations. 

\begin{figure}
%\begin{center}
  \includegraphics[width=8.5cm]{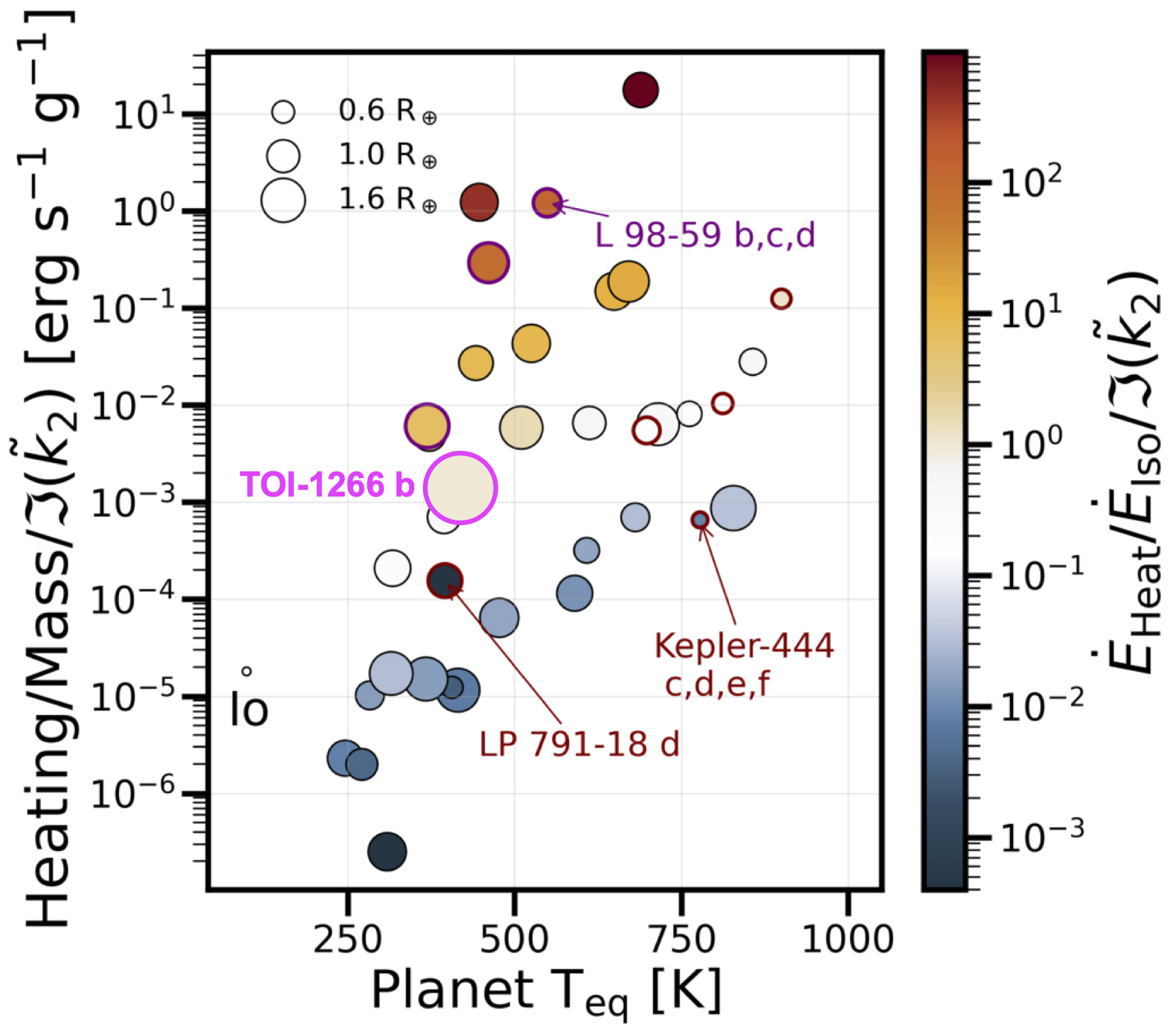}
  \caption{Planetary equilibrium temperature versus tidal heat flux per unit mass, adapted from Figure 5 of \cite{Seligman2023}, showing the most promising rocky planets for tidal volcanism. Heat fluxes are normalized by the uncertain tidal quality parameters, point sizes are scaled by planet size, and color indicates the ratio of tidal heat flux to insolation flux. TOI-1266b has an estimated tidal heat flux nearly equal to its insolation flux. If TOI-1266 c has a damped free eccentricity of 0.001, then it falls outside of this plot (Heating/Mass/$\Im(\tilde{k}_2) \simeq 10^{-8}$, $\dot{E}_{\rm Heat}/\dot{E}_{\rm Iso}/\Im(\tilde{k}_2) \simeq 5\times10^{-5}$).}
  \label{fig:volcanism population plot}
%\end{center}
\end{figure}

Next, we assess the potential importance of tidal heating caused by the nonzero eccentricity for the overall energy budgets of planets b and c. For TOI-1266 b, we assume a lower bound of 0.0001 on the eccentricity from our simulations of tidal damping and otherwise draw eccentricity values directly from our TTV+RV posterior distribution. For TOI-1266 c, we also assume a lower bound of 0.0001 from our tidal damping simulations. But since the TTV+RV posterior distribution does not indicate any preference for a nonzero eccentricity of TOI-1266 c, we instead impose an upper bound of 0.001 representing the largest values from our tidal damping simulations. We then calculated the range of tidal heat fluxes from Equation 4 of \cite{Jackson2008} for both planets and plot the resulting distribution in Figure \ref{fig:tidal heat plot}. We find that for $Q$ and $k_2$ combinations that yield a tidal circularization timescale larger than 2 Gyr and are therefore consistent with the larger eccentricity values preferred for planet b, the upper end of the distribution for tidal heat flux per unit mass is $\sim1\%$ that of Io \citep[TOI-1266 b: $F_{\mathrm{tidal}} \in 5\times10^{-15} - 5\times10^{-12}$  W/kg, Io: $F_{\mathrm{tidal}}\simeq10^{-10}$ W/kg,][]{Veeder2012}, while the tidal heat flux in W/m$^2$ at the surface of TOI-1266 b is nearly equal to the insolation flux (Figure \ref{fig:volcanism population plot}). This suggests that tidal heating is likely to be important for interpreting the internal and atmospheric composition of TOI-1266 b, with potential implications for long-term atmospheric evolution as well. 

Lastly, we compare TOI-1266 b to the broader population of small planets with measured nonzero eccentricities, implying significant tidal heat fluxes. \cite{Seligman2023} analyzed the population of rocky planets with measured eccentricities to identify top candidates for volcanic outgassing by comparing the tidal heat flux per unit mass and the corresponding ratio of tidal heat flux to insolation flux -- normalized against the uncertain tidal parameters $Q$/$k_2$. In Figure \ref{fig:volcanism population plot} we show that TOI-1266 b is also a top candidate for tidally driven outgassing, although the implications of this tidal heating are complicated by the presence of a thick atmospheric envelope. TOI-1266 b may have a molten surface negating traditional volcanism through a solid crust, but the significant tidal heat could still inflate the radius by heating the base of the atmosphere and reduce atmospheric escape rates by increasing the mean molecular weight of the atmosphere via outgassing. We find that TOI-1266 b's tidal heat flux is comparable to its insolation flux ($F_{\mathrm{tidal}}/F_{\mathrm{insolation}} \simeq 0.9$), and is more than four orders of magnitude larger than the tidal heat flux of TOI-1266 c. This provides another promising opportunity for comparative planetology in the TOI-1266 system.   

\section{Discussion}  \label{sec:planetary compositions}

\subsection{Bulk Density Constraints}

\begin{figure}
%\begin{center}
  \includegraphics[width=8.5cm]{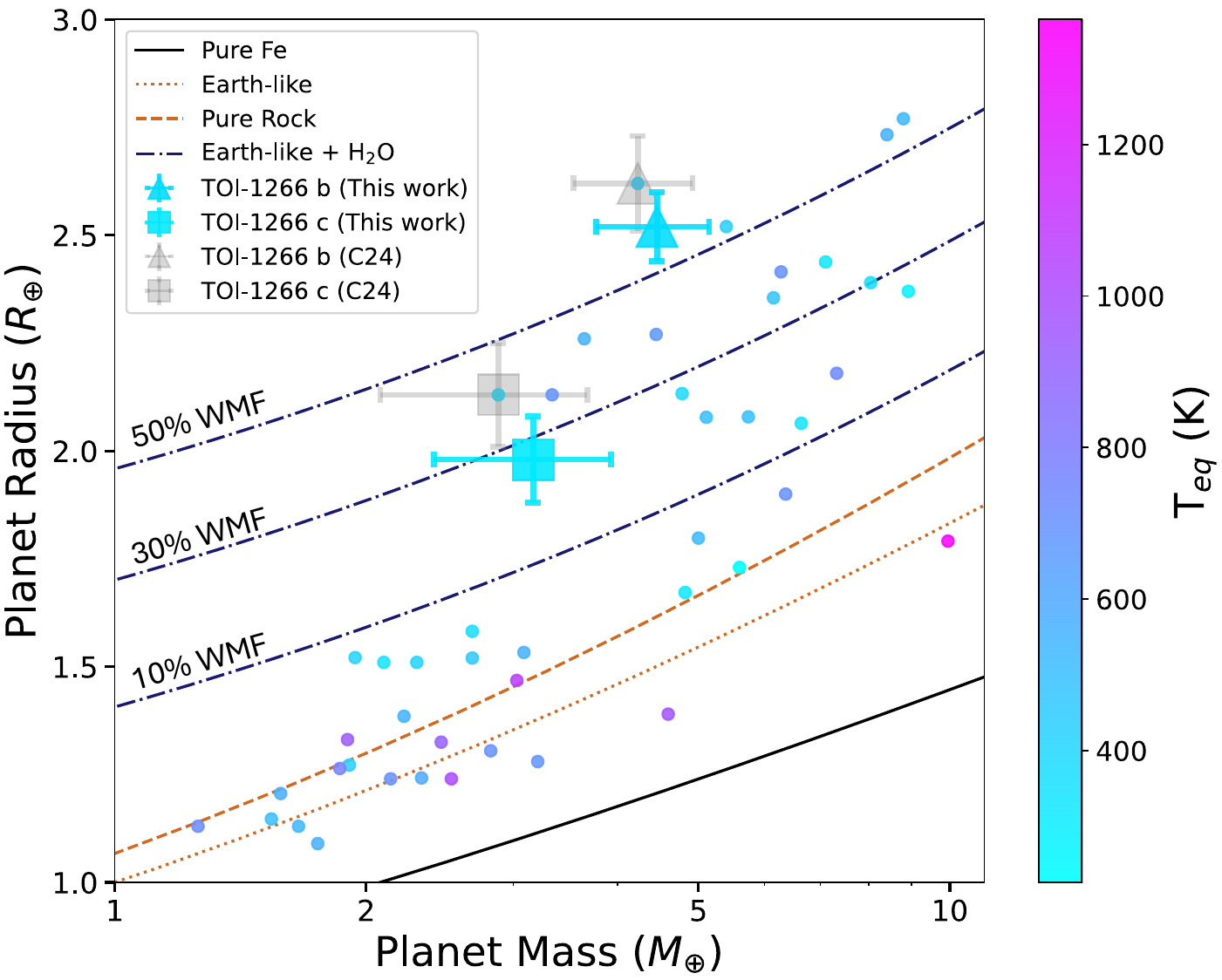}
  \caption{Mass-radius diagram for TOI-1266 b (triangle) and c (square), with updated measurements from this work (blue markers) compared to C24 (gray markers). Filled circles represent all small ($<$ 3 R$_{\oplus}$) planets orbiting M dwarfs (T$_*$ $<$ 3900 K) with masses and radii measured to better than 3$\sigma$, based on the NASA Exoplanet Archive list of confirmed planets as of Aug 23 2024. The predicted equilibrium temperatures of the planets are indicated by the point color. For comparison, we also plot Earth-like water-rich mass-radius curves from \cite{Aguichine2021} and pure iron, Earth-like, and rocky iso-density curves from \cite{zeng_2016}.
  }
  \label{fig:mass_radius}
%\end{center}
\end{figure}

Our updated planetary radii (see \S\ref{sec:radius analysis}) and our updated masses (see \S\ref{sec:Joint TTV RV Modeling}) for TOI-1266 b and c yield slightly smaller radii and larger masses than those reported by C24, and we additionally make small improvements ($\sim$1-3\%) to the fractional mass and radius uncertainties. Figure \ref{fig:mass_radius} compares the new mass and radius constraints for TOI-1266 b and c to the previous constraints, and places this system in the context of the broader population of small planets around M dwarfs with well-measured masses and radii. 

Our updated planet masses and radii do not change the fundamental interpretations of the planet compositions based on the bulk density constraints from C24.  The bulk density of TOI-1266 b is likely too low to be explained solely by rock-water mixtures, given the expected upper limit on water mass fraction (WMF) of $\sim$50\% beyond the snow line for solar composition nebulae \citep{Lodders2003}, although the expected upper limit on WMF for sub-Neptune exoplanets is still debated \citep[e.g.,][]{Bitsch2021,Burn2024,Rogers2025}.  This implies that TOI-1266 b's atmosphere likely contains some hydrogen and helium. TOI-1266 c's smaller radius and larger mass relative to the values reported in C24 now make it easier to match with water-rich scenarios, although it is also consistent with a modest hydrogen-rich envelope  (see \S\ref{sec:interior modeling}). Our updated mass constraint for TOI-1266 c increases the bulk density from 1.6 $\pm$ 0.6 g/cm$^3$ to 2.0 $\pm$ 0.6 g/cm$^3$, reducing the water-to-rock ratio required to explain the observed mass and radius to $\sim$30\%. Part of the shift to smaller radii and larger masses for TOI-1266 b and c is also explained by our choice to adopt the stellar parameters from S20, which are based on effective temperature and surface gravity constraints from spectroscopic observations, while C24 adopted stellar parameters from the TIC v8.2 catalog. S20's stellar mass is larger and their stellar radius is smaller than the values reported in C24.

\subsection{Interior and Atmospheric Compositions} \label{sec:interior modeling}

In principle, we can leverage refractory abundance measurements for the host star to break degeneracies in interior structure models derived from planetary mass and radius measurements \citep{Dorn2017}. This method has been applied to small M dwarf planets in a number of previous studies \citep[e.g.,][]{Mortier2020,Leleu2021,Delrez2021,Demangeon2021,Bonfanti2024,Rosario2024,Cointepas2024}.  However, the relation between star and planet composition is not straightforward \citep[e.g.,][]{Plotnykov2020}, and our measured stellar [Fe/H], [Mg/H], and [Si/H] values are all close to solar values (see \S\ref{sec:Stellar Characterization}).
This suggests that interior modeling tools that assume Earth-like core and mantle compositions should be sufficient to describe the TOI-1266 system. 
%We report Fe, Mg, Si, and many additional stellar abundance measurements for TOI-1266 in Table \ref{Stellar Characterization} to enable potential future study with more detailed consideration of possible interior compositions.

We use the ExoMDN code \cite{Baumeister2023} to model the structure of TOI-1266 b and c using a four-layer model with an iron core, Earth-like silicate mantle, water/high pressure ice layer, and solar metallicity H/He atmosphere. ExoMDN uses mixture density networks trained on millions of synthetic planet models computed using the TATOOINE code \citep{Baumeister2020,MacKenzie2023}. The planets in the training set have masses $<$ 25 M$_{\oplus}$ and T$_{\mathrm{eff}}$ from 100 to 1000 K. Inputs to the model are the planet mass, radius, and equilibrium temperature. This framework has been applied to constrain the range of possible compositions which are degenerate in mass and radius alone for several other super-Earth and sub-Neptune planets \citep[e.g.,][]{Murgas2024,SuarezMascareno2024,Hobson2024}.

We display the parameter distributions from ExoMDN modeling of the four-layer mass fractions for TOI-1266 b and c in Figure \ref{fig:ExoMDN} in the Appendix. We also model the radius fractions for each layer, and find that TOI-1266 b has a radius fraction of 0.29$^{+0.15}_{-0.14}$ for its potential H/He layer, while TOI-1266 c has a potential H/He radius fraction of 0.14$^{+0.20}_{-0.11}$.

Despite slightly refined planetary mass and radius measurements, the interpretation of the planetary bulk compositions are unchanged relative to C24. Our water and atmospheric mass fractions for both planets are consistent within 1$\sigma$. TOI-1266 b is likely a gas-enveloped terrestrial with an envelope mass fraction $<$ 12\% (95\% upper limit), though we cannot rule out a water-rich interior. TOI-1266 c is more consistent with a water-rich interior without a significant H/He envelope ($m_{env} = 0.0^{+0.03}_{-0.00}\%$), though we cannot rule out a terrestrial composition with a small H/He atmospheric mass fraction. The bulk densities of TOI-1266 b and c are discrepant from each other by 1$\sigma$ but the compositions for both planets remain degenerate using mass, radius, and equilibrium temperature alone. 

\subsection{Possible Explanations for the Inverted Architecture of TOI-1266}  \label{sec:discussion}

Although TOI-1266 b and c both lie on the larger side of the radius gap, they have different densities and correspondingly distinct constraints on their mantle compositions. Relative to the proposed rocky, water-rich, and gaseous sub-Neptune populations proposed by \cite{Luque_2022}, TOI-1266 c lies on the border between water worlds and sub-Neptunes, while TOI-1266 b is firmly aligned with the sub-Neptune population. If TOI-1266 c has a significantly higher water mass fraction than TOI-1266 b, it might suggest that planet c formed outside the water ice line while b originated from inside this line. However, this raises the question of why TOI-1266 c did not also accrete significant quantities of hydrogen and helium given its more distant formation location. If both planets instead host hydrogen-rich atmospheres, some mechanism is required to explain why TOI-1266 b was able to accrete and retain a more massive primordial atmosphere than TOI-1266 c. TOI-1266 b has a larger mass than TOI-1266 c, which increases the gravitational potential well and reduces the predicted mass loss rate. However, planet b is also more highly irradiated than planet c, and atmospheric escape modeling in C24 indicates that the predicted mass loss rates for the two planets are not different enough to explain their differing radii if both planets had similar primordial atmospheric compositions. This suggests that TOI-1266 c is more consistent with a water-rich composition.

This picture is additionally complicated by the discovery that TOI-1266~b has a significant tidal heat flux (see \S\ref{sec:eccs and tides}). There are very few compact multi-planet systems with eccentrities as large as our preferred value for TOI-1266 b \citep[e.g.,][]{VanEylen19, He2020}. TOI-1266~b's large tidal heat flux means that its atmosphere may be less massive and/or more metal-rich than implied by the models described in \S\ref{sec:planetary compositions}, and might help to reduce the disparity in inferred envelope masses relative to TOI-1266 c.
%If this results in significant volcanic outgassing, it could enrich the atmospheric metallicity of TOI-1266 b to create a high mean molecular weight atmosphere robust against photoevaporation, while the lower expected tidal heat flux of TOI-1266 c could not. 

%Modeling in C24 finds that both planets are capable of accreting massive envelopes in excess of what their present-day masses and radii suggest, but that any model of atmospheric escape including XUV-driven photoevaporation is difficult to reconcile with the observed radii of both TOI-1266 b and c in the case where both planets start as H$_2$-rich Sub-Neptunes. This is because models that reproduce the observed radius of TOI-1266 c result in too much mass loss to explain the radius of b, when assuming solar metallacity envelopes. If TOI-1266 b has a highly metal-rich atmosphere, however, it may be possible to reconcile the observed radii atmospheric mass fractions under XUV irradiation. The significant tidal heat flux and potential volcanic outgassing for TOI-1266 b is one possible method to enhance the atmospheric metallicity and make b less susceptible to escape than c, consistent with the observed radii and masses.

We can also leverage the unique dynamical configuration of this system to constrain the formation and migration histories of the three planets. The classic picture of close-in sub-Neptune formation assumes that these systems undergo convergent migration that places them in mean-motion resonant chains. These resonant chains then destabilize on $\sim$100 Myr timescales, leaving the planets in orbits that are just wide of commensurability \citep{Batygin_Morbi_2023}. The planet formation simulations of \cite{Batygin_Morbi_2023} produce systems that are on average more mass-uniform than the Kepler sample overall, while post-nebular instabilities and collisions can degrade the mass-uniformity enough to match the Kepler sample for most systems \citep{Goldberg_2022}. But the TOI-1266 system is very uniform in mass (consistent with the resonant subset of Kepler multi-planet systems described in \cite{Goldberg_2021}) and close to resonance, which implies that the current system may be a pristine remnant of planet formation, where the post-nebular relaxation away from resonance happened through some soft instability that did not lead to orbit crossing. This would imply that the differences in atmospheric mass fraction for TOI-1266 b and c are primordial.

Alternatively, \cite{Li2024} recently suggested that some planets can experience major collisions after the destabilization of the original resonance, resulting in a broader range of period ratios. If the three TOI-1266 planets were originally located in a resonant chain (e.g., adjacent pairs in the 3:2 resonance), our simulations predict that the resonance would be destabilized by tidal damping on relatively short timescales (\S\ref{sec:eccs and tides}). If this led to mergers, it could have shifted the orbital period ratios of the b/c and c/d pairs wide of the original resonances (e.g., a chain of 1.5 period ratios) to their current values near 1.7. If TOI-1266 c underwent a giant impact but TOI-1266 b did not, it might also explain the different atmospheric mass fractions of the two planets in the case where the inflated radius of TOI-1266 b relative to c is not due to its larger tidal heat flux.

\section{Future Observations} \label{sec:future obs}

\subsection{TTV Follow-up}

There is still much to be gained by additional TTV monitoring of the TOI-1266 system. As shown in Figure \ref{fig:ttvplot_joint}, we have yet to observe a full TTV super-period for TOI-1266 b. This super-period is caused by its proximity to the 3:1 resonance with TOI-1266 d and has a predicted duration of $\sim$2000 days \citep[][Equation 5]{Lithwick_2012}, while the current observational baseline is only 1700 days.  This suggests that future TESS and ground-based transit observations over the next decade could significantly improve on current TTV-based mass and eccentricity constraints.  To that end, we provide a list of predicted transit times for all three planets from our dynamical fit in Table \ref{tab:Predicted transits} in the Appendix.

For planets with circular orbits, there is a well-known degeneracy between the planet masses and eccentricities when fitting the observed TTV signal \citep{Lithwick_2012}. The nonzero orbital eccentricities of planets b and/or d create a short-period chopping signature in the measured TTVs for planet b, allowing us to break this degeneracy in our TTV fits \citep[e.g.,][]{Nesvorny2014,Deck_2015}.  
 
In the future, repeated high-precision timing measurements for TOI-1266 b over shorter timescales could be used to more fully sample this $\sim$2-3 minute chopping signature and further improve the planetary mass and eccentricity constraints. The median TESS timing uncertainty for planet b is larger than the amplitude of the chopping TTV signature, but our most precise WIRC observations have uncertainties of less than two minutes. Transit observations with this level of timing precision could significantly improve the eccentricity and therefore tidal parameter constraints.     

TOI-1266 c's location between the near-resonant pair of planets b and d means that its estimated mass and eccentricity are sensitive to the corresponding values of these parameters for TOI-1266 b and d in our current fits. As a result, additional high-precision timing measurements of TOI-1266 c will also improve the mass and eccentricity constraints of all three planets. Improved eccentricity measurements are essential in order to better constrain the tidal heat fluxes of the two transiting planets, enabling a more detailed investigation of the likely planetary interior states, mantle melt fractions, atmospheric outgassing rates, and corresponding atmospheric chemistry.

\subsection{Improved Interior and Atmospheric Mass Loss Modeling}

TOI-1266 has upcoming Cycle 23 XMM-Newton observations. These observations will measure the XUV flux of TOI-1266, providing an improved constraint on the planetary radiation environment. An estimate of the stellar high-energy flux is needed in order to calculate mass loss timescales for solar metallicity atmospheres, potentially yielding improved constraints on the likely envelope compositions of these two planets \citep[e.g.,][]{Diamond-Lowe2022}. A high predicted mass loss rate for a solar metallicity envelope around TOI-1266 b would indicate that it likely has a high ($>100\times$ solar) atmospheric metallicity that acts to suppress photoevaporative mass loss \citep{Zhang2022}. If TOI-1266 c is unable to retain a solar metallicity gas envelope, it could either have a high atmospheric metallicity or a water-dominated envelope as modeled in \cite{Harman_2022} and \cite{Yoshida2022}.

\subsection{Atmospheric Characterization}

A comparative study of the atmospheric compositions of TOI-1266 b and c would tell us a great deal about possible differences in their formation environment, internal heat fluxes, and atmospheric mass loss histories.  As discussed in \S\ref{sec:intro}, published atmospheric absorption detections for K2-18 b and TOI-270 d \citep{Madhusudhan2023,Benneke2024} indicate that colder planets may have less opacity from photochemical hazes in their upper atmospheres, and should therefore be high priority targets for atmospheric characterization via transmission spectroscopy. The low equilibrium temperatures of TOI-1266 b and c therefore make them especially valuable targets for transmission spectroscopy studies with JWST. 
%Is TOI-1266 c a water-rich planet reflecting the pristine conditions of formation beyond the ice line? Or is the reason why the more highly irradiated TOI-1266 b retained its primordial H-rich atmosphere and TOI-1266 c did not because of differences in atmospheric metallicity, potentially related to volcanic outgassing from tidal heat? The answers to these questions are readily accessible in the TOI-1266 system.
TOI-1266 c was previously identified by \cite{Hord2024} as one of the most favorable planets in its size and temperature range for transmission spectroscopy. Our updated masses and radii yield Transmission Spectroscopy Metric \citep[TSM,][]{Kempton2018} values of 120 and 69 for TOI-1266 b and c, respectively. For planets in the sub-Neptune size category ($1.5 - 2.75$ R$_{\oplus}$), this makes TOI-1266 b a more favorable target than all others with scheduled JWST observations except LP 791-18 c and L 98-59 d, and TOI-1266 c one of the five most favorable targets cooler than 400 K. 

If TOI-1266 b has a relatively low atmospheric metallicity (e.g., $1-10\times$ solar, measured by the abundance of CH$_4$), then the trace abundances of CO and CO$_2$ in the upper atmosphere can be used to constrain the planet’s internal heat flux due to tides \citep[e.g.,][]{Fortney2020,Hu2021}. If TOI-1266 b has a more metal-rich envelope (e.g. $>100\times$ solar), it would have detectable amounts of CO$_2$ even without tidal heating \citep[e.g.][]{Wogan2024}, but the additional tidal heat flux would produce more CO$_2$ and deplete NH$_3$ in the part of the atmosphere probed by transmission spectroscopy. This means that precise abundance measurements for these molecules can be used to constrain the internal heat flux of TOI-1266 b \citep[e.g.][]{Yang2024}.

If transmission spectroscopy reveals that TOI-1266 c hosts a water-rich envelope, it would also be valuable to explore how much water might be sequestered into the planetary core versus the atmosphere. \cite{Dorn2021} presented an interior model for water-rich rocky planets that incorporates the effects of rock melting and the redistribution of water between magma ocean and atmosphere on the planet radius. This study found that accounting for these effects can lead to deviations in planet radius of up to 16\%. Our fractional radius uncertainty is already lower than this value, suggesting that it may be possible to use the bulk density and tidal heat flux constraints to assess possible melt fractions in the core and quantify the corresponding distribution of water in TOI-1266 c's interior. 

\section{Summary and Conclusions} \label{sec:summary}

In this study, we present updated constraints on the properties of the two temperate sub-Neptune-sized transiting planets in the TOI-1266 system. We analyze new TESS data that extend the TTV baseline to $\sim$4 years, add ten new ground-based transit observations, and incorporate information from archival HARPS-N spectrograph RV measurements in a joint TTV+RV fit. Our joint TTV+RV modeling yields updated mass, radius, and eccentricity measurements that allow us to refine constraints on the planetary compositions and tidal heat fluxes. We summarize our main conclusions below.

First, we use our new transit timing measurements to confirm the existence of a third planet in the system, TOI-1266 d. This companion was previously identified as a planet candidate by \cite{Cloutier_2024} using radial velocity observations. Our dynamical fit reveals that this planet is located on an exterior orbit near the 3:1 MMR with TOI-1266 b and has an orbital period of 32.5 days.
We use the dynamical model from our joint TTV+RV fit to phase up the TESS photometry around its expected transit times and confirm that it is not currently transiting.  However, our dynamical simulations suggest that it may become transiting on timescales of tens to hundreds of years. Our joint TTV+RV fit provides much tighter constraints on the orbital properties of all three planets. We find that although planets b and d are located close to the 3:1 MMR, there are no librating two- or three-body resonant angles in the system. Some combinations of planetary parameters yield a librating two-body resonance between TOI-1266 b and d, but long-term dynamical simulations show that this resonance state is not stable.

Our joint TTV+RV fit also provides improved constraints on the masses and radii of the two inner transiting planets, and on the mass of the outer non-transiting planet. First, we use two additional sectors of TESS data and new ground-based transit observations to resolve the tension in transit depth for TOI-1266 c between the TESS prime and extended missions reported in \cite{Cloutier_2024}. We conclude that the scatter in observed transit depths is consistent with stochastic variations. We find an updated mass and radius measurement of $M_b$ = 4.46 $\pm$ 0.69 $M_{\oplus}$ and $R_b$ = 2.52 $\pm$ 0.08 $R_{\oplus}$ for TOI-1266~b, and $M_c$ = 3.17 $\pm$ 0.76 $M_{\oplus}$ and $R_c$ = 1.98 $\pm$ 0.10 $R_{\oplus}$ for TOI-1266~c. Separately, we measure a mass of $M_d$ = 3.68$^{+1.05}_{-1.11}$ for TOI-1266 d in joint fits where we assume that all three planets are coplanar. Finally, we confirm that the planet masses and orbital parameters from our joint TTV+RV fit are consistent with results obtained by fitting each data set individually.

Our updated planetary interior and composition modeling confirms that TOI-1266 b requires a hydrogen-rich envelope to explain its observed mass and radius. If we assume that this planet has an Earth-like rocky core and a solar composition envelope, this corresponds to a H/He atmospheric mass fraction of $0.02^{+0.05}_{-0.02}$\%. TOI-1266 c has a modestly higher bulk density ($\rho_b$ = 2.24 g/cc, $\rho_c$ = 1.54 g/cc) and can therefore be matched with either a small hydrogen-rich envelope with an atmospheric mass fraction ($< 0.06 \%$ at 95\% confidence), or a water-rich envelope with a mass fraction of $0.50^{+0.36}_{-0.42}$\% from interior structure models, though we caution that much of this range is unsupported by formation models which predict maximum water mass fractions for terrestrial planets $< 50\%$. This picture is additionally complicated by our discovery that planet b has a significantly nonzero orbital eccentricity, corresponding to a large tidal heat flux. We simulate the long-term evolution of the system including the effects of tides and find that TOI-1266~b must have relatively weak tidal dissipation ($Q$/$k_2>1200$) in order to maintain this eccentricity over Gyr timescales. In the future, transmission spectroscopy observations of TOI-1266~b with JWST could constrain its internal heat flux, while equivalent observations of TOI~1266~c are needed in order to determine the bulk water content of its envelope.
Both planets are among the most favorable temperate sub-Neptune candidates for transmission spectroscopy with JWST, and a detection of atmospheric absorption for either planet would significantly improve our understanding of the TOI-1266 system. 

\section{Acknowledgments}

We thank the Palomar Observatory telescope operators, support astronomers, hospitality, and administrative staff for their support. We are especially grateful to Monastery keeper Jeff. We thank all of the Palomar chefs -- gone from the Monastery but never forgotten -- the ones who made observing at Palomar a uniquely pleasant experience amongst astronomers. Part of this program was supported by JPL Hale telescope time allocations. We are thankful to the PARVI team and Palomar Observatory directorate, especially Chas Beichman, Aurora Kesseli, and Andy Boden for their gracious support of the Palomar TTV survey program during periods requiring quick readjustment of the 200-inch observing schedule. We benefited from useful conversations with Fei Dai, George King, Luke Bouma, and Julie Inglis. Portions of code used to create and edit figures and tables in this manuscript were generated with assistance from Anthropic's Claude 3.5 Sonnet.

This paper is based on observations made with the MuSCAT3 instrument, developed by the Astrobiology Center and under financial supports by JSPS KAKENHI (JP18H05439) and JST PRESTO (JPMJPR1775), at Faulkes Telescope North on Maui, HI, operated by the Las Cumbres Observatory. This work is partly supported by JSPS KAKENHI Grant Numbers JP21K13955, JP24H00017, JP24K00689, JP24K17083, JP24K17082, JP24H00248, JSPS Bilateral Program Number JPJSBP120249910, JSPS Grant-in-Aid for JSPS Fellows Grant Number JP24KJ0241, Astrobiology Center SATELLITE Research project AB022006, and JST SPRING, Grant Number JPMJSP2108.

This research was supported from the Wilf Family Discovery Fund in Space and Planetary Science established by the Wilf Family Foundation.  This research has made use of the NASA Exoplanet Archive and the Exoplanet Follow-up Observation Program website, which are operated by the California Institute of Technology, under contract with the National Aeronautics and Space Administration under the Exoplanet Exploration Program. The research made use of the Swarthmore transit finder online tool \citep{SwarthmoreTTF}. We acknowledge the use of public TESS data from pipelines
at the TESS Science Office and at the TESS Science Processing
Operations Center. This paper includes data collected by the
TESS mission that are publicly available from the Mikulski
Archive for Space Telescopes (MAST) at the Space Telescope Science Institute (STScI). STScI is operated by the Association of
Universities for Research in Astronomy, Inc., under NASA contract NAS5-26555. Support for MAST
for non-HST data is provided by the NASA Office of
Space Science via grant NNX13AC07G and by other
grants and contracts. The specific TESS sectors used in this work can be accessed
via doi:10.17909/d4f2-t519. Funding for the TESS mission
is provided by NASA’s Science Mission Directorate.

This research made use of \textsf{exoplanet} \citep{exoplanet:joss,
exoplanet:zenodo} and its dependencies \citep{exoplanet:foremanmackey17,
exoplanet:foremanmackey18, exoplanet:agol20, exoplanet:arviz,
exoplanet:astropy13, exoplanet:astropy18, exoplanet:luger18, exoplanet:pymc3,
exoplanet:theano}. 

\facilities{ADS, NASA Exoplanet Archive, TESS, Palomar 200-inch (WIRC), NAOJ 188cm (MuSCAT), LCOGT (MuSCAT3 \& Sinistro)}

\software{\texttt{astropy} \citep{astropy},
\texttt{scipy} \citep{scipy},
\texttt{numpy} \citep{numpy},
\texttt{matplotlib} \citep{matplotlib},
\texttt{rebound} \citep{rebound}, 
\texttt{BATMAN} \citep{batman},
\texttt{emcee} \citep{emcee},
\texttt{corner} \citep{corner},
\texttt{exoplanet} \citep{exoplanet:joss},
\texttt{RadVel} \citep{Radvel},
\texttt{TTVFast} \citep{TTVFast},
\texttt{reboundx} \citep{Tamayo2020},
\texttt{corner} \citep{corner}, and
\texttt{lightkurve} \citep{lightkurve},
\texttt{The Cannon} (Behmard et al. in prep),
\texttt{Claude 3.5 Sonnet.}}


%\vspace{5mm}




\appendix

\section{Transit Times and Posterior Probability Distributions} \label{appendix}

\begin{table*}[htbp]
\centering
\caption{Observed Transits of TOI-1266 b and c}
\small
\begin{tabular}{lcccc}
\hline
Planet & Source\textsuperscript{a} & Transit Number & Transit Time\textsuperscript{b} & Uncertainty \\
\hline
TOI-1266 b & TESS PM & 0 & 1691.004273 & 0.004655648 \\
TOI-1266 b & TESS PM & 1 & 1701.898951 & 0.005405627 \\
TOI-1266 b & TESS PM & 2 & 1712.796673 & 0.004117942 \\
TOI-1266 b & TESS PM & 4 & 1734.586684 & 0.002735273 \\
TOI-1266 b & D20 & 14 & 1843.535750 & 0.001580000 \\
TOI-1266 b & TESS PM & 17 & 1876.219970 & 0.001899578 \\
TOI-1266 b & TESS PM & 18 & 1887.115710 & 0.002404801 \\
TOI-1266 b & D20 & 20 & 1908.901470 & 0.001030000 \\
TOI-1266 b & TESS PM & 21 & 1919.798369 & 0.001747413 \\
TOI-1266 b & D20 & 22 & 1930.692530 & 0.001270000 \\
TOI-1266 b & D20 & 23 & 1941.590140 & 0.001480000 \\
TOI-1266 b & D20 & 25 & 1963.377320 & 0.002670000 \\
TOI-1266 b & MuSCAT & 57 & 2312.013329 & 0.000771316 \\
TOI-1266 b & MuSCAT3 & 58 & 2322.907835 & 0.000357439 \\
TOI-1266 b & TESS EM & 67 & 2420.963088 & 0.001549753 \\
TOI-1266 b & TESS EM & 68 & 2431.856823 & 0.001426380 \\
TOI-1266 b & TESS EM & 69 & 2442.751478 & 0.001556668 \\
TOI-1266 b & MuSCAT3 & 84 & 2606.172940 & 0.000626247 \\
TOI-1266 b & TESS EM & 85 & 2617.067416 & 0.001701500 \\
TOI-1266 b & TESS EM & 86 & 2627.962889 & 0.001367946 \\
TOI-1266 b & WIRC & 87 & 2638.857341 & 0.000514139 \\
TOI-1266 b & MuSCAT3 & 87 & 2638.859811 & 0.001039443 \\
TOI-1266 b & TESS EM & 88 & 2649.750889 & 0.001525124 \\
TOI-1266 b & TESS EM & 89 & 2660.645312 & 0.001736380 \\
TOI-1266 b & WIRC & 96 & 2736.907201 & 0.002131367 \\
TOI-1266 b & TESS SEM & 152 & 3347.007129 & 0.002401229 \\
TOI-1266 b & TESS SEM & 154 & 3368.796296 & 0.001593378 \\
TOI-1266 b & TESS SEM & 155 & 3379.692673 & 0.002864712 \\
TOI-1266 b & TESS SEM & 156 & 3390.588036 & 0.002639571 \\
\hline
TOI-1266 c & TESS PM & 0 & 1689.960319 & 0.042524505 \\
TOI-1266 c & TESS PM & 2 & 1727.560744 & 0.007774661 \\
TOI-1266 c & D20 & 10 & 1877.975750 & 0.003210000 \\
TOI-1266 c & TESS PM & 11 & 1896.775891 & 0.009553679 \\
TOI-1266 c & MuSCAT3 & 35 & 2348.011768 & 0.001941440 \\
TOI-1266 c & TESS EM & 39 & 2423.229205 & 0.005522536 \\
TOI-1266 c & TESS EM & 40 & 2442.021347 & 0.003062748 \\
TOI-1266 c & MuSCAT & 49 & 2611.240567 & 0.001092947 \\
TOI-1266 c & TESS EM & 49 & 2611.248289 & 0.003351727 \\
TOI-1266 c & TESS EM & 50 & 2630.041436 & 0.003002736 \\
TOI-1266 c & Sinistro & 51 & 2648.843530 & 0.002544507 \\
TOI-1266 c & WIRC & 51 & 2648.843729 & 0.001267564 \\
TOI-1266 c & TESS EM & 51 & 2648.845373 & 0.003014198 \\
TOI-1266 c & WIRC & 56 & 2742.873075 & 0.015162918 \\
TOI-1266 c & TESS SEM & 89 & 3363.304827 & 0.003415930 \\
TOI-1266 c & TESS SEM & 90 & 3382.112135 & 0.002607625 \\
\hline
\end{tabular}
\label{tab:Observed transits}
\vspace{0.5cm}
\footnotesize
\begin{flushleft}
Transit observation sources are D20 (\cite{Demory_2020}), TESS (PM: Primary Mission, EM: Extended Mission, SEM: Second Extended Mission), WIRC, MuSCAT, MuSCAT3, or Sinistro. Transit times are in units of BJD - 2457000 and uncertainties are in units of days.
\end{flushleft}
\end{table*}

\begin{table*}[htbp]
\centering
\caption{Predicted Transit Times for TOI-1266 b and c}
\small
\begin{tabular}{lccccc}
\hline
Planet & Transit Number & Midtime\textsuperscript{a} & +1$\sigma$ Uncertainty & -1$\sigma$ Uncertainty \\
\hline
TOI-1266 b & 0 & 1691.0067 & 0.0017 & 0.0017 \\
TOI-1266 b & 1 & 1701.9012 & 0.0019 & 0.0024 \\
TOI-1266 b & 2 & 1712.7960 & 0.0021 & 0.0030 \\
TOI-1266 b & 3 & 1723.6924 & 0.0028 & 0.0037 \\
TOI-1266 b & 4 & 1734.5869 & 0.0032 & 0.0042 \\
... & ... & ... & ... & ... \\
% ... [Additional rows for TOI-1266 b] ...
TOI-1266 b & 288 & 4828.7047 & 0.1394 & 0.2920 \\
TOI-1266 b & 289 & 4839.5993 & 0.1399 & 0.2925 \\
TOI-1266 b & 290 & 4850.4937 & 0.1401 & 0.2933 \\
\hline
TOI-1266 c & 0 & 1689.9588 & 0.0026 & 0.0028 \\
TOI-1266 c & 1 & 1708.7610 & 0.0048 & 0.0038 \\
TOI-1266 c & 2 & 1727.5597 & 0.0063 & 0.0082 \\
TOI-1266 c & 3 & 1746.3620 & 0.0081 & 0.0099 \\
TOI-1266 c & 4 & 1765.1629 & 0.0093 & 0.0102 \\
... & ... & ... & ... & ... \\
% ... [Additional rows for TOI-1266 c] ...
TOI-1266 c & 166 & 4811.0398 & 0.2963 & 0.2074 \\
TOI-1266 c & 167 & 4829.8429 & 0.2988 & 0.2081 \\
TOI-1266 c & 168 & 4848.6433 & 0.2995 & 0.2131 \\
\hline
\end{tabular}
\label{tab:Predicted transits}
\vspace{0.5cm}
\footnotesize
\begin{flushleft}
\textsuperscript{a}Midtimes are in units of BJD - 2457000. Uncertainties are in units of days. Only a subset of rows are depicted here for conciseness. The entirety of this table is provided in the arXiv source code.
\end{flushleft}
\end{table*}

%\section{Corner Plots}

\begin{figure*}
\begin{center}
  \includegraphics[width=15.5cm]{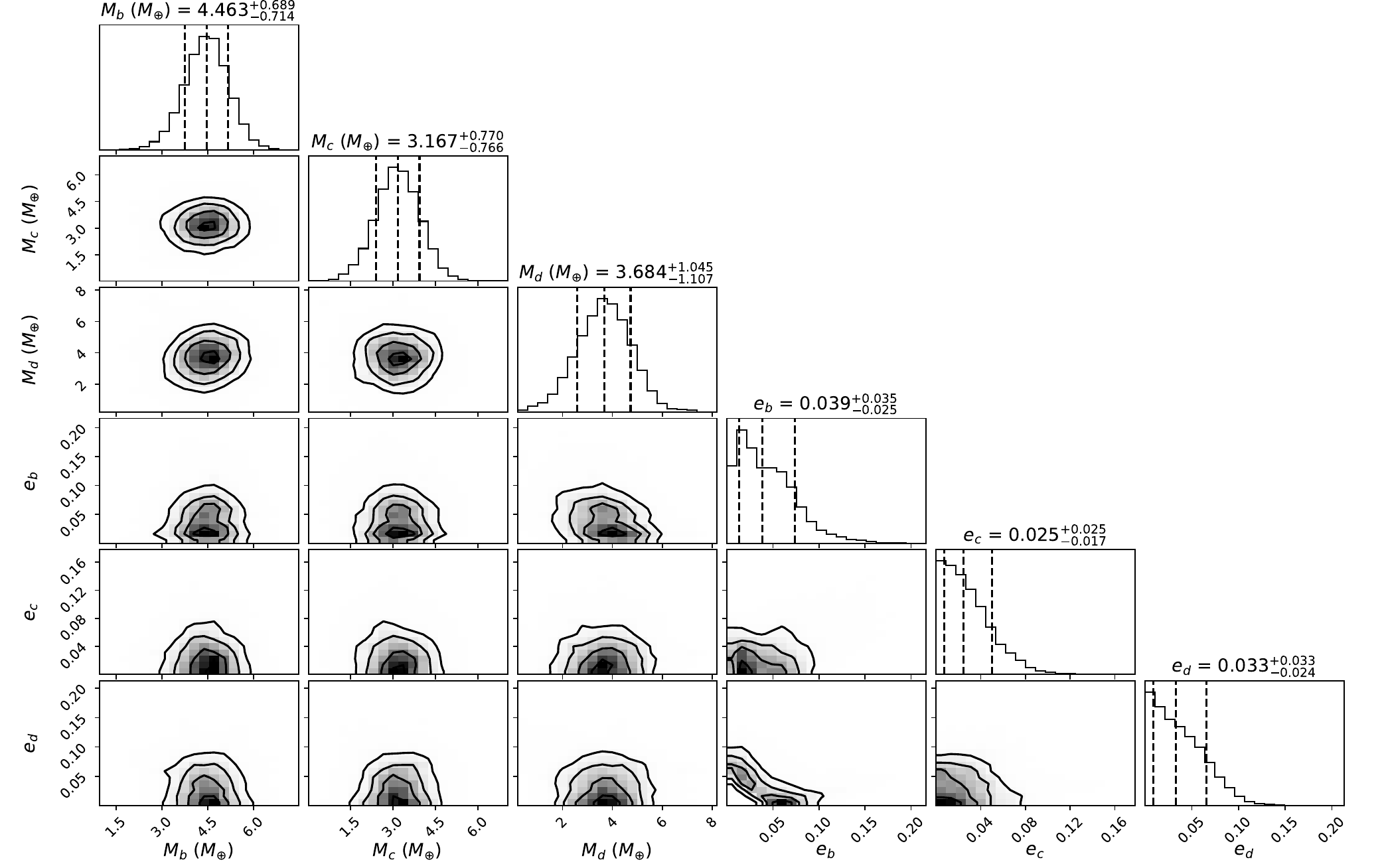}
  \caption{Corner plot of posteriors for planetary masses and eccentricities for TOI-1266 b, c, and d from the TTV+RV joint fit, made with the \texttt{corner} package \citep{corner}}
  \label{fig:Joint_MassEcc_Corner}
\end{center}
\end{figure*}

\begin{figure*}
\begin{center}
  \includegraphics[width=\textwidth]{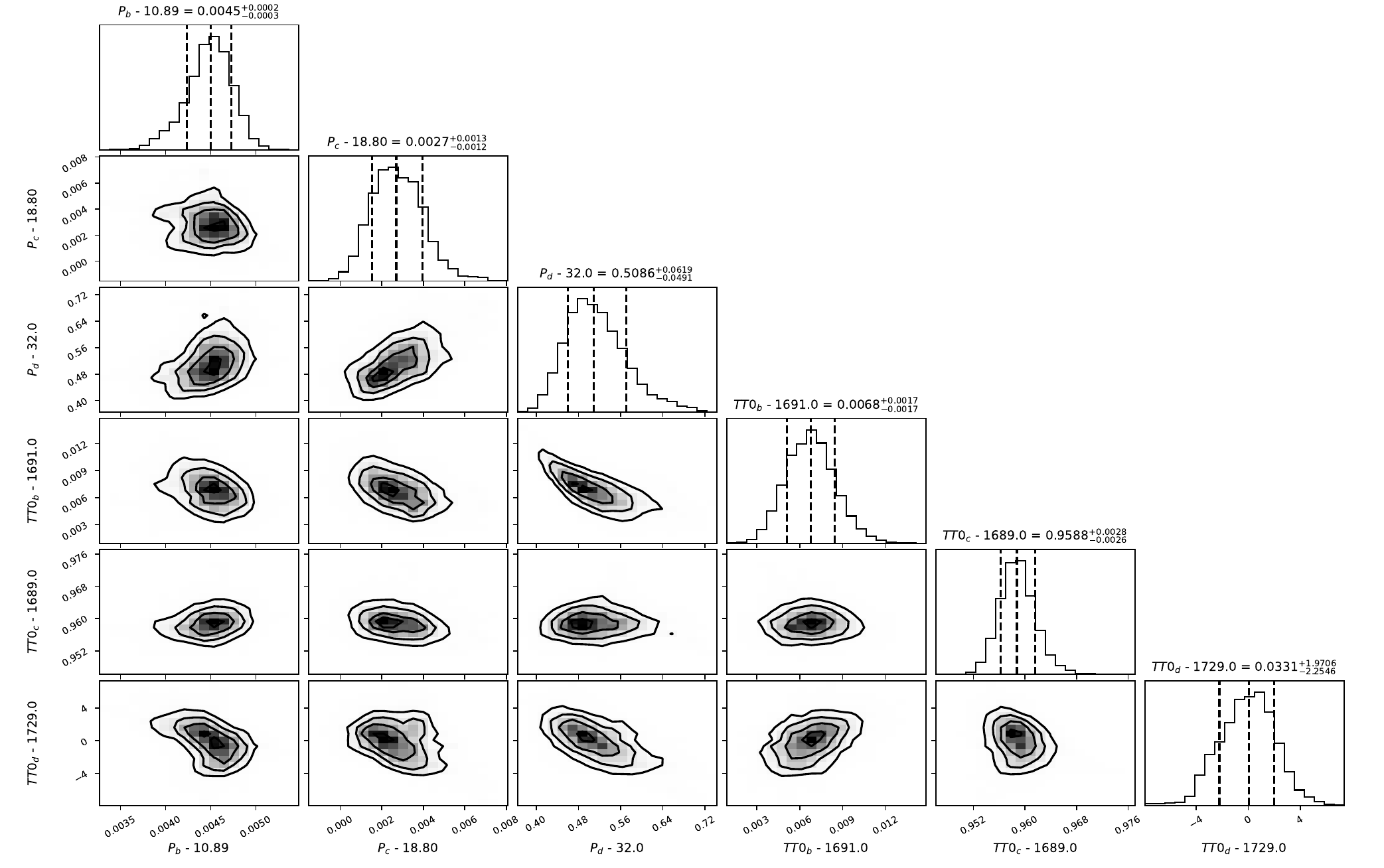}
  \caption{Corner plot of posteriors for planetary orbital period and initial transit time for TOI-1266 b, c, and d from the TTV+RV joint fit, made with the \texttt{corner} package. Modified periods are reported in units of days, while moditified initial transit times are in units of BJD -2457000.}
  \label{fig:Joint_Ephem_Corner}
\end{center}
\end{figure*}

Figures \ref{fig:Joint_MassEcc_Corner} and \ref{fig:Joint_Ephem_Corner} show corner plots of the posterior distributions from our joint fit to the TTV and RV data with a 3-planet model. The parameters in our fits were the planet-to-star mass ratio and $\sqrt{e}\cos(\omega)$, $\sqrt{e}\sin(\omega)$, but we have converted these distributions into units of Earth masses and eccentricity for ease of reference in Figure \ref{fig:Joint_MassEcc_Corner}. Detrended light curves for all new ground-based transit observations presented in this work, and our stacked TESS transit profiles, are available in the TeX source code documents for this paper.

\begin{figure*}
\begin{center}
  \includegraphics[width=11.0cm]{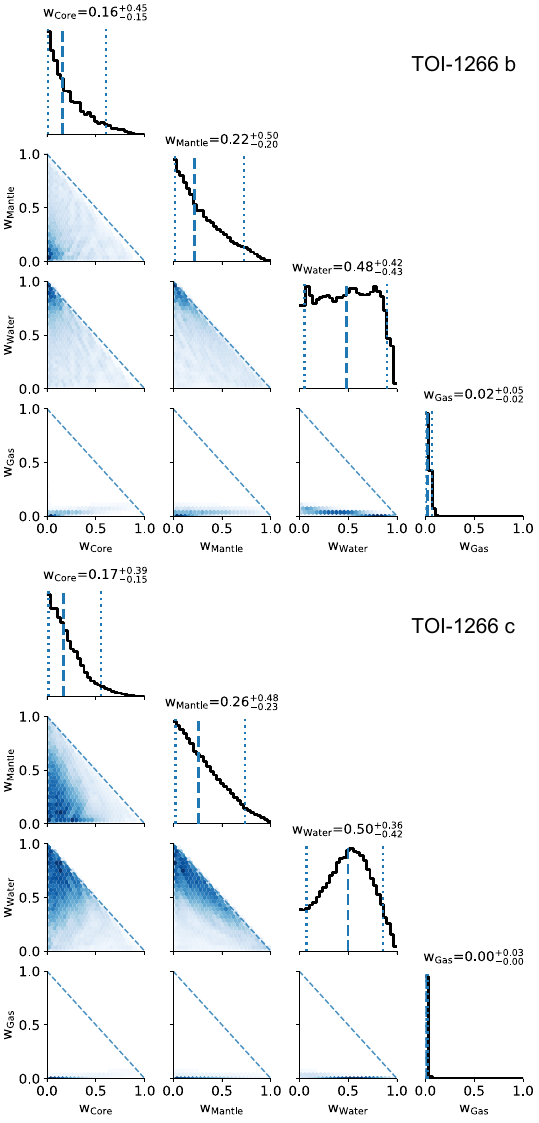}
  \caption{Corner plots of the mass fractions from the four-layer ExoMDN interior structure models of TOI-1266 b (top) and TOI-1266 c (bottom).}
  \label{fig:ExoMDN}
\end{center}
\end{figure*}


\bibliography{citations1266}{}

\begin{thebibliography}{}
\expandafter\ifx\csname natexlab\endcsname\relax\def\natexlab#1{#1}\fi
\providecommand{\url}[1]{\href{#1}{#1}}
\providecommand{\dodoi}[1]{doi:~\href{http://doi.org/#1}{\nolinkurl{#1}}}
\providecommand{\doeprint}[1]{\href{http://ascl.net/#1}{\nolinkurl{http://ascl.net/#1}}}
\providecommand{\doarXiv}[1]{\href{https://arxiv.org/abs/#1}{\nolinkurl{https://arxiv.org/abs/#1}}}

\bibitem[{{Agol} {et~al.}(2020){Agol}, {Luger}, \& {Foreman-Mackey}}]{exoplanet:agol20}
{Agol}, E., {Luger}, R., \& {Foreman-Mackey}, D. 2020, \aj, 159, 123, \dodoi{10.3847/1538-3881/ab4fee}

\bibitem[{{Agol} {et~al.}(2021){Agol}, {Dorn}, {Grimm}, {Turbet}, {Ducrot}, {Delrez}, {Gillon}, {Demory}, {Burdanov}, {Barkaoui}, {Benkhaldoun}, {Bolmont}, {Burgasser}, {Carey}, {de Wit}, {Fabrycky}, {Foreman-Mackey}, {Haldemann}, {Hernandez}, {Ingalls}, {Jehin}, {Langford}, {Leconte}, {Lederer}, {Luger}, {Malhotra}, {Meadows}, {Morris}, {Pozuelos}, {Queloz}, {Raymond}, {Selsis}, {Sestovic}, {Triaud}, \& {Van Grootel}}]{Agol_2021}
{Agol}, E., {Dorn}, C., {Grimm}, S.~L., {et~al.} 2021, \psj, 2, 1, \dodoi{10.3847/PSJ/abd022}

\bibitem[{{Aguichine} {et~al.}(2021){Aguichine}, {Mousis}, {Deleuil}, \& {Marcq}}]{Aguichine2021}
{Aguichine}, A., {Mousis}, O., {Deleuil}, M., \& {Marcq}, E. 2021, \apj, 914, 84, \dodoi{10.3847/1538-4357/abfa99}

\bibitem[{{Alibert} \& {Benz}(2017)}]{Alibert_bez_2017}
{Alibert}, Y., \& {Benz}, W. 2017, \aap, 598, L5, \dodoi{10.1051/0004-6361/201629671}

\bibitem[{{Astropy Collaboration} {et~al.}(2013){Astropy Collaboration}, {Robitaille}, {Tollerud}, {Greenfield}, {Droettboom}, {Bray}, {Aldcroft}, {Davis}, {Ginsburg}, {Price-Whelan}, {Kerzendorf}, {Conley}, {Crighton}, {Barbary}, {Muna}, {Ferguson}, {Grollier}, {Parikh}, {Nair}, {Unther}, {Deil}, {Woillez}, {Conseil}, {Kramer}, {Turner}, {Singer}, {Fox}, {Weaver}, {Zabalza}, {Edwards}, {Azalee Bostroem}, {Burke}, {Casey}, {Crawford}, {Dencheva}, {Ely}, {Jenness}, {Labrie}, {Lim}, {Pierfederici}, {Pontzen}, {Ptak}, {Refsdal}, {Servillat}, \& {Streicher}}]{exoplanet:astropy13}
{Astropy Collaboration}, {Robitaille}, T.~P., {Tollerud}, E.~J., {et~al.} 2013, \aap, 558, A33, \dodoi{10.1051/0004-6361/201322068}

\bibitem[{{Astropy Collaboration} {et~al.}(2018{\natexlab{a}}){Astropy Collaboration}, {Price-Whelan}, {Sip{\H o}cz}, {G{\"u}nther}, {Lim}, {Crawford}, {Conseil}, {Shupe}, {Craig}, {Dencheva}, {Ginsburg}, {VanderPlas}, {Bradley}, {P{\'e}rez-Su{\'a}rez}, {de Val-Borro}, {Aldcroft}, {Cruz}, {Robitaille}, {Tollerud}, {Ardelean}, {Babej}, {Bach}, {Bachetti}, {Bakanov}, {Bamford}, {Barentsen}, {Barmby}, {Baumbach}, {Berry}, {Biscani}, {Boquien}, {Bostroem}, {Bouma}, {Brammer}, {Bray}, {Breytenbach}, {Buddelmeijer}, {Burke}, {Calderone}, {Cano Rodr{\'{\i}}guez}, {Cara}, {Cardoso}, {Cheedella}, {Copin}, {Corrales}, {Crichton}, {D'Avella}, {Deil}, {Depagne}, {Dietrich}, {Donath}, {Droettboom}, {Earl}, {Erben}, {Fabbro}, {Ferreira}, {Finethy}, {Fox}, {Garrison}, {Gibbons}, {Goldstein}, {Gommers}, {Greco}, {Greenfield}, {Groener}, {Grollier}, {Hagen}, {Hirst}, {Homeier}, {Horton}, {Hosseinzadeh}, {Hu}, {Hunkeler}, {Ivezi{\'c}}, {Jain}, {Jenness}, {Kanarek}, {Kendrew}, {Kern}, {Kerzendorf}, {Khvalko}, {King}, {Kirkby},
  {Kulkarni}, {Kumar}, {Lee}, {Lenz}, {Littlefair}, {Ma}, {Macleod}, {Mastropietro}, {McCully}, {Montagnac}, {Morris}, {Mueller}, {Mumford}, {Muna}, {Murphy}, {Nelson}, {Nguyen}, {Ninan}, {N{\"o}the}, {Ogaz}, {Oh}, {Parejko}, {Parley}, {Pascual}, {Patil}, {Patil}, {Plunkett}, {Prochaska}, {Rastogi}, {Reddy Janga}, {Sabater}, {Sakurikar}, {Seifert}, {Sherbert}, {Sherwood-Taylor}, {Shih}, {Sick}, {Silbiger}, {Singanamalla}, {Singer}, {Sladen}, {Sooley}, {Sornarajah}, {Streicher}, {Teuben}, {Thomas}, {Tremblay}, {Turner}, {Terr{\'o}n}, {van Kerkwijk}, {de la Vega}, {Watkins}, {Weaver}, {Whitmore}, {Woillez}, {Zabalza}, \& {Astropy Contributors}}]{exoplanet:astropy18}
{Astropy Collaboration}, {Price-Whelan}, A.~M., {Sip{\H o}cz}, B.~M., {et~al.} 2018{\natexlab{a}}, \aj, 156, 123, \dodoi{10.3847/1538-3881/aabc4f}

\bibitem[{{Astropy Collaboration} {et~al.}(2018{\natexlab{b}}){Astropy Collaboration}, {Price-Whelan}, {Sip{\H{o}}cz}, {G{\"u}nther}, {Lim}, {Crawford}, {Conseil}, {Shupe}, {Craig}, {Dencheva}, {Ginsburg}, {VanderPlas}, {Bradley}, {P{\'e}rez-Su{\'a}rez}, {de Val-Borro}, {Aldcroft}, {Cruz}, {Robitaille}, {Tollerud}, {Ardelean}, {Babej}, {Bach}, {Bachetti}, {Bakanov}, {Bamford}, {Barentsen}, {Barmby}, {Baumbach}, {Berry}, {Biscani}, {Boquien}, {Bostroem}, {Bouma}, {Brammer}, {Bray}, {Breytenbach}, {Buddelmeijer}, {Burke}, {Calderone}, {Cano Rodr{\'\i}guez}, {Cara}, {Cardoso}, {Cheedella}, {Copin}, {Corrales}, {Crichton}, {D'Avella}, {Deil}, {Depagne}, {Dietrich}, {Donath}, {Droettboom}, {Earl}, {Erben}, {Fabbro}, {Ferreira}, {Finethy}, {Fox}, {Garrison}, {Gibbons}, {Goldstein}, {Gommers}, {Greco}, {Greenfield}, {Groener}, {Grollier}, {Hagen}, {Hirst}, {Homeier}, {Horton}, {Hosseinzadeh}, {Hu}, {Hunkeler}, {Ivezi{\'c}}, {Jain}, {Jenness}, {Kanarek}, {Kendrew}, {Kern}, {Kerzendorf}, {Khvalko}, {King}, {Kirkby},
  {Kulkarni}, {Kumar}, {Lee}, {Lenz}, {Littlefair}, {Ma}, {Macleod}, {Mastropietro}, {McCully}, {Montagnac}, {Morris}, {Mueller}, {Mumford}, {Muna}, {Murphy}, {Nelson}, {Nguyen}, {Ninan}, {N{\"o}the}, {Ogaz}, {Oh}, {Parejko}, {Parley}, {Pascual}, {Patil}, {Patil}, {Plunkett}, {Prochaska}, {Rastogi}, {Reddy Janga}, {Sabater}, {Sakurikar}, {Seifert}, {Sherbert}, {Sherwood-Taylor}, {Shih}, {Sick}, {Silbiger}, {Singanamalla}, {Singer}, {Sladen}, {Sooley}, {Sornarajah}, {Streicher}, {Teuben}, {Thomas}, {Tremblay}, {Turner}, {Terr{\'o}n}, {van Kerkwijk}, {de la Vega}, {Watkins}, {Weaver}, {Whitmore}, {Woillez}, {Zabalza}, \& {Astropy Contributors}}]{astropy}
{Astropy Collaboration}, {Price-Whelan}, A.~M., {Sip{\H{o}}cz}, B.~M., {et~al.} 2018{\natexlab{b}}, \aj, 156, 123, \dodoi{10.3847/1538-3881/aabc4f}

\bibitem[{{Bailer-Jones} {et~al.}(2018){Bailer-Jones}, {Rybizki}, {Fouesneau}, {Mantelet}, \& {Andrae}}]{Bailer-Jones2018}
{Bailer-Jones}, C.~A.~L., {Rybizki}, J., {Fouesneau}, M., {Mantelet}, G., \& {Andrae}, R. 2018, \aj, 156, 58, \dodoi{10.3847/1538-3881/aacb21}

\bibitem[{{Ballard}(2019)}]{ballard_2019}
{Ballard}, S. 2019, \aj, 157, 113, \dodoi{10.3847/1538-3881/aaf477}

\bibitem[{{Batalha} {et~al.}(2017){Batalha}, {Kempton}, \& {Mbarek}}]{Batalha2017}
{Batalha}, N.~E., {Kempton}, E. M.~R., \& {Mbarek}, R. 2017, \apjl, 836, L5, \dodoi{10.3847/2041-8213/aa5c7d}

\bibitem[{{Batalha} {et~al.}(2019){Batalha}, {Lewis}, {Fortney}, {Batalha}, {Kempton}, {Lewis}, \& {Line}}]{Batalha2019}
{Batalha}, N.~E., {Lewis}, T., {Fortney}, J.~J., {et~al.} 2019, \apjl, 885, L25, \dodoi{10.3847/2041-8213/ab4909}

\bibitem[{{Batygin} \& {Morbidelli}(2023)}]{Batygin_Morbi_2023}
{Batygin}, K., \& {Morbidelli}, A. 2023, Nature Astronomy, 7, 330, \dodoi{10.1038/s41550-022-01850-5}

\bibitem[{{Baumeister} {et~al.}(2020){Baumeister}, {Padovan}, {Tosi}, {Montavon}, {Nettelmann}, {MacKenzie}, \& {Godolt}}]{Baumeister2020}
{Baumeister}, P., {Padovan}, S., {Tosi}, N., {et~al.} 2020, \apj, 889, 42, \dodoi{10.3847/1538-4357/ab5d32}

\bibitem[{{Baumeister} \& {Tosi}(2023)}]{Baumeister2023}
{Baumeister}, P., \& {Tosi}, N. 2023, \aap, 676, A106, \dodoi{10.1051/0004-6361/202346216}

\bibitem[{{Behmard} {et~al.}(2025){Behmard}, {Ness}, {Casey}, {Angus}, {Cunha}, {Souto}, {Yuxi}, {Lu}, \& {Johnson}}]{Behmard2025}
{Behmard}, A., {Ness}, M.~K., {Casey}, A.~R., {et~al.} 2025, arXiv e-prints, arXiv:2501.14955, \dodoi{10.48550/arXiv.2501.14955}

\bibitem[{{Benneke} {et~al.}(2024){Benneke}, {Roy}, {Coulombe}, {Radica}, {Piaulet}, {Ahrer}, {Pierrehumbert}, {Krissansen-Totton}, {Schlichting}, {Hu}, {Yang}, {Christie}, {Thorngren}, {Young}, {Pelletier}, {Knutson}, {Miguel}, {Evans-Soma}, {Dorn}, {Gagnebin}, {Fortney}, {Komacek}, {MacDonald}, {Raul}, {Cloutier}, {Acuna}, {Lafreni{\`e}re}, {Cadieux}, {Doyon}, {Welbanks}, \& {Allart}}]{Benneke2024}
{Benneke}, B., {Roy}, P.-A., {Coulombe}, L.-P., {et~al.} 2024, arXiv e-prints, arXiv:2403.03325, \dodoi{10.48550/arXiv.2403.03325}

\bibitem[{{Bitsch} \& {Battistini}(2020)}]{bitsch_2020}
{Bitsch}, B., \& {Battistini}, C. 2020, \aap, 633, A10, \dodoi{10.1051/0004-6361/201936463}

\bibitem[{{Bitsch} {et~al.}(2021){Bitsch}, {Raymond}, {Buchhave}, {Bello-Arufe}, {Rathcke}, \& {Schneider}}]{Bitsch2021}
{Bitsch}, B., {Raymond}, S.~N., {Buchhave}, L.~A., {et~al.} 2021, \aap, 649, L5, \dodoi{10.1051/0004-6361/202140793}

\bibitem[{{Bonfanti} {et~al.}(2024){Bonfanti}, {Brady}, {Wilson}, {Venturini}, {Egger}, {Brandeker}, {Sousa}, {Lendl}, {Simon}, {Queloz}, {Olofsson}, {Adibekyan}, {Alibert}, {Fossati}, {Hooton}, {Kubyshkina}, {Luque}, {Murgas}, {Mustill}, {Santos}, {Van Grootel}, {Alonso}, {Asquier}, {Bandy}, {B{\'a}rczy}, {Barrado Navascues}, {Barros}, {Baumjohann}, {Bean}, {Beck}, {Beck}, {Benz}, {Bergomi}, {Billot}, {Borsato}, {Broeg}, {Collier Cameron}, {Csizmadia}, {Cubillos}, {Davies}, {Deleuil}, {Deline}, {Delrez}, {Demangeon}, {Demory}, {Ehrenreich}, {Erikson}, {Fortier}, {Fridlund}, {Gandolfi}, {Gillon}, {G{\"u}del}, {G{\"u}nther}, {Heitzmann}, {Helling}, {Hoyer}, {Isaak}, {Kasper}, {Kiss}, {Lam}, {Laskar}, {Lecavelier des Etangs}, {Magrin}, {Maxted}, {Mordasini}, {Nascimbeni}, {Ottensamer}, {Pagano}, {Pall{\'e}}, {Peter}, {Piotto}, {Pollacco}, {Ragazzoni}, {Rando}, {Rauer}, {Ribas}, {Scandariato}, {S{\'e}gransan}, {Seifahrt}, {Smith}, {Stalport}, {Stef{\'a}nsson}, {Steinberger}, {St{\"u}rmer}, {Szab{\'o}}, {Thomas},
  {Udry}, {Villaver}, {Walton}, {Westerdorff}, \& {Zingales}}]{Bonfanti2024}
{Bonfanti}, A., {Brady}, M., {Wilson}, T.~G., {et~al.} 2024, \aap, 682, A66, \dodoi{10.1051/0004-6361/202348180}

\bibitem[{{Brande} {et~al.}(2024){Brande}, {Crossfield}, {Kreidberg}, {Morley}, {Barman}, {Benneke}, {Christiansen}, {Dragomir}, {Fortney}, {Greene}, {Hardegree-Ullman}, {Howard}, {Knutson}, {Lothringer}, \& {Mikal-Evans}}]{Brande2024}
{Brande}, J., {Crossfield}, I. J.~M., {Kreidberg}, L., {et~al.} 2024, \apjl, 961, L23, \dodoi{10.3847/2041-8213/ad1b5c}

\bibitem[{{Brown} {et~al.}(2013){Brown}, {Baliber}, {Bianco}, {Bowman}, {Burleson}, {Conway}, {Crellin}, {Depagne}, {De Vera}, {Dilday}, {Dragomir}, {Dubberley}, {Eastman}, {Elphick}, {Falarski}, {Foale}, {Ford}, {Fulton}, {Garza}, {Gomez}, {Graham}, {Greene}, {Haldeman}, {Hawkins}, {Haworth}, {Haynes}, {Hidas}, {Hjelstrom}, {Howell}, {Hygelund}, {Lister}, {Lobdill}, {Martinez}, {Mullins}, {Norbury}, {Parrent}, {Paulson}, {Petry}, {Pickles}, {Posner}, {Rosing}, {Ross}, {Sand}, {Saunders}, {Shobbrook}, {Shporer}, {Street}, {Thomas}, {Tsapras}, {Tufts}, {Valenti}, {Vander Horst}, {Walker}, {White}, \& {Willis}}]{Brown2013}
{Brown}, T.~M., {Baliber}, N., {Bianco}, F.~B., {et~al.} 2013, \pasp, 125, 1031, \dodoi{10.1086/673168}

\bibitem[{{Burn} {et~al.}(2024){Burn}, {Mordasini}, {Mishra}, {Haldemann}, {Venturini}, {Emsenhuber}, \& {Henning}}]{Burn2024}
{Burn}, R., {Mordasini}, C., {Mishra}, L., {et~al.} 2024, Nature Astronomy, 8, 463, \dodoi{10.1038/s41550-023-02183-7}

\bibitem[{{Cadieux} {et~al.}(2024){Cadieux}, {Plotnykov}, {Doyon}, {Valencia}, {Jahandar}, {Dang}, {Turbet}, {Fauchez}, {Cloutier}, {Cherubim}, {Artigau}, {Cook}, {Edwards}, {Hallatt}, {Charnay}, {Bouchy}, {Allart}, {Mignon}, {Baron}, {Barros}, {Benneke}, {Canto Martins}, {Cowan}, {De Medeiros}, {Delfosse}, {Delgado-Mena}, {Dumusque}, {Ehrenreich}, {Frensch}, {Gonz{\'a}lez Hern{\'a}ndez}, {Hara}, {Lafreni{\`e}re}, {Lo Curto}, {Malo}, {Melo}, {Mounzer}, {Passeger}, {Pepe}, {Poulin-Girard}, {Santos}, {Sosnowska}, {Su{\'a}rez Mascare{\~n}o}, {Thibault}, {Vaulato}, {Wade}, \& {Wildi}}]{Cadieux2024}
{Cadieux}, C., {Plotnykov}, M., {Doyon}, R., {et~al.} 2024, \apjl, 960, L3, \dodoi{10.3847/2041-8213/ad1691}

\bibitem[{{Castro-Gonz{\'a}lez} {et~al.}(2023){Castro-Gonz{\'a}lez}, {Demangeon}, {Lillo-Box}, {Lovis}, {Lavie}, {Adibekyan}, {Acu{\~n}a}, {Deleuil}, {Aguichine}, {Zapatero Osorio}, {Tabernero}, {Davoult}, {Alibert}, {Santos}, {Sousa}, {Antoniadis-Karnavas}, {Borsa}, {Winn}, {Allende Prieto}, {Figueira}, {Jenkins}, {Sozzetti}, {Damasso}, {Silva}, {Astudillo-Defru}, {Barros}, {Bonfils}, {Cristiani}, {Di Marcantonio}, {Gonz{\'a}lez Hern{\'a}ndez}, {Curto}, {Martins}, {Nunes}, {Palle}, {Pepe}, {Seager}, \& {Su{\'a}rez Mascare{\~n}o}}]{Castro-Gonzalez2023}
{Castro-Gonz{\'a}lez}, A., {Demangeon}, O.~D.~S., {Lillo-Box}, J., {et~al.} 2023, \aap, 675, A52, \dodoi{10.1051/0004-6361/202346550}

\bibitem[{{Choksi} \& {Chiang}(2023)}]{Choksi2024}
{Choksi}, N., \& {Chiang}, E. 2023, \mnras, 522, 1914, \dodoi{10.1093/mnras/stad835}

\bibitem[{{Clausen} \& {Tilgner}(2015)}]{Clausen2015}
{Clausen}, N., \& {Tilgner}, A. 2015, \aap, 584, A60, \dodoi{10.1051/0004-6361/201526082}

\bibitem[{{Cloutier} {et~al.}(2024){Cloutier}, {Greklek-McKeon}, {Wurmser}, {Cherubim}, {Gillis}, {Vanderburg}, {Hadden}, {Cadieux}, {Artigau}, {Vissapragada}, {Mortier}, {L{\'o}pez-Morales}, {Latham}, {Knutson}, {Haywood}, {Pall{\'e}}, {Doyon}, {Cook}, {Andreuzzi}, {Cecconi}, {Cosentino}, {Ghedina}, {Harutyunyan}, {Pinamonti}, {Stalport}, {Damasso}, {Rescigno}, {Wilson}, {Buchhave}, {Charbonneau}, {Cameron}, {Dumusque}, {Lovis}, {Mayor}, {Molinari}, {Pepe}, {Piotto}, {Rice}, {Sasselov}, {S{\'e}gransan}, {Sozzetti}, {Udry}, \& {Watson}}]{Cloutier_2024}
{Cloutier}, R., {Greklek-McKeon}, M., {Wurmser}, S., {et~al.} 2024, \mnras, 527, 5464, \dodoi{10.1093/mnras/stad3450}

\bibitem[{{Cointepas} {et~al.}(2024){Cointepas}, {Bouchy}, {Almenara}, {Bonfils}, {Astudillo-Defru}, {Knierim}, {Stalport}, {Mignon}, {Grieves}, {Bean}, {Brady}, {Burt}, {Canto Martins}, {Collins}, {Collins}, {Delfosse}, {de Medeiros}, {Demory}, {Dorn}, {Forveille}, {Fukui}, {Gan}, {G{\'o}mez Maqueo Chew}, {Halverson}, {Helled}, {Helm}, {Hirano}, {Horne}, {Howell}, {Isogai}, {Kasper}, {Kawauchi}, {Livingston}, {Massey}, {Matson}, {Murgas}, {Narita}, {Palle}, {Relles}, {Sabin}, {Schanche}, {Schwarz}, {Seifahrt}, {Shporer}, {Stefansson}, {Sturmer}, {Tamura}, {Tan}, {Twicken}, {Watanabe}, {Wells}, {Wilkin}, {Ricker}, {Seager}, {Winn}, \& {Jenkins}}]{Cointepas2024}
{Cointepas}, M., {Bouchy}, F., {Almenara}, J.~M., {et~al.} 2024, \aap, 685, A19, \dodoi{10.1051/0004-6361/202346899}

\bibitem[{{Cutri} {et~al.}(2021){Cutri}, {Wright}, {Conrow}, {Fowler}, {Eisenhardt}, {Grillmair}, {Kirkpatrick}, {Masci}, {McCallon}, {Wheelock}, {Fajardo-Acosta}, {Yan}, {Benford}, {Harbut}, {Jarrett}, {Lake}, {Leisawitz}, {Ressler}, {Stanford}, {Tsai}, {Liu}, {Helou}, {Mainzer}, {Gettngs}, {Gonzalez}, {Hoffman}, {Marsh}, {Padgett}, {Skrutskie}, {Beck}, {Papin}, \& {Wittman}}]{Cutri2014}
{Cutri}, R.~M., {Wright}, E.~L., {Conrow}, T., {et~al.} 2021, {VizieR Online Data Catalog: AllWISE Data Release (Cutri+ 2013)}, VizieR On-line Data Catalog: II/328. Originally published in: IPAC/Caltech (2013)

\bibitem[{Dai {et~al.}(2023)Dai, Masuda, Beard, Robertson, Goldberg, Batygin, Bouma, Lissauer, Knudstrup, Albrecht, Howard, Knutson, Petigura, Weiss, Isaacson, Kristiansen, Osborn, Wang, Wang, Behmard, Greklek-McKeon, Vissapragada, Batalha, Brinkman, Chontos, Crossfield, Dressing, Fetherolf, Fulton, Hill, Huber, Kane, Lubin, MacDougall, Mayo, Močnik, Akana~Murphy, Rubenzahl, Scarsdale, Tyler, Zandt, Polanski, Schwengeler, Terentev, Benni, Bieryla, Ciardi, Falk, Furlan, Girardin, Guerra, Hesse, Howell, Lillo-Box, Matthews, Twicken, Villaseñor, Latham, Jenkins, Ricker, Seager, Vanderspek, \& Winn}]{TOI-1136}
Dai, F., Masuda, K., Beard, C., {et~al.} 2023, The Astronomical Journal, 165, 33, \dodoi{10.3847/1538-3881/aca327}

\bibitem[{{Damiano} {et~al.}(2024){Damiano}, {Bello-Arufe}, {Yang}, \& {Hu}}]{Damiano2024}
{Damiano}, M., {Bello-Arufe}, A., {Yang}, J., \& {Hu}, R. 2024, \apjl, 968, L22, \dodoi{10.3847/2041-8213/ad5204}

\bibitem[{Deck \& Agol(2015)}]{Deck_2015}
Deck, K.~M., \& Agol, E. 2015, The Astrophysical Journal, 802, 116, \dodoi{10.1088/0004-637x/802/2/116}

\bibitem[{Deck {et~al.}(2014)Deck, Agol, Holman, \& Nesvorný}]{TTVFast}
Deck, K.~M., Agol, E., Holman, M.~J., \& Nesvorný, D. 2014, The Astrophysical Journal, 787, 132, \dodoi{10.1088/0004-637x/787/2/132}

\bibitem[{{Delrez} {et~al.}(2021){Delrez}, {Ehrenreich}, {Alibert}, {Bonfanti}, {Borsato}, {Fossati}, {Hooton}, {Hoyer}, {Pozuelos}, {Salmon}, {Sulis}, {Wilson}, {Adibekyan}, {Bourrier}, {Brandeker}, {Charnoz}, {Deline}, {Guterman}, {Haldemann}, {Hara}, {Oshagh}, {Sousa}, {Van Grootel}, {Alonso}, {Anglada-Escud{\'e}}, {B{\'a}rczy}, {Barrado}, {Barros}, {Baumjohann}, {Beck}, {Bekkelien}, {Benz}, {Billot}, {Bonfils}, {Broeg}, {Cabrera}, {Collier Cameron}, {Davies}, {Deleuil}, {Delisle}, {Demangeon}, {Demory}, {Erikson}, {Fortier}, {Fridlund}, {Futyan}, {Gandolfi}, {Garcia Mu{\~n}oz}, {Gillon}, {Guedel}, {Heng}, {Kiss}, {Laskar}, {Lecavelier des Etangs}, {Lendl}, {Lovis}, {Maxted}, {Nascimbeni}, {Olofsson}, {Osborn}, {Pagano}, {Pall{\'e}}, {Piotto}, {Pollacco}, {Queloz}, {Rauer}, {Ragazzoni}, {Ribas}, {Santos}, {Scandariato}, {S{\'e}gransan}, {Simon}, {Smith}, {Steller}, {Szab{\'o}}, {Thomas}, {Udry}, \& {Walton}}]{Delrez2021}
{Delrez}, L., {Ehrenreich}, D., {Alibert}, Y., {et~al.} 2021, Nature Astronomy, 5, 775, \dodoi{10.1038/s41550-021-01381-5}

\bibitem[{{Demangeon} {et~al.}(2021){Demangeon}, {Zapatero Osorio}, {Alibert}, {Barros}, {Adibekyan}, {Tabernero}, {Antoniadis-Karnavas}, {Camacho}, {Su{\'a}rez Mascare{\~n}o}, {Oshagh}, {Micela}, {Sousa}, {Lovis}, {Pepe}, {Rebolo}, {Cristiani}, {Santos}, {Allart}, {Allende Prieto}, {Bossini}, {Bouchy}, {Cabral}, {Damasso}, {Di Marcantonio}, {D'Odorico}, {Ehrenreich}, {Faria}, {Figueira}, {G{\'e}nova Santos}, {Haldemann}, {Hara}, {Gonz{\'a}lez Hern{\'a}ndez}, {Lavie}, {Lillo-Box}, {Lo Curto}, {Martins}, {M{\'e}gevand}, {Mehner}, {Molaro}, {Nunes}, {Pall{\'e}}, {Pasquini}, {Poretti}, {Sozzetti}, \& {Udry}}]{Demangeon2021}
{Demangeon}, O.~D.~S., {Zapatero Osorio}, M.~R., {Alibert}, Y., {et~al.} 2021, \aap, 653, A41, \dodoi{10.1051/0004-6361/202140728}

\bibitem[{{Demory} {et~al.}(2020){Demory}, {Pozuelos}, {G{\'o}mez Maqueo Chew}, {Sabin}, {Petrucci}, {Schroffenegger}, {Grimm}, {Sestovic}, {Gillon}, {McCormac}, {Barkaoui}, {Benz}, {Bieryla}, {Bouchy}, {Burdanov}, {Collins}, {de Wit}, {Dressing}, {Garcia}, {Giacalone}, {Guerra}, {Haldemann}, {Heng}, {Jehin}, {Jofr{\'e}}, {Kane}, {Lillo-Box}, {Maign{\'e}}, {Mordasini}, {Morris}, {Niraula}, {Queloz}, {Rackham}, {Savel}, {Soubkiou}, {Srdoc}, {Stassun}, {Triaud}, {Zambelli}, {Ricker}, {Latham}, {Seager}, {Winn}, {Jenkins}, {Calvario-Vel{\'a}squez}, {Franco Herrera}, {Colorado}, {Cadena Zepeda}, {Figueroa}, {Watson}, {Lugo-Ibarra}, {Carigi}, {Guisa}, {Herrera}, {Sierra D{\'\i}az}, {Su{\'a}rez}, {Barrado}, {Batalha}, {Benkhaldoun}, {Chontos}, {Dai}, {Essack}, {Ghachoui}, {Huang}, {Huber}, {Isaacson}, {Lissauer}, {Morales-Calder{\'o}n}, {Robertson}, {Roy}, {Twicken}, {Vanderburg}, \& {Weiss}}]{Demory_2020}
{Demory}, B.~O., {Pozuelos}, F.~J., {G{\'o}mez Maqueo Chew}, Y., {et~al.} 2020, \aap, 642, A49, \dodoi{10.1051/0004-6361/202038616}

\bibitem[{{Di Maio} {et~al.}(2023){Di Maio}, {Changeat}, {Benatti}, \& {Micela}}]{DiMaio2023}
{Di Maio}, C., {Changeat}, Q., {Benatti}, S., \& {Micela}, G. 2023, \aap, 669, A150, \dodoi{10.1051/0004-6361/202244881}

\bibitem[{{Diamond-Lowe} {et~al.}(2022){Diamond-Lowe}, {Kreidberg}, {Harman}, {Kempton}, {Rogers}, {Joyce}, {Eastman}, {King}, {Kopparapu}, {Youngblood}, {Kosiarek}, {Livingston}, {Hardegree-Ullman}, \& {Crossfield}}]{Diamond-Lowe2022}
{Diamond-Lowe}, H., {Kreidberg}, L., {Harman}, C.~E., {et~al.} 2022, \aj, 164, 172, \dodoi{10.3847/1538-3881/ac7807}

\bibitem[{{Dorn} \& {Lichtenberg}(2021)}]{Dorn2021}
{Dorn}, C., \& {Lichtenberg}, T. 2021, \apjl, 922, L4, \dodoi{10.3847/2041-8213/ac33af}

\bibitem[{{Dorn} {et~al.}(2017){Dorn}, {Venturini}, {Khan}, {Heng}, {Alibert}, {Helled}, {Rivoldini}, \& {Benz}}]{Dorn2017}
{Dorn}, C., {Venturini}, J., {Khan}, A., {et~al.} 2017, \aap, 597, A37, \dodoi{10.1051/0004-6361/201628708}

\bibitem[{Eastman {et~al.}(2013)Eastman, Gaudi, \& Agol}]{exofast}
Eastman, J., Gaudi, B.~S., \& Agol, E. 2013, Publications of the Astronomical Society of the Pacific, 125, 83–112, \dodoi{10.1086/669497}

\bibitem[{{Eggleton} {et~al.}(1998){Eggleton}, {Kiseleva}, \& {Hut}}]{Eggleton1998}
{Eggleton}, P.~P., {Kiseleva}, L.~G., \& {Hut}, P. 1998, \apj, 499, 853, \dodoi{10.1086/305670}

\bibitem[{{Engle}(2024)}]{Engle2024}
{Engle}, S.~G. 2024, \apj, 960, 62, \dodoi{10.3847/1538-4357/ad0840}

\bibitem[{{Engle} \& {Guinan}(2023)}]{Engle_Guinan_2023}
{Engle}, S.~G., \& {Guinan}, E.~F. 2023, \apjl, 954, L50, \dodoi{10.3847/2041-8213/acf472}

\bibitem[{{Fleming} {et~al.}(2020){Fleming}, {Barnes}, {Luger}, \& {VanderPlas}}]{Fleming_2020}
{Fleming}, D.~P., {Barnes}, R., {Luger}, R., \& {VanderPlas}, J.~T. 2020, \apj, 891, 155, \dodoi{10.3847/1538-4357/ab77ad}

\bibitem[{Foreman-Mackey(2016)}]{corner}
Foreman-Mackey, D. 2016, The Journal of Open Source Software, 24, \dodoi{10.21105/joss.00024}

\bibitem[{{Foreman-Mackey}(2018)}]{exoplanet:foremanmackey18}
{Foreman-Mackey}, D. 2018, Research Notes of the American Astronomical Society, 2, 31, \dodoi{10.3847/2515-5172/aaaf6c}

\bibitem[{{Foreman-Mackey} {et~al.}(2017){Foreman-Mackey}, {Agol}, {Ambikasaran}, \& {Angus}}]{exoplanet:foremanmackey17}
{Foreman-Mackey}, D., {Agol}, E., {Ambikasaran}, S., \& {Angus}, R. 2017, \aj, 154, 220, \dodoi{10.3847/1538-3881/aa9332}

\bibitem[{Foreman-Mackey {et~al.}(2013)Foreman-Mackey, Hogg, Lang, \& Goodman}]{emcee}
Foreman-Mackey, D., Hogg, D.~W., Lang, D., \& Goodman, J. 2013, Publications of the Astronomical Society of the Pacific, 125, 306–312, \dodoi{10.1086/670067}

\bibitem[{{Foreman-Mackey} {et~al.}(2019){Foreman-Mackey}, {Farr}, {Sinha}, {Archibald}, {Hogg}, {Sanders}, {Zuntz}, {Williams}, {Nelson}, {de Val-Borro}, {Erhardt}, {Pashchenko}, \& {Pla}}]{ForemanMackey2019}
{Foreman-Mackey}, D., {Farr}, W., {Sinha}, M., {et~al.} 2019, The Journal of Open Source Software, 4, 1864, \dodoi{10.21105/joss.01864}

\bibitem[{{Foreman-Mackey} {et~al.}(2021){Foreman-Mackey}, {Luger}, {Agol}, {Barclay}, {Bouma}, {Brandt}, {Czekala}, {David}, {Dong}, {Gilbert}, {Gordon}, {Hedges}, {Hey}, {Morris}, {Price-Whelan}, \& {Savel}}]{exoplanet:joss}
{Foreman-Mackey}, D., {Luger}, R., {Agol}, E., {et~al.} 2021, arXiv e-prints, arXiv:2105.01994.
\newblock \doarXiv{2105.01994}

\bibitem[{Foreman-Mackey {et~al.}(2021)Foreman-Mackey, Savel, Luger, Agol, Czekala, Price-Whelan, Hedges, Gilbert, Bouma, Brandt, \& Barclay}]{exoplanet:zenodo}
Foreman-Mackey, D., Savel, A., Luger, R., {et~al.} 2021, exoplanet-dev/exoplanet v0.5.1, \dodoi{10.5281/zenodo.1998447}

\bibitem[{{Fortney} {et~al.}(2020){Fortney}, {Visscher}, {Marley}, {Hood}, {Line}, {Thorngren}, {Freedman}, \& {Lupu}}]{Fortney2020}
{Fortney}, J.~J., {Visscher}, C., {Marley}, M.~S., {et~al.} 2020, \aj, 160, 288, \dodoi{10.3847/1538-3881/abc5bd}

\bibitem[{{Fukui} {et~al.}(2011){Fukui}, {Narita}, {Tristram}, {Sumi}, {Abe}, {Itow}, {Sullivan}, {Bond}, {Hirano}, {Tamura}, {Bennett}, {Furusawa}, {Hayashi}, {Hearnshaw}, {Hosaka}, {Kamiya}, {Kobara}, {Korpela}, {Kilmartin}, {Lin}, {Ling}, {Makita}, {Masuda}, {Matsubara}, {Miyake}, {Muraki}, {Nagaya}, {Nishimoto}, {Ohnishi}, {Omori}, {Perrott}, {Rattenbury}, {Saito}, {Skuljan}, {Suzuki}, {Sweatman}, \& {Wada}}]{2011PASJ...63..287F}
{Fukui}, A., {Narita}, N., {Tristram}, P.~J., {et~al.} 2011, \pasj, 63, 287, \dodoi{10.1093/pasj/63.1.287}

\bibitem[{{Fulton} {et~al.}(2018){Fulton}, {Petigura}, {Blunt}, \& {Sinukoff}}]{Radvel}
{Fulton}, B.~J., {Petigura}, E.~A., {Blunt}, S., \& {Sinukoff}, E. 2018, \pasp, 130, 044504, \dodoi{10.1088/1538-3873/aaaaa8}

\bibitem[{{Gaia Collaboration}(2018)}]{gaia2018}
{Gaia Collaboration}. 2018, {VizieR Online Data Catalog: Gaia DR2 (Gaia Collaboration, 2018)}, VizieR On-line Data Catalog: I/345. Originally published in: 2018A\&A...616A...1G; doi:10.5270/esa-ycs, \dodoi{10.26093/cds/vizier.1345}

\bibitem[{{Gan} {et~al.}(2022){Gan}, {Lin}, {Wang}, {Mao}, {Fouqu{\'e}}, {Fan}, {Bedell}, {Stassun}, {Giacalone}, {Fukui}, {Murgas}, {Ciardi}, {Howell}, {Collins}, {Shporer}, {Arnold}, {Barclay}, {Charbonneau}, {Christiansen}, {Crossfield}, {Dressing}, {Elliott}, {Esparza-Borges}, {Evans}, {Gnilka}, {Gonzales}, {Howard}, {Isogai}, {Kawauchi}, {Kurita}, {Liu}, {Livingston}, {Matson}, {Narita}, {Palle}, {Parviainen}, {Rackham}, {Rodriguez}, {Rose}, {Rudat}, {Schlieder}, {Scott}, {Vezie}, {Ricker}, {Vanderspek}, {Latham}, {Seager}, {Winn}, \& {Jenkins}}]{Gan2022}
{Gan}, T., {Lin}, Z., {Wang}, S.~X., {et~al.} 2022, \mnras, 511, 83, \dodoi{10.1093/mnras/stab3708}

\bibitem[{{Gao} {et~al.}(2023){Gao}, {Piette}, {Steinrueck}, {Nixon}, {Zhang}, {Kempton}, {Bean}, {Rauscher}, {Parmentier}, {Batalha}, {Savel}, {Arnold}, {Roman}, {Malsky}, \& {Taylor}}]{Gao2023}
{Gao}, P., {Piette}, A. A.~A., {Steinrueck}, M.~E., {et~al.} 2023, \apj, 951, 96, \dodoi{10.3847/1538-4357/acd16f}

\bibitem[{{Garc{\'\i}a-Melendo} \& {L{\'o}pez-Morales}(2011)}]{ttv_detection_bias}
{Garc{\'\i}a-Melendo}, E., \& {L{\'o}pez-Morales}, M. 2011, \mnras, 417, L16, \dodoi{10.1111/j.1745-3933.2011.01111.x}

\bibitem[{{Gladman} {et~al.}(1996){Gladman}, {Quinn}, {Nicholson}, \& {Rand}}]{Gladman1996}
{Gladman}, B., {Quinn}, D.~D., {Nicholson}, P., \& {Rand}, R. 1996, \icarus, 122, 166, \dodoi{10.1006/icar.1996.0117}

\bibitem[{Goldberg \& Batygin(2021)}]{Goldberg_2021}
Goldberg, M., \& Batygin, K. 2021, The Astronomical Journal, 162, 16, \dodoi{10.3847/1538-3881/abfb78}

\bibitem[{{Goldberg} \& {Batygin}(2022)}]{Goldberg_2022}
{Goldberg}, M., \& {Batygin}, K. 2022, arXiv e-prints, arXiv:2203.00801.
\newblock \doarXiv{2203.00801}

\bibitem[{{Goldberg} \& {Batygin}(2023)}]{Goldberg2023}
---. 2023, \apj, 948, 12, \dodoi{10.3847/1538-4357/acc9ae}

\bibitem[{{Goldreich} \& {Soter}(1966)}]{Goldreich1996}
{Goldreich}, P., \& {Soter}, S. 1966, \icarus, 5, 375, \dodoi{10.1016/0019-1035(66)90051-0}

\bibitem[{{Goyal} {et~al.}(2023){Goyal}, {Dai}, \& {Wang}}]{Goyal2023}
{Goyal}, A.~V., {Dai}, F., \& {Wang}, S. 2023, \apj, 955, 118, \dodoi{10.3847/1538-4357/acebe2}

\bibitem[{{Goyal} \& {Wang}(2024)}]{Goyal_Wang2024}
{Goyal}, A.~V., \& {Wang}, S. 2024, \apjl, 968, L4, \dodoi{10.3847/2041-8213/ad4f6e}

\bibitem[{Goździewski {et~al.}(2015)Goździewski, Migaszewski, Panichi, \& Szuszkiewicz}]{Godziewski15}
Goździewski, K., Migaszewski, C., Panichi, F., \& Szuszkiewicz, E. 2015, Monthly Notices of the Royal Astronomical Society: Letters, 455, L104–L108, \dodoi{10.1093/mnrasl/slv156}

\bibitem[{{Greklek-McKeon} {et~al.}(2023){Greklek-McKeon}, {Knutson}, {Vissapragada}, {Jontof-Hutter}, {Chachan}, {Thorngren}, \& {Vasisht}}]{GreklekMcKeon2023}
{Greklek-McKeon}, M., {Knutson}, H.~A., {Vissapragada}, S., {et~al.} 2023, \aj, 165, 48, \dodoi{10.3847/1538-3881/ac8553}

\bibitem[{{Hadden}(2019)}]{Hadden_2019}
{Hadden}, S. 2019, \aj, 158, 238, \dodoi{10.3847/1538-3881/ab5287}

\bibitem[{{Hadden} \& {Lithwick}(2016)}]{hadden_lithwick_2016}
{Hadden}, S., \& {Lithwick}, Y. 2016, \apj, 828, 44, \dodoi{10.3847/0004-637X/828/1/44}

\bibitem[{{Harman} {et~al.}(2022){Harman}, {Kopparapu}, {Stef{\'a}nsson}, {Lin}, {Mahadevan}, {Hedges}, \& {Batalha}}]{Harman_2022}
{Harman}, C.~E., {Kopparapu}, R.~K., {Stef{\'a}nsson}, G., {et~al.} 2022, \psj, 3, 45, \dodoi{10.3847/PSJ/ac38ac}

\bibitem[{Harris {et~al.}(2020)Harris, Millman, van~der Walt, Gommers, Virtanen, Cournapeau, Wieser, Taylor, Berg, Smith, Kern, Picus, Hoyer, van Kerkwijk, Brett, Haldane, Fernández~del Río, Wiebe, Peterson, Gérard-Marchant, Sheppard, Reddy, Weckesser, Abbasi, Gohlke, \& Oliphant}]{numpy}
Harris, C.~R., Millman, K.~J., van~der Walt, S.~J., {et~al.} 2020, Nature, 585, 357–362, \dodoi{10.1038/s41586-020-2649-2}

\bibitem[{{He} {et~al.}(2020){He}, {Ford}, {Ragozzine}, \& {Carrera}}]{He2020}
{He}, M.~Y., {Ford}, E.~B., {Ragozzine}, D., \& {Carrera}, D. 2020, \aj, 160, 276, \dodoi{10.3847/1538-3881/abba18}

\bibitem[{{Henden} {et~al.}(2015){Henden}, {Levine}, {Terrell}, \& {Welch}}]{Henden2015APASS}
{Henden}, A.~A., {Levine}, S., {Terrell}, D., \& {Welch}, D.~L. 2015, in American Astronomical Society Meeting Abstracts, Vol. 225, American Astronomical Society Meeting Abstracts \#225, 336.16

\bibitem[{{Higson} {et~al.}(2019{\natexlab{a}}){Higson}, {Handley}, {Hobson}, \& {Lasenby}}]{dynesty}
{Higson}, E., {Handley}, W., {Hobson}, M., \& {Lasenby}, A. 2019{\natexlab{a}}, Statistics and Computing, 29, 891, \dodoi{10.1007/s11222-018-9844-0}

\bibitem[{{Higson} {et~al.}(2019{\natexlab{b}}){Higson}, {Handley}, {Hobson}, \& {Lasenby}}]{nestcheck}
---. 2019{\natexlab{b}}, \mnras, 483, 2044, \dodoi{10.1093/mnras/sty3090}

\bibitem[{Hippke \& Heller(2019)}]{tls}
Hippke, M., \& Heller, R. 2019, Astronomy \& Astrophysics, 623, A39, \dodoi{10.1051/0004-6361/201834672}

\bibitem[{{Hobson} {et~al.}(2024){Hobson}, {Bouchy}, {Lavie}, {Lovis}, {Adibekyan}, {Allende Prieto}, {Alibert}, {Barros}, {Castro-Gonz{\'a}lez}, {Cristiani}, {D'Odorico}, {Damasso}, {Di Marcantonio}, {Dumusque}, {Ehrenreich}, {Figueira}, {G{\'e}nova Santos}, {Gilbert}, {Gonz{\'a}lez Hern{\'a}ndez}, {Lillo-Box}, {Lo Curto}, {Martins}, {Mehner}, {Micela}, {Molaro}, {Nunes}, {Palle}, {Pepe}, {Rebolo}, {Rodrigues}, {Santos}, {Sousa}, {Sozzetti}, {Su{\'a}rez Mascare{\~n}o}, {Tabernero}, {Udry}, {Zapatero Osorio}, {Armstrong}, {Ciardi}, {Collins}, {Collins}, {Everett}, {Gandolfi}, {Howell}, {Jenkins}, {Kielkopf}, {Livingston}, {Lund}, {Mireles}, {Ricker}, {Schwarz}, {Seager}, {Tan}, {Ting}, \& {Winn}}]{Hobson2024}
{Hobson}, M.~J., {Bouchy}, F., {Lavie}, B., {et~al.} 2024, \aap, 688, A216, \dodoi{10.1051/0004-6361/202450505}

\bibitem[{{Hord} {et~al.}(2024){Hord}, {Kempton}, {Evans-Soma}, {Latham}, {Ciardi}, {Dragomir}, {Col{\'o}n}, {Ross}, {Vanderburg}, {de Beurs}, {Collins}, {Watkins}, {Bean}, {Cowan}, {Daylan}, {Morley}, {Ih}, {Baker}, {Barkaoui}, {Batalha}, {Behmard}, {Belinski}, {Benkhaldoun}, {Benni}, {Bernacki}, {Bieryla}, {Binnenfeld}, {Bosch-Cabot}, {Bouchy}, {Bozza}, {Brahm}, {Buchhave}, {Calkins}, {Chontos}, {Clark}, {Cloutier}, {Cointepas}, {Collins}, {Conti}, {Crossfield}, {Dai}, {de Leon}, {Dransfield}, {Dressing}, {Dustor}, {Esquerdo}, {Evans}, {Fajardo-Acosta}, {Fio{\l}ka}, {For{\'e}s-Toribio}, {Frasca}, {Fukui}, {Fulton}, {Furlan}, {Gan}, {Gandolfi}, {Ghachoui}, {Giacalone}, {Gilbert}, {Gillon}, {Girardin}, {Gonzales}, {Grau Horta}, {Gregorio}, {Greklek-McKeon}, {Guerra}, {Hartman}, {Hellier}, {Helm}, {He{\l}miniak}, {Henning}, {Hill}, {Horne}, {Howard}, {Howell}, {Huber}, {Isopi}, {Jehin}, {Jenkins}, {Jensen}, {Johnson}, {Jord{\'a}n}, {Kane}, {Kielkopf}, {Krushinsky}, {Lasota}, {Lee}, {Lewin}, {Livingston},
  {Lubin}, {Lund}, {Mallia}, {Mann}, {Marino}, {Maslennikova}, {Massey}, {Matson}, {Matthews}, {Mayo}, {Mazeh}, {McLeod}, {Michaels}, {Mo{\v{c}}nik}, {Mori}, {Mraz}, {Mu{\~n}oz}, {Narita}, {Natarajan}, {Dyregaard Nielsen}, {Osborn}, {Palle}, {Panahi}, {Papini}, {Plavchan}, {Polanski}, {Popowicz}, {Pozuelos}, {Quinn}, {Radford}, {Reed}, {Relles}, {Rice}, {Robertson}, {Rodriguez}, {Rosenthal}, {Rubenzahl}, {Schanche}, {Schlieder}, {Schwarz}, {Sefako}, {Shporer}, {Sozzetti}, {Srdoc}, {Stockdale}, {Tarasenkov}, {Tan}, {Timmermans}, {Ting}, {Van Zandt}, {Vignes}, {Waite}, {Watanabe}, {Weiss}, {Wittrock}, {Zhou}, {Ziegler}, \& {Zucker}}]{Hord2024}
{Hord}, B.~J., {Kempton}, E. M.~R., {Evans-Soma}, T.~M., {et~al.} 2024, \aj, 167, 233, \dodoi{10.3847/1538-3881/ad3068}

\bibitem[{{Hori} \& {Ogihara}(2020)}]{Hori_2020}
{Hori}, Y., \& {Ogihara}, M. 2020, \apj, 889, 77, \dodoi{10.3847/1538-4357/ab6168}

\bibitem[{{Hu}(2021)}]{Hu2021}
{Hu}, R. 2021, \apj, 921, 27, \dodoi{10.3847/1538-4357/ac1789}

\bibitem[{Hunter(2007)}]{matplotlib}
Hunter, J.~D. 2007, Computing In Science \& Engineering, 9, 90

\bibitem[{{Husser} {et~al.}(2013){Husser}, {Wende-von Berg}, {Dreizler}, {Homeier}, {Reiners}, {Barman}, \& {Hauschildt}}]{Husser2013}
{Husser}, T.~O., {Wende-von Berg}, S., {Dreizler}, S., {et~al.} 2013, \aap, 553, A6, \dodoi{10.1051/0004-6361/201219058}

\bibitem[{{Jackson} {et~al.}(2008){Jackson}, {Barnes}, \& {Greenberg}}]{Jackson2008}
{Jackson}, B., {Barnes}, R., \& {Greenberg}, R. 2008, \mnras, 391, 237, \dodoi{10.1111/j.1365-2966.2008.13868.x}

\bibitem[{{Jensen}(2013)}]{SwarthmoreTTF}
{Jensen}, E. 2013, {Tapir: A web interface for transit/eclipse observability}, Astrophysics Source Code Library, record ascl:1306.007

\bibitem[{Judkovsky {et~al.}(2023)Judkovsky, Ofir, \& Aharonson}]{Judkovsky2023}
Judkovsky, Y., Ofir, A., \& Aharonson, O. 2023, The Advantages of Global Photometric Models in Fitting Transit Variations.
\newblock \doarXiv{2311.06948}

\bibitem[{{Kellermann} {et~al.}(2018){Kellermann}, {Becker}, \& {Redmer}}]{Kellermann2018}
{Kellermann}, C., {Becker}, A., \& {Redmer}, R. 2018, \aap, 615, A39, \dodoi{10.1051/0004-6361/201731775}

\bibitem[{{Kempton} {et~al.}(2018){Kempton}, {Bean}, {Louie}, {Deming}, {Koll}, {Mansfield}, {Christiansen}, {L{\'o}pez-Morales}, {Swain}, {Zellem}, {Ballard}, {Barclay}, {Barstow}, {Batalha}, {Beatty}, {Berta-Thompson}, {Birkby}, {Buchhave}, {Charbonneau}, {Cowan}, {Crossfield}, {de Val-Borro}, {Doyon}, {Dragomir}, {Gaidos}, {Heng}, {Hu}, {Kane}, {Kreidberg}, {Mallonn}, {Morley}, {Narita}, {Nascimbeni}, {Pall{\'e}}, {Quintana}, {Rauscher}, {Seager}, {Shkolnik}, {Sing}, {Sozzetti}, {Stassun}, {Valenti}, \& {von Essen}}]{Kempton2018}
{Kempton}, E. M.~R., {Bean}, J.~L., {Louie}, D.~R., {et~al.} 2018, \pasp, 130, 114401, \dodoi{10.1088/1538-3873/aadf6f}

\bibitem[{{Kimura} \& {Ikoma}(2022)}]{kimura_2022}
{Kimura}, T., \& {Ikoma}, M. 2022, Nature Astronomy, 6, 1296, \dodoi{10.1038/s41550-022-01781-1}

\bibitem[{{Kochanek} {et~al.}(2017){Kochanek}, {Shappee}, {Stanek}, {Holoien}, {Thompson}, {Prieto}, {Dong}, {Shields}, {Will}, {Britt}, {Perzanowski}, \& {Pojma{\'n}ski}}]{Kochanek2017}
{Kochanek}, C.~S., {Shappee}, B.~J., {Stanek}, K.~Z., {et~al.} 2017, \pasp, 129, 104502, \dodoi{10.1088/1538-3873/aa80d9}

\bibitem[{Kovács {et~al.}(2002)Kovács, Zucker, \& Mazeh}]{bls}
Kovács, G., Zucker, S., \& Mazeh, T. 2002, Astronomy \&; Astrophysics, 391, 369–377, \dodoi{10.1051/0004-6361:20020802}

\bibitem[{Kreidberg(2015)}]{batman}
Kreidberg, L. 2015, Publications of the Astronomical Society of the Pacific, 127, 1161

\bibitem[{Kumar {et~al.}(2019)Kumar, Carroll, Hartikainen, \& Martin}]{exoplanet:arviz}
Kumar, R., Carroll, C., Hartikainen, A., \& Martin, O.~A. 2019, The Journal of Open Source Software, \dodoi{10.21105/joss.01143}

\bibitem[{{Kuzuhara} {et~al.}(2024){Kuzuhara}, {Fukui}, {Livingston}, {Caballero}, {de Leon}, {Hirano}, {Kasagi}, {Murgas}, {Narita}, {Omiya}, {Orell-Miquel}, {Palle}, {Changeat}, {Esparza-Borges}, {Harakawa}, {Hellier}, {Hori}, {Ikuta}, {Ishikawa}, {Kodama}, {Kotani}, {Kudo}, {Morales}, {Mori}, {Nagel}, {Parviainen}, {Perdelwitz}, {Reiners}, {Ribas}, {Sanz-Forcada}, {Sato}, {Schweitzer}, {Tabernero}, {Takarada}, {Uyama}, {Watanabe}, {Zechmeister}, {Garc{\'\i}a}, {Aoki}, {Beichman}, {B{\'e}jar}, {Brandt}, {Calatayud-Borras}, {Carleo}, {Charbonneau}, {Collins}, {Currie}, {Doty}, {Dreizler}, {Fern{\'a}ndez-Rodr{\'\i}guez}, {Fukuda}, {Gal{\'a}n}, {Gerald{\'\i}a-Gonz{\'a}lez}, {Gonz{\'a}lez-Rodr{\'\i}guez}, {Hayashi}, {Hedges}, {Henning}, {Hodapp}, {Ikoma}, {Isogai}, {Jacobson}, {Janson}, {Jenkins}, {Kagetani}, {Kambe}, {Kawai}, {Kawauchi}, {Kokubo}, {Konishi}, {Korth}, {Krishnamurthy}, {Kurokawa}, {Kusakabe}, {Kwon}, {Laza-Ramos}, {Libotte}, {Luque}, {Madrigal-Aguado}, {Matsumoto}, {Mawet}, {McElwain}, {Meni
  Gallardo}, {Morello}, {Mu{\~n}oz Torres}, {Nishikawa}, {Nugroho}, {Ogihara}, {Pel{\'a}ez-Torres}, {Rapetti}, {S{\'a}nchez-Benavente}, {Schlecker}, {Seager}, {Serabyn}, {Serizawa}, {Stangret}, {Takahashi}, {Teng}, {Tamura}, {Terada}, {Ueda}, {Usuda}, {Vanderspek}, {Vievard}, {Watanabe}, {Winn}, \& {Zapatero Osorio}}]{Kuzuhara2024}
{Kuzuhara}, M., {Fukui}, A., {Livingston}, J.~H., {et~al.} 2024, \apjl, 967, L21, \dodoi{10.3847/2041-8213/ad3642}

\bibitem[{Leleu {et~al.}(2021)Leleu, Chatel, Udry, Alibert, Delisle, \& Mardling}]{Leleu_2021}
Leleu, A., Chatel, G., Udry, S., {et~al.} 2021, Astronomy \&; Astrophysics, 655, A66, \dodoi{10.1051/0004-6361/202141471}

\bibitem[{{Leleu} {et~al.}(2021){Leleu}, {Alibert}, {Hara}, {Hooton}, {Wilson}, {Robutel}, {Delisle}, {Laskar}, {Hoyer}, {Lovis}, {Bryant}, {Ducrot}, {Cabrera}, {Delrez}, {Acton}, {Adibekyan}, {Allart}, {Allende Prieto}, {Alonso}, {Alves}, {Anderson}, {Angerhausen}, {Anglada Escud{\'e}}, {Asquier}, {Barrado}, {Barros}, {Baumjohann}, {Bayliss}, {Beck}, {Beck}, {Bekkelien}, {Benz}, {Billot}, {Bonfanti}, {Bonfils}, {Bouchy}, {Bourrier}, {Bou{\'e}}, {Brandeker}, {Broeg}, {Buder}, {Burdanov}, {Burleigh}, {B{\'a}rczy}, {Cameron}, {Chamberlain}, {Charnoz}, {Cooke}, {Corral Van Damme}, {Correia}, {Cristiani}, {Damasso}, {Davies}, {Deleuil}, {Demangeon}, {Demory}, {Di Marcantonio}, {Di Persio}, {Dumusque}, {Ehrenreich}, {Erikson}, {Figueira}, {Fortier}, {Fossati}, {Fridlund}, {Futyan}, {Gandolfi}, {Garc{\'\i}a Mu{\~n}oz}, {Garcia}, {Gill}, {Gillen}, {Gillon}, {Goad}, {Gonz{\'a}lez Hern{\'a}ndez}, {Guedel}, {G{\"u}nther}, {Haldemann}, {Henderson}, {Heng}, {Hogan}, {Isaak}, {Jehin}, {Jenkins}, {Jord{\'a}n}, {Kiss},
  {Kristiansen}, {Lam}, {Lavie}, {Lecavelier des Etangs}, {Lendl}, {Lillo-Box}, {Lo Curto}, {Magrin}, {Martins}, {Maxted}, {McCormac}, {Mehner}, {Micela}, {Molaro}, {Moyano}, {Murray}, {Nascimbeni}, {Nunes}, {Olofsson}, {Osborn}, {Oshagh}, {Ottensamer}, {Pagano}, {Pall{\'e}}, {Pedersen}, {Pepe}, {Persson}, {Peter}, {Piotto}, {Polenta}, {Pollacco}, {Poretti}, {Pozuelos}, {Queloz}, {Ragazzoni}, {Rando}, {Ratti}, {Rauer}, {Raynard}, {Rebolo}, {Reimers}, {Ribas}, {Santos}, {Scandariato}, {Schneider}, {Sebastian}, {Sestovic}, {Simon}, {Smith}, {Sousa}, {Sozzetti}, {Steller}, {Su{\'a}rez Mascare{\~n}o}, {Szab{\'o}}, {S{\'e}gransan}, {Thomas}, {Thompson}, {Tilbrook}, {Triaud}, {Turner}, {Udry}, {Van Grootel}, {Venus}, {Verrecchia}, {Vines}, {Walton}, {West}, {Wheatley}, {Wolter}, \& {Zapatero Osorio}}]{Leleu2021}
{Leleu}, A., {Alibert}, Y., {Hara}, N.~C., {et~al.} 2021, \aap, 649, A26, \dodoi{10.1051/0004-6361/202039767}

\bibitem[{{Leleu} {et~al.}(2024){Leleu}, {Delisle}, {Burn}, {Izidoro}, {Udry}, {Dumusque}, {Lovis}, {Millholland}, {Parc}, {Bouchy}, {Bourrier}, {Alibert}, {Faria}, {Mordasini}, \& {S{\'e}gransan}}]{Leleu2024}
{Leleu}, A., {Delisle}, J.-B., {Burn}, R., {et~al.} 2024, \aap, 687, L1, \dodoi{10.1051/0004-6361/202450587}

\bibitem[{Li {et~al.}(2024)Li, Chiang, Choksi, \& Dai}]{Li2024}
Li, R., Chiang, E., Choksi, N., \& Dai, F. 2024, The Resonant Remains of Broken Chains from Major and Minor Mergers.
\newblock \doarXiv{2408.10206}

\bibitem[{{Lightkurve Collaboration} {et~al.}(2018){Lightkurve Collaboration}, {Cardoso}, {Hedges}, {Gully-Santiago}, {Saunders}, {Cody}, {Barclay}, {Hall}, {Sagear}, {Turtelboom}, {Zhang}, {Tzanidakis}, {Mighell}, {Coughlin}, {Bell}, {Berta-Thompson}, {Williams}, {Dotson}, \& {Barentsen}}]{lightkurve}
{Lightkurve Collaboration}, {Cardoso}, J.~V.~d.~M., {Hedges}, C., {et~al.} 2018, {Lightkurve: Kepler and TESS time series analysis in Python}, Astrophysics Source Code Library.
\newblock \doeprint{1812.013}

\bibitem[{{Linssen} {et~al.}(2024){Linssen}, {Shih}, {MacLeod}, \& {Oklop{\v{c}}i{\'c}}}]{Linssen2024}
{Linssen}, D., {Shih}, J., {MacLeod}, M., \& {Oklop{\v{c}}i{\'c}}, A. 2024, \aap, 688, A43, \dodoi{10.1051/0004-6361/202450240}

\bibitem[{{Lissauer} {et~al.}(2011){Lissauer}, {Ragozzine}, {Fabrycky}, {Steffen}, {Ford}, {Jenkins}, {Shporer}, {Holman}, {Rowe}, {Quintana}, {Batalha}, {Borucki}, {Bryson}, {Caldwell}, {Carter}, {Ciardi}, {Dunham}, {Fortney}, {Gautier}, {Howell}, {Koch}, {Latham}, {Marcy}, {Morehead}, \& {Sasselov}}]{mutual_incs}
{Lissauer}, J.~J., {Ragozzine}, D., {Fabrycky}, D.~C., {et~al.} 2011, \apjs, 197, 8, \dodoi{10.1088/0067-0049/197/1/8}

\bibitem[{Lithwick {et~al.}(2012)Lithwick, Xie, \& Wu}]{Lithwick_2012}
Lithwick, Y., Xie, J., \& Wu, Y. 2012, The Astrophysical Journal, 761, 122, \dodoi{10.1088/0004-637x/761/2/122}

\bibitem[{{Lodders}(2003)}]{Lodders2003}
{Lodders}, K. 2003, \apj, 591, 1220, \dodoi{10.1086/375492}

\bibitem[{{Louden} {et~al.}(2023){Louden}, {Laughlin}, \& {Millholland}}]{Louden2023}
{Louden}, E.~M., {Laughlin}, G.~P., \& {Millholland}, S.~C. 2023, \apjl, 958, L21, \dodoi{10.3847/2041-8213/ad0843}

\bibitem[{Lu {et~al.}(2023)Lu, Rein, Tamayo, Hadden, Mardling, Millholland, \& Laughlin}]{Lu_2023}
Lu, T., Rein, H., Tamayo, D., {et~al.} 2023, The Astrophysical Journal, 948, 41, \dodoi{10.3847/1538-4357/acc06d}

\bibitem[{{Luger} {et~al.}(2019){Luger}, {Agol}, {Foreman-Mackey}, {Fleming}, {Lustig-Yaeger}, \& {Deitrick}}]{exoplanet:luger18}
{Luger}, R., {Agol}, E., {Foreman-Mackey}, D., {et~al.} 2019, \aj, 157, 64, \dodoi{10.3847/1538-3881/aae8e5}

\bibitem[{{Luque} \& {Pall{\'e}}(2022)}]{Luque_2022}
{Luque}, R., \& {Pall{\'e}}, E. 2022, Science, 377, 1211, \dodoi{10.1126/science.abl7164}

\bibitem[{MacDonald {et~al.}(2016)MacDonald, Ragozzine, Fabrycky, Ford, Holman, Isaacson, Lissauer, Lopez, Mazeh, Rogers, Rowe, Steffen, \& Torres}]{MacDonald16}
MacDonald, M.~G., Ragozzine, D., Fabrycky, D.~C., {et~al.} 2016, The Astronomical Journal, 152, 105, \dodoi{10.3847/0004-6256/152/4/105}

\bibitem[{{MacKenzie} {et~al.}(2023){MacKenzie}, {Grenfell}, {Baumeister}, {Tosi}, {Cabrera}, \& {Rauer}}]{MacKenzie2023}
{MacKenzie}, J., {Grenfell}, J.~L., {Baumeister}, P., {et~al.} 2023, \aap, 671, A65, \dodoi{10.1051/0004-6361/202141784}

\bibitem[{{Madhusudhan} {et~al.}(2023){Madhusudhan}, {Sarkar}, {Constantinou}, {Holmberg}, {Piette}, \& {Moses}}]{Madhusudhan2023}
{Madhusudhan}, N., {Sarkar}, S., {Constantinou}, S., {et~al.} 2023, \apjl, 956, L13, \dodoi{10.3847/2041-8213/acf577}

\bibitem[{{Maldonado} {et~al.}(2015){Maldonado}, {Affer}, {Micela}, {Scandariato}, {Damasso}, {Stelzer}, {Barbieri}, {Bedin}, {Biazzo}, {Bignamini}, {Borsa}, {Claudi}, {Covino}, {Desidera}, {Esposito}, {Gratton}, {Gonz{\'a}lez Hern{\'a}ndez}, {Lanza}, {Maggio}, {Molinari}, {Pagano}, {Perger}, {Pillitteri}, {Piotto}, {Poretti}, {Prisinzano}, {Rebolo}, {Ribas}, {Shkolnik}, {Southworth}, {Sozzetti}, \& {Su{\'a}rez Mascare{\~n}o}}]{Maldonado2015}
{Maldonado}, J., {Affer}, L., {Micela}, G., {et~al.} 2015, \aap, 577, A132, \dodoi{10.1051/0004-6361/201525797}

\bibitem[{{Masci} {et~al.}(2019){Masci}, {Laher}, {Rusholme}, {Shupe}, {Groom}, {Surace}, {Jackson}, {Monkewitz}, {Beck}, {Flynn}, {Terek}, {Landry}, {Hacopians}, {Desai}, {Howell}, {Brooke}, {Imel}, {Wachter}, {Ye}, {Lin}, {Cenko}, {Cunningham}, {Rebbapragada}, {Bue}, {Miller}, {Mahabal}, {Bellm}, {Patterson}, {Juri{\'c}}, {Golkhou}, {Ofek}, {Walters}, {Graham}, {Kasliwal}, {Dekany}, {Kupfer}, {Burdge}, {Cannella}, {Barlow}, {Van Sistine}, {Giomi}, {Fremling}, {Blagorodnova}, {Levitan}, {Riddle}, {Smith}, {Helou}, {Prince}, \& {Kulkarni}}]{Masci2019}
{Masci}, F.~J., {Laher}, R.~R., {Rusholme}, B., {et~al.} 2019, \pasp, 131, 018003, \dodoi{10.1088/1538-3873/aae8ac}

\bibitem[{{McCully} {et~al.}(2018){McCully}, {Volgenau}, {Harbeck}, {Lister}, {Saunders}, {Turner}, {Siiverd}, \& {Bowman}}]{McCully:2018}
{McCully}, C., {Volgenau}, N.~H., {Harbeck}, D.-R., {et~al.} 2018, in Society of Photo-Optical Instrumentation Engineers (SPIE) Conference Series, Vol. 10707, \procspie, 107070K, \dodoi{10.1117/12.2314340}

\bibitem[{Millholland {et~al.}(2017)Millholland, Wang, \& Laughlin}]{Millholland_2017}
Millholland, S., Wang, S., \& Laughlin, G. 2017, The Astrophysical Journal Letters, 849, L33, \dodoi{10.3847/2041-8213/aa9714}

\bibitem[{{Millholland} {et~al.}(2022){Millholland}, {He}, \& {Zink}}]{Millholland2022}
{Millholland}, S.~C., {He}, M.~Y., \& {Zink}, J.~K. 2022, \aj, 164, 72, \dodoi{10.3847/1538-3881/ac7c67}

\bibitem[{{Millholland} {et~al.}(2024){Millholland}, {Lara}, \& {Toomlaid}}]{Millholland2024}
{Millholland}, S.~C., {Lara}, T., \& {Toomlaid}, J. 2024, \apj, 961, 203, \dodoi{10.3847/1538-4357/ad10a0}

\bibitem[{{Mills} {et~al.}(2016){Mills}, {Fabrycky}, {Migaszewski}, {Ford}, {Petigura}, \& {Isaacson}}]{Mills16}
{Mills}, S.~M., {Fabrycky}, D.~C., {Migaszewski}, C., {et~al.} 2016, \nat, 533, 509, \dodoi{10.1038/nature17445}

\bibitem[{{Mortier} {et~al.}(2020){Mortier}, {Zapatero Osorio}, {Malavolta}, {Alibert}, {Rice}, {Lillo-Box}, {Vanderburg}, {Oshagh}, {Buchhave}, {Adibekyan}, {Delgado Mena}, {Lopez-Morales}, {Charbonneau}, {Sousa}, {Lovis}, {Affer}, {Allende Prieto}, {Barros}, {Benatti}, {Bonomo}, {Boschin}, {Bouchy}, {Cabral}, {Collier Cameron}, {Cosentino}, {Cristiani}, {Demangeon}, {Di Marcantonio}, {D'Odorico}, {Dumusque}, {Ehrenreich}, {Figueira}, {Fiorenzano}, {Ghedina}, {Gonz{\'a}lez Hern{\'a}ndez}, {Haldemann}, {Harutyunyan}, {Haywood}, {Latham}, {Lavie}, {Lo Curto}, {Maldonado}, {Manescau}, {Martins}, {Mayor}, {M{\'e}gevand}, {Mehner}, {Micela}, {Molaro}, {Molinari}, {Nunes}, {Pepe}, {Palle}, {Phillips}, {Piotto}, {Pinamonti}, {Poretti}, {Riva}, {Rebolo}, {Santos}, {Sasselov}, {Sozzetti}, {Su{\'a}rez Mascare{\~n}o}, {Udry}, {West}, {Watson}, \& {Wilson}}]{Mortier2020}
{Mortier}, A., {Zapatero Osorio}, M.~R., {Malavolta}, L., {et~al.} 2020, \mnras, 499, 5004, \dodoi{10.1093/mnras/staa3144}

\bibitem[{{Murgas} {et~al.}(2024){Murgas}, {Pall{\'e}}, {Orell-Miquel}, {Carleo}, {Pe{\~n}a-Mo{\~n}ino}, {P{\'e}rez-Torres}, {Watkins}, {Jeffers}, {Azzaro}, {Barkaoui}, {Belinski}, {Caballero}, {Charbonneau}, {Cheryasov}, {Ciardi}, {Collins}, {Cort{\'e}s-Contreras}, {de Leon}, {Duque-Arribas}, {Enoc}, {Esparza-Borges}, {Fukui}, {Gerald{\'\i}a-Gonz{\'a}lez}, {Gilbert}, {Hatzes}, {Hayashi}, {Henning}, {Herrero}, {Jenkins}, {Lillo-Box}, {Lodieu}, {Lund}, {Luque}, {Montes}, {Nagel}, {Narita}, {Parviainen}, {Polanski}, {Reffert}, {Schlecker}, {Sch{\"o}fer}, {Schwarz}, {Schweitzer}, {Seager}, {Stassun}, {Tabernero}, {Terada}, {Twicken}, {Vanaverbeke}, {Winn}, {Zambelli}, {Amado}, {Quirrenbach}, {Reiners}, \& {Ribas}}]{Murgas2024}
{Murgas}, F., {Pall{\'e}}, E., {Orell-Miquel}, J., {et~al.} 2024, \aap, 684, A83, \dodoi{10.1051/0004-6361/202348813}

\bibitem[{{Narita} {et~al.}(2015){Narita}, {Fukui}, {Kusakabe}, {Onitsuka}, {Ryu}, {Yanagisawa}, {Izumiura}, {Tamura}, \& {Yamamuro}}]{2015JATIS...1d5001N}
{Narita}, N., {Fukui}, A., {Kusakabe}, N., {et~al.} 2015, Journal of Astronomical Telescopes, Instruments, and Systems, 1, 045001, \dodoi{10.1117/1.JATIS.1.4.045001}

\bibitem[{{Narita} {et~al.}(2020){Narita}, {Fukui}, {Yamamuro}, {Harbeck}, {Bowman}, {Elphick}, {Nation}, {Armstrong}, {Han}, {Abe}, {Ikoma}, {Isogai}, {Kawauchi}, {Kurita}, {Kusakabe}, {de Leon}, {Livingston}, {Mori}, {Nishiumi}, {Tamura}, {Watanabe}, {Volgenau}, {Heinrich-Josties}, {Foale}, {Daily}, {McCully}, {Kirby}, {Smith}, {Haworth}, {Conway}, {Storrie-Lombardi}, {Rosing}, {Chatelain}, {Bachelet}, {Johnson}, \& {Rabus}}]{2020SPIE11447E..5KN}
{Narita}, N., {Fukui}, A., {Yamamuro}, T., {et~al.} 2020, in Society of Photo-Optical Instrumentation Engineers (SPIE) Conference Series, Vol. 11447, Society of Photo-Optical Instrumentation Engineers (SPIE) Conference Series, 114475K, \dodoi{10.1117/12.2559947}

\bibitem[{Nelson {et~al.}(2013)Nelson, Ford, \& Payne}]{DEMove}
Nelson, B., Ford, E.~B., \& Payne, M.~J. 2013, The Astrophysical Journal Supplement Series, 210, 11, \dodoi{10.1088/0067-0049/210/1/11}

\bibitem[{{Nesvorn{\'y}} \& {Vokrouhlick{\'y}}(2014{\natexlab{a}})}]{nesvorny_2014}
{Nesvorn{\'y}}, D., \& {Vokrouhlick{\'y}}, D. 2014{\natexlab{a}}, \apj, 790, 58, \dodoi{10.1088/0004-637X/790/1/58}

\bibitem[{{Nesvorn{\'y}} \& {Vokrouhlick{\'y}}(2014{\natexlab{b}})}]{Nesvorny2014}
---. 2014{\natexlab{b}}, \apj, 790, 58, \dodoi{10.1088/0004-637X/790/1/58}

\bibitem[{{Nettelmann} {et~al.}(2011){Nettelmann}, {Fortney}, {Kramm}, \& {Redmer}}]{Nettelmann2011}
{Nettelmann}, N., {Fortney}, J.~J., {Kramm}, U., \& {Redmer}, R. 2011, \apj, 733, 2, \dodoi{10.1088/0004-637X/733/1/2}

\bibitem[{{Ormel} {et~al.}(2017){Ormel}, {Liu}, \& {Schoonenberg}}]{ormel_2017}
{Ormel}, C.~W., {Liu}, B., \& {Schoonenberg}, D. 2017, \aap, 604, A1, \dodoi{10.1051/0004-6361/201730826}

\bibitem[{{Otegi} {et~al.}(2022){Otegi}, {Helled}, \& {Bouchy}}]{Otegi2022}
{Otegi}, J.~F., {Helled}, R., \& {Bouchy}, F. 2022, \aap, 658, A107, \dodoi{10.1051/0004-6361/202142110}

\bibitem[{{Parc} {et~al.}(2024){Parc}, {Bouchy}, {Venturini}, {Dorn}, \& {Helled}}]{Parc2024}
{Parc}, L., {Bouchy}, F., {Venturini}, J., {Dorn}, C., \& {Helled}, R. 2024, \aap, 688, A59, \dodoi{10.1051/0004-6361/202449911}

\bibitem[{Parviainen \& Aigrain(2015)}]{Parviainen_2015}
Parviainen, H., \& Aigrain, S. 2015, Monthly Notices of the Royal Astronomical Society, 453, 3822–3827, \dodoi{10.1093/mnras/stv1857}

\bibitem[{{P{\'e}rez-Gonz{\'a}lez} {et~al.}(2024){P{\'e}rez-Gonz{\'a}lez}, {Greklek-McKeon}, {Vissapragada}, {Saidel}, {Knutson}, {Linssen}, \& {Oklop{\v{c}}i{\'c}}}]{Jorge2024}
{P{\'e}rez-Gonz{\'a}lez}, J., {Greklek-McKeon}, M., {Vissapragada}, S., {et~al.} 2024, \aj, 167, 214, \dodoi{10.3847/1538-3881/ad34b6}

\bibitem[{{Piaulet} {et~al.}(2023){Piaulet}, {Benneke}, {Almenara}, {Dragomir}, {Knutson}, {Thorngren}, {Peterson}, {Crossfield}, {Kempton}, {Kubyshkina}, {Howard}, {Angus}, {Isaacson}, {Weiss}, {Beichman}, {Fortney}, {Fossati}, {Lammer}, {McCullough}, {Morley}, \& {Wong}}]{Piaulet2023}
{Piaulet}, C., {Benneke}, B., {Almenara}, J.~M., {et~al.} 2023, Nature Astronomy, 7, 206, \dodoi{10.1038/s41550-022-01835-4}

\bibitem[{{Piaulet-Ghorayeb} {et~al.}(2024){Piaulet-Ghorayeb}, {Benneke}, {Radica}, {Raul}, {Coulombe}, {Ahrer}, {Kubyshkina}, {Howard}, {Krissansen-Totton}, {MacDonald}, {Roy}, {Louca}, {Christie}, {Fournier-Tondreau}, {Allart}, {Miguel}, {Schlichting}, {Welbanks}, {Cadieux}, {Dorn}, {Evans-Soma}, {Fortney}, {Pierrehumbert}, {Lafreni{\`e}re}, {Acu{\~n}a}, {Komacek}, {Innes}, {Beatty}, {Cloutier}, {Doyon}, {Gagnebin}, {Gapp}, \& {Knutson}}]{PiauletGhorayeb2024}
{Piaulet-Ghorayeb}, C., {Benneke}, B., {Radica}, M., {et~al.} 2024, \apjl, 974, L10, \dodoi{10.3847/2041-8213/ad6f00}

\bibitem[{{Pierrehumbert}(2023)}]{Pierrehumbert2023}
{Pierrehumbert}, R.~T. 2023, \apj, 944, 20, \dodoi{10.3847/1538-4357/acafdf}

\bibitem[{{Plotnykov} \& {Valencia}(2020)}]{Plotnykov2020}
{Plotnykov}, M., \& {Valencia}, D. 2020, \mnras, 499, 932, \dodoi{10.1093/mnras/staa2615}

\bibitem[{{Raftery}(1995)}]{raftery1995}
{Raftery}, A.~E. 1995, JSTOR, 25, 111–63, \dodoi{10.2307/271063}

\bibitem[{Rein \& Tamayo(2015)}]{rebound}
Rein, H., \& Tamayo, D. 2015, Monthly Notices of the Royal Astronomical Society, 452, 376–388, \dodoi{10.1093/mnras/stv1257}

\bibitem[{{Rice} {et~al.}(2024){Rice}, {Steffen}, \& {Vazan}}]{Rice2024}
{Rice}, D.~R., {Steffen}, J.~H., \& {Vazan}, A. 2024, arXiv e-prints, arXiv:2406.12239, \dodoi{10.48550/arXiv.2406.12239}

\bibitem[{Ricker {et~al.}(2014)Ricker, Winn, Vanderspek, Latham, Bakos, Bean, Berta-Thompson, Brown, Buchhave, Butler, \& et~al.}]{Ricker2014}
Ricker, G.~R., Winn, J.~N., Vanderspek, R., {et~al.} 2014, Journal of Astronomical Telescopes, Instruments, and Systems, 1, 014003, \dodoi{10.1117/1.jatis.1.1.014003}

\bibitem[{{Roettenbacher} \& {Kane}(2017)}]{Roettenbacher_2017}
{Roettenbacher}, R.~M., \& {Kane}, S.~R. 2017, \apj, 851, 77, \dodoi{10.3847/1538-4357/aa991e}

\bibitem[{{Rogers} {et~al.}(2025){Rogers}, {Dorn}, {Aditya Raj}, {Schlichting}, \& {Young}}]{Rogers2025}
{Rogers}, J.~G., {Dorn}, C., {Aditya Raj}, V., {Schlichting}, H.~E., \& {Young}, E.~D. 2025, \apj, 979, 79, \dodoi{10.3847/1538-4357/ad9f61}

\bibitem[{{Rogers} {et~al.}(2023){Rogers}, {Schlichting}, \& {Owen}}]{rogers_2023}
{Rogers}, J.~G., {Schlichting}, H.~E., \& {Owen}, J.~E. 2023, \apjl, 947, L19, \dodoi{10.3847/2041-8213/acc86f}

\bibitem[{{Ros{\'a}rio} {et~al.}(2024){Ros{\'a}rio}, {Demangeon}, {Barros}, {Gandolfi}, {Egger}, {Serrano}, {Osborn}, {Beck}, {Benz}, {Flor{\'e}n}, {Guterman}, {Wilson}, {Alibert}, {Fossati}, {Hooton}, {Delrez}, {Santos}, {Sousa}, {Bonfanti}, {Salmon}, {Adibekyan}, {Nigioni}, {Venturini}, {Alonso}, {Anglada}, {Asquier}, {B{\'a}rczy}, {Barrado Navascues}, {Barrag{\'a}n}, {Baumjohann}, {Beck}, {Billot}, {Biondi}, {Bonfils}, {Borsato}, {Brandeker}, {Broeg}, {Cessa}, {Charnoz}, {Collier Cameron}, {Csizmadia}, {Cubillos}, {Davies}, {Deleuil}, {Deline}, {Demory}, {Ehrenreich}, {Erikson}, {Esposito}, {Fortier}, {Fridlund}, {Gillon}, {G{\"u}del}, {G{\"u}nther}, {Helling}, {Hoyer}, {Isaak}, {Kiss}, {Lam}, {Laskar}, {Lecavelier des Etangs}, {Lendl}, {Luntzer}, {Magrin}, {Maxted}, {Mordasini}, {Nascimbeni}, {Olofsson}, {Osborne}, {Ottensamer}, {Pagano}, {Pall{\'e}}, {Peter}, {Piotto}, {Pollacco}, {Queloz}, {Ragazzoni}, {Rando}, {Rauer}, {Ribas}, {Scandariato}, {S{\'e}gransan}, {Simon}, {Smith}, {Stalport}, {Szab{\'o}},
  {Thomas}, {Udry}, {Van Eylen}, {Van Grootel}, {Villaver}, {Walter}, \& {Walton}}]{Rosario2024}
{Ros{\'a}rio}, N.~M., {Demangeon}, O.~D.~S., {Barros}, S.~C.~C., {et~al.} 2024, \aap, 686, A282, \dodoi{10.1051/0004-6361/202347759}

\bibitem[{Salvatier {et~al.}(2016)Salvatier, Wiecki, \& Fonnesbeck}]{exoplanet:pymc3}
Salvatier, J., Wiecki, T.~V., \& Fonnesbeck, C. 2016, PeerJ Computer Science, 2, e55

\bibitem[{{Schwarz}(1978)}]{Schwarz1978}
{Schwarz}, G. 1978, Annals of Statistics, 6, 461

\bibitem[{SciPy(2001)}]{scipy}
SciPy. 2001, {SciPy}: Open source scientific tools for Python.
\newblock \url{http://www.scipy.org/}

\bibitem[{Seligman {et~al.}(2023)Seligman, Feinstein, Lai, Welbanks, Taylor, Becker, Adams, Morgan, \& Bergner}]{Seligman2023}
Seligman, D.~Z., Feinstein, A.~D., Lai, D., {et~al.} 2023, Potential Melting of Extrasolar Planets by Tidal Dissipation.
\newblock \doarXiv{2311.01187}

\bibitem[{{Stassun} {et~al.}(2017){Stassun}, {Collins}, \& {Gaudi}}]{Stassun2017}
{Stassun}, K.~G., {Collins}, K.~A., \& {Gaudi}, B.~S. 2017, \aj, 153, 136, \dodoi{10.3847/1538-3881/aa5df3}

\bibitem[{{Stassun} \& {Torres}(2016)}]{Stassun2016}
{Stassun}, K.~G., \& {Torres}, G. 2016, \aj, 152, 180, \dodoi{10.3847/0004-6256/152/6/180}

\bibitem[{{Stassun} {et~al.}(2018){Stassun}, {Oelkers}, {Pepper}, {Paegert}, {De Lee}, {Torres}, {Latham}, {Charpinet}, {Dressing}, {Huber}, {Kane}, {L{\'e}pine}, {Mann}, {Muirhead}, {Rojas-Ayala}, {Silvotti}, {Fleming}, {Levine}, \& {Plavchan}}]{Stassun2018}
{Stassun}, K.~G., {Oelkers}, R.~J., {Pepper}, J., {et~al.} 2018, \aj, 156, 102, \dodoi{10.3847/1538-3881/aad050}

\bibitem[{{Stassun} {et~al.}(2019){Stassun}, {Oelkers}, {Paegert}, {Torres}, {Pepper}, {De Lee}, {Collins}, {Latham}, {Muirhead}, {Chittidi}, {Rojas-Ayala}, {Fleming}, {Rose}, {Tenenbaum}, {Ting}, {Kane}, {Barclay}, {Bean}, {Brassuer}, {Charbonneau}, {Ge}, {Lissauer}, {Mann}, {McLean}, {Mullally}, {Narita}, {Plavchan}, {Ricker}, {Sasselov}, {Seager}, {Sharma}, {Shiao}, {Sozzetti}, {Stello}, {Vanderspek}, {Wallace}, \& {Winn}}]{Stassun_2019}
{Stassun}, K.~G., {Oelkers}, R.~J., {Paegert}, M., {et~al.} 2019, \aj, 158, 138, \dodoi{10.3847/1538-3881/ab3467}

\bibitem[{{Stefansson} {et~al.}(2017){Stefansson}, {Mahadevan}, {Hebb}, {Wisniewski}, {Huehnerhoff}, {Morris}, {Halverson}, {Zhao}, {Wright}, {O'rourke}, {Knutson}, {Hawley}, {Kanodia}, {Li}, {Hagen}, {Liu}, {Beatty}, {Bender}, {Robertson}, {Dembicky}, {Gray}, {Ketzeback}, {McMillan}, \& {Rudyk}}]{Stefansson2017}
{Stefansson}, G., {Mahadevan}, S., {Hebb}, L., {et~al.} 2017, \apj, 848, 9, \dodoi{10.3847/1538-4357/aa88aa}

\bibitem[{{Stef{\'a}nsson} {et~al.}(2020){Stef{\'a}nsson}, {Kopparapu}, {Lin}, {Mahadevan}, {Ca{\~n}as}, {Kanodia}, {Ninan}, {Cochran}, {Endl}, {Hebb}, {Wisniewski}, {Gupta}, {Everett}, {Bender}, {Diddams}, {Ford}, {Fredrick}, {Halverson}, {Hearty}, {Levi}, {Maney}, {Metcalf}, {Monson}, {Ramsey}, {Robertson}, {Roy}, {Schwab}, {Terrien}, \& {Wright}}]{Stefansson_2020}
{Stef{\'a}nsson}, G., {Kopparapu}, R., {Lin}, A., {et~al.} 2020, \aj, 160, 259, \dodoi{10.3847/1538-3881/abbe19}

\bibitem[{{Stefansson} {et~al.}(2020){Stefansson}, {Ca{\~n}as}, {Wisniewski}, {Robertson}, {Mahadevan}, {Maney}, {Kanodia}, {Beard}, {Bender}, {Brunt}, {Clemens}, {Cochran}, {Diddams}, {Endl}, {Ford}, {Fredrick}, {Halverson}, {Hearty}, {Hebb}, {Huehnerhoff}, {Jennings}, {Kaplan}, {Levi}, {Lubar}, {Metcalf}, {Monson}, {Morris}, {Ninan}, {Nitroy}, {Ramsey}, {Roy}, {Schwab}, {Sigurdsson}, {Terrien}, \& {Wright}}]{Stefansson2020a}
{Stefansson}, G., {Ca{\~n}as}, C., {Wisniewski}, J., {et~al.} 2020, \aj, 159, 100, \dodoi{10.3847/1538-3881/ab5f15}

\bibitem[{{Su{\'a}rez Mascare{\~n}o} {et~al.}(2024){Su{\'a}rez Mascare{\~n}o}, {Passegger}, {Gonz{\'a}lez Hern{\'a}ndez}, {Armstrong}, {Nielsen}, {Lovis}, {Lavie}, {Sousa}, {Silva}, {Allart}, {Rebolo}, {Pepe}, {Santos}, {Cristiani}, {Sozzetti}, {Zapatero Osorio}, {Tabernero}, {Dumusque}, {Udry}, {Adibekyan}, {Allende Prieto}, {Alibert}, {Barros}, {Bouchy}, {Castro-Gonz{\'a}lez}, {Collins}, {Damasso}, {D'Odorico}, {Demangeon}, {Di Marcantonio}, {Ehrenreich}, {Hadjigeorghiou}, {Hara}, {Hawthorn}, {Jenkins}, {Lillo-Box}, {Lo Curto}, {Martins}, {Mehner}, {Micela}, {Molaro}, {Nunes}, {Nari}, {Osborn}, {Pall{\'e}}, {Ricker}, {Rodrigues}, {Rowden}, {Seager}, {Stefanov}, {Str{\o}m}, {Villase{\~n}or}, {Watkins}, {Winn}, {Wohler}, \& {Zambelli}}]{SuarezMascareno2024}
{Su{\'a}rez Mascare{\~n}o}, A., {Passegger}, V.~M., {Gonz{\'a}lez Hern{\'a}ndez}, J.~I., {et~al.} 2024, \aap, 685, A56, \dodoi{10.1051/0004-6361/202348958}

\bibitem[{{Tamayo} {et~al.}(2020){Tamayo}, {Rein}, {Shi}, \& {Hernandez}}]{Tamayo2020}
{Tamayo}, D., {Rein}, H., {Shi}, P., \& {Hernandez}, D.~M. 2020, \mnras, 491, 2885, \dodoi{10.1093/mnras/stz2870}

\bibitem[{ter Braak \& Vrugt(2008)}]{SnookerMove}
ter Braak, C., \& Vrugt, J. 2008, Stat Comput, 18, 435–446, \dodoi{10.1007/s11222-008-9104-9}

\bibitem[{{Theano Development Team}(2016)}]{exoplanet:theano}
{Theano Development Team}. 2016, arXiv e-prints, abs/1605.02688.
\newblock \url{http://arxiv.org/abs/1605.02688}

\bibitem[{{Tobie} {et~al.}(2019){Tobie}, {Grasset}, {Dumoulin}, \& {Mocquet}}]{Tobie2019}
{Tobie}, G., {Grasset}, O., {Dumoulin}, C., \& {Mocquet}, A. 2019, \aap, 630, A70, \dodoi{10.1051/0004-6361/201935297}

\bibitem[{{Van Eylen} {et~al.}(2019){Van Eylen}, {Albrecht}, {Huang}, {MacDonald}, {Dawson}, {Cai}, {Foreman-Mackey}, {Lundkvist}, {Silva Aguirre}, {Snellen}, \& {Winn}}]{VanEylen19}
{Van Eylen}, V., {Albrecht}, S., {Huang}, X., {et~al.} 2019, \aj, 157, 61, \dodoi{10.3847/1538-3881/aaf22f}

\bibitem[{{Veeder} {et~al.}(2012){Veeder}, {Davies}, {Matson}, {Johnson}, {Williams}, \& {Radebaugh}}]{Veeder2012}
{Veeder}, G.~J., {Davies}, A.~G., {Matson}, D.~L., {et~al.} 2012, \icarus, 219, 701, \dodoi{10.1016/j.icarus.2012.04.004}

\bibitem[{Vissapragada {et~al.}(2020)Vissapragada, Jontof-Hutter, Shporer, Knutson, Liu, Thorngren, Lee, Chachan, Mawet, Millar-Blanchaer, \& et~al.}]{Vissapragada2020}
Vissapragada, S., Jontof-Hutter, D., Shporer, A., {et~al.} 2020, The Astronomical Journal, 159, 108, \dodoi{10.3847/1538-3881/ab65c8}

\bibitem[{{Wallack} {et~al.}(2024){Wallack}, {Batalha}, {Alderson}, {Scarsdale}, {Adams Redai}, {Aguichine}, {Alam}, {Gao}, {Wolfgang}, {Batalha}, {Kirk}, {L{\'o}pez-Morales}, {Moran}, {Teske}, {Wakeford}, \& {Wogan}}]{Wallack2024}
{Wallack}, N.~L., {Batalha}, N.~E., {Alderson}, L., {et~al.} 2024, \aj, 168, 77, \dodoi{10.3847/1538-3881/ad3917}

\bibitem[{{Weiss} {et~al.}(2018){Weiss}, {Marcy}, {Petigura}, {Fulton}, {Howard}, {Winn}, {Isaacson}, {Morton}, {Hirsch}, {Sinukoff}, {Cumming}, {Hebb}, \& {Cargile}}]{Weiss2018}
{Weiss}, L.~M., {Marcy}, G.~W., {Petigura}, E.~A., {et~al.} 2018, \aj, 155, 48, \dodoi{10.3847/1538-3881/aa9ff6}

\bibitem[{{Wilson} {et~al.}(2003){Wilson}, {Eikenberry}, {Henderson}, {Hayward}, {Carson}, {Pirger}, {Barry}, {Brandl}, {Houck}, {Fitzgerald}, \& {Stolberg}}]{Wilson2003}
{Wilson}, J.~C., {Eikenberry}, S.~S., {Henderson}, C.~P., {et~al.} 2003, in Society of Photo-Optical Instrumentation Engineers (SPIE) Conference Series, Vol. 4841, Instrument Design and Performance for Optical/Infrared Ground-based Telescopes, ed. M.~{Iye} \& A.~F.~M. {Moorwood}, 451--458, \dodoi{10.1117/12.460336}

\bibitem[{{Wogan} {et~al.}(2024){Wogan}, {Batalha}, {Zahnle}, {Krissansen-Totton}, {Tsai}, \& {Hu}}]{Wogan2024}
{Wogan}, N.~F., {Batalha}, N.~E., {Zahnle}, K.~J., {et~al.} 2024, \apjl, 963, L7, \dodoi{10.3847/2041-8213/ad2616}

\bibitem[{{Yang} \& {Hu}(2024)}]{Yang2024}
{Yang}, J., \& {Hu}, R. 2024, \apjl, 971, L48, \dodoi{10.3847/2041-8213/ad6b25}

\bibitem[{{Yee} {et~al.}(2017){Yee}, {Petigura}, \& {von Braun}}]{Yee2017}
{Yee}, S.~W., {Petigura}, E.~A., \& {von Braun}, K. 2017, \apj, 836, 77, \dodoi{10.3847/1538-4357/836/1/77}

\bibitem[{{Yoshida} {et~al.}(2022){Yoshida}, {Terada}, {Ikoma}, \& {Kuramoto}}]{Yoshida2022}
{Yoshida}, T., {Terada}, N., {Ikoma}, M., \& {Kuramoto}, K. 2022, \apj, 934, 137, \dodoi{10.3847/1538-4357/ac7be7}

\bibitem[{Zeng {et~al.}(2016)Zeng, Sasselov, \& Jacobsen}]{zeng_2016}
Zeng, L., Sasselov, D.~D., \& Jacobsen, S.~B. 2016, The Astrophysical Journal, 819, 127, \dodoi{10.3847/0004-637x/819/2/127}

\bibitem[{{Zhang} {et~al.}(2022){Zhang}, {Knutson}, {Wang}, {Dai}, \& {Barrag{\'a}n}}]{Zhang2022}
{Zhang}, M., {Knutson}, H.~A., {Wang}, L., {Dai}, F., \& {Barrag{\'a}n}, O. 2022, \aj, 163, 67, \dodoi{10.3847/1538-3881/ac3fa7}

\end{thebibliography}
\bibliographystyle{aasjournal}



\end{document}